\documentclass[11pt]{article}

\usepackage[a4paper,margin=1in]{geometry}
\usepackage[T1]{fontenc}

\usepackage{mathtools} 
\mathtoolsset{showonlyrefs=true}
\usepackage{amssymb}
\usepackage{bm}
\usepackage{dsfont} 
\usepackage{mathrsfs}
\usepackage{upgreek}
\allowdisplaybreaks

\usepackage{algorithm}
\usepackage{algpseudocode}

\usepackage{graphicx}
\usepackage[font={small}]{caption}
\usepackage{subcaption} 
\usepackage{booktabs}
\usepackage[flushleft]{threeparttable}
\usepackage{multirow}
\usepackage{tabularx}
\usepackage{float}
\usepackage{enumitem} 
\usepackage{algorithm}
\usepackage{algpseudocode}

\usepackage[dvipsnames]{xcolor}
\usepackage[normalem]{ulem}
\usepackage{soul}
       
\usepackage{amsthm}
\theoremstyle{plain}
\newtheorem{theorem}{Theorem}

\newtheorem{proposition}{Proposition}
\newtheorem{corollary}[theorem]{Corollary}
\theoremstyle{definition}

\newtheorem{remark}{Remark}
\newtheorem{assumption}{Assumption}

\usepackage[style=apa, natbib, uniquelist=false, backend=biber]{biblatex}
\usepackage{hyperref}
\hypersetup{
    colorlinks   = true,
    citecolor    = RoyalBlue,
    linkcolor    = RubineRed, 
    urlcolor     = RubineRed
}      

\usepackage{todonotes}

\usepackage{appendix}

\DeclareMathOperator*{\argsup}{arg\,sup}

\newcommand\Eb{\mathds{E}}
\newcommand\Fb{\mathds{F}}

\newcommand\Nb{\mathds{N}}

\newcommand\Pb{\mathds{P}}

\newcommand\Rb{\mathds{R}}

\newcommand\Ac{\mathcal{A}}
\newcommand\Bc{\mathcal{B}}
\newcommand\Cc{\mathcal{C}}
\newcommand\Dc{\mathcal{D}}
\newcommand\Ec{\mathcal{E}}
\newcommand\Fc{\mathcal{F}}
\newcommand\Gc{\mathcal{G}}
\newcommand\Hc{\mathcal{H}}
\newcommand\Ic{\mathcal{I}}
\newcommand\Jc{\mathcal{J}}
\newcommand\Kc{\mathcal{K}}
\newcommand\Lc{\mathcal{L}}

\newcommand\Nc{\mathcal{N}}

\newcommand\Pc{\mathcal{P}}

\newcommand\Sc{\mathcal{S}}
\newcommand\Tc{\mathcal{T}}
\newcommand\Uc{\mathcal{U}}
\newcommand\Vc{\mathcal{V}}
\newcommand\Wc{\mathcal{W}}

\newcommand\ind{\mathds{1}}
\newcommand{\ee}{\mathrm{e}}
\newcommand{\dd}{\mathrm{d}}

\newcommand{\balp}{\bm{\alpha}}
\newcommand{\bbet}{\bm{\beta}}
\newcommand{\bgam}{\bm{\gamma}}
\newcommand{\bp}{\bm{p}}

\newcommand{\bc}{\bm{c}}

\newcommand{\tGam}{\widetilde{\Gamma}}

\newcommand{\pbar}{\bar{p}}
\newcommand{\Hbar}{\bar{\Hc}}
\newcommand{\Pbar}{\bar{\Pc}}
\newcommand{\Vbar}{\bar{\Vc}}

\newcommand{\hp}{\widehat{p}}
\newcommand{\hbp}{\widehat{\bm{p}}}
\newcommand{\hgam}{\widehat{\gamma}}
\newcommand{\hbgam}{\widehat{\bm{\gamma}}}
\newcommand{\hGam}{\widehat{\Gamma}}
\newcommand{\hP}{\widehat{\Pc}}
\newcommand{\hV}{\widehat{\Vc}}

\newcommand{\hbgami}{\widehat{\bm{\gamma}}^{i, *}}
\newcommand{\hgamei}{\widehat{\gamma}_e^{i, *}}
\newcommand{\hgameii}{\widehat{\gamma}_e^{i, \mathrm{int}, *}}

\newcommand\EPL{\mathrm{EPL}}
\newcommand\PL{\mathrm{PL}}
\newcommand\DP{\mathrm{DP}}

\newcommand\ATM{\mathrm{ATM}}

\newcommand\SSP{\mathrm{SSP}}

\newcommand\cov{\mathrm{Cov}}

\makeatletter
\renewcommand\footnotesize{%
   \@setfontsize\footnotesize{10pt}{11.5pt}
}

\newcommand{\eqlnostar}[2]{\begin{align}\label{#1}#2\end{align}}
\newcommand{\eqstar}[1]{\begin{align*}#1\end{align*}}
\newcommand{\eq}[1]{\ifthenelse{\equal{#1}{*}}
  {\eqstar}
  {\eqlnostar{#1}}
 }
\makeatother

\makeatletter
\newcommand*{\rom}[1]{\expandafter\@slowromancap\romannumeral #1@}
\makeatother

\title{Modelling and Calibration of Climate-Related Credit Risk Under the Emissions Trading System}
\author{Y.~Liao\thanks{University of Glasgow (\href{mailto:2636546L@student.gla.ac.uk}{2636546L@student.gla.ac.uk})}
\qquad
A.~Agarwal\thanks{Western University (\href{mailto:aagarw93@uwo.ca}{aagarw93@uwo.ca})}
}
\date{}

\begin{document}

\maketitle

\begin{abstract}
Emission trading systems (ETSs) price a growing share of global emissions, and some of their caps are set to reach zero within the maturity of corporate debt. We develop a structural model in which firms choose their emissions from several energy sources under an anticipated carbon price, while the realized carbon price clears the carbon market of each sector-region cell, with a scenario-dependent demand shock and physical losses mitigated by adaptation. We show that the equilibrium carbon price exists, characterize when the cap binds, and derive the optimal emission of each firm in feedback form. In a linearized case of the model, the carbon price is deterministic, and the default probability is explicit. Unlike a reward-penalty scheme, the ETS imposes no cost while the cap is slack and makes firms meet it exactly once it binds. Calibrated to 35 European power producers under the EU ETS, the model reproduces the average level of the European Union Allowance price over 2018 to 2024. 
\end{abstract}

\section{Introduction}
\label{sec:chap3_part1_introduction}

Climate risk, arising from the increasing severity of climate change and the corresponding stricter climate policies, has become an increasingly important source of financial risk for firms and financial institutions. Commonly, climate risk is divided into two main categories: transition risk and physical risk. Transition risk refers to the financial risks that arise from the transition to a low-carbon economy, including policy and regulatory changes, technological advancements, and shifts in market preferences. These risks can lead to significant changes in asset values, increased costs of compliance, and potential stranded assets for firms that are not aligned with the transition \citep[see][for more details]{tcfdRecommendationsTaskForce2017}. Physical risk refers to the economic and financial consequences of extreme weather and climate events, such as tornadoes, tropical cyclones, droughts, and heatwaves, as well as of chronic changes, such as sea level rise. In this paper, we focus on the impact of climate risk on the credit risk of firms, which is a crucial aspect of financial risk management and has significant implications for the stability of financial institutions and the broader economy.

In the transition to a low-carbon economy, carbon pricing plays an increasingly important role, covering around 28\% of global Greenhouse Gas (GHG) emissions in 2025, compared to only 5\% in 2005 \citep{theworldbankStateTrendsCarbon2025}. The three main direct carbon pricing instruments are emission trading systems (ETSs), carbon taxes, and carbon crediting mechanisms. Among them, ETSs have overtaken carbon taxes in terms of global coverage \citep{theworldbankStateTrendsCarbon2025}. Nowaday, 41 ETSs are in force worldwide, covering 26\% of global GHG emissions, from the European Union Emission Trading System (EU ETS) and China's national ETS to the Korean ETS and the systems in North America \citep{ICAP2026}. Unlike a carbon tax, an ETS fixes the quantity of emissions and lets the price adjust.
Under the current legislation of the EU ETS, for instance, the cap is set to reach zero around 2040 \citep{EU2023Directive959}, a date that lies within the maturity of the long-term debt of the covered firms. The effect of carbon pricing on the credit risk of these firms is therefore not a question about the distant future. The answer is, however, less obvious than it may seem. A tighter cap forces a firm to cut its emissions and to pay a higher carbon price, but how much it pays depends on what all the other firms under the same cap emit, and part of the cost may be passed on to its customers through a higher selling price, or recovered through the replacement of the lost production by a cleaner energy mix. Hence, a tighter cap does not translate one-to-one into a higher default probability, and the same transition may raise the credit risk of some firms in a sector while lowering that of others.

The link between climate risk and credit risk has been documented in a growing empirical literature. \citet{boltonInvestorsCareCarbon2021} show that the GHG emissions of firms are priced in stock returns, and \citet{capassoClimateChangeCredit2020} find that firms with higher GHG emissions have a shorter distance to default. At the level of the financial system, \citet{battistonClimateStresstestFinancial2017} show that the exposures of financial institutions to climate-policy-relevant sectors are large, so that the timing of climate policy may have systemic implications. To translate climate scenarios into default probabilities, \citet{bourgeyBridgingSocioeconomicPathways2024,bourgeyEfficientSharedSocioeconomic2026} combine the emission pathways of the Shared Socioeconomic Pathways (SSPs) with a structural model of credit risk \citep{mertonPricingCorporateDebt1974,blackValuingCorporateSecurities1976}. In their framework, the firm chooses its energy usage to maximize its expected discounted profit under a reward-penalty scheme around the emission pathway of the scenario, and defaults when its firm value falls below a threshold. However, under such a scheme, the policy cost of a firm does not depend on what the other firms emit, and the firm is penalized for any deviation from the scenario, so that it reduces its emission even when the scenario allows it to emit more. Under an ETS, the picture is different. A cap that does not bind costs nothing, whereas a binding cap is met exactly, at a price that depends on the aggregate emission of all covered firms and rises sharply as the cap approaches zero. The literature on the carbon market derives this price from the optimal abatement of the covered firms in a competitive equilibrium \citep{aidOptimalDynamicRegulation2023, carmonaOptimalStochasticControl2009, hitzemannEquilibriumPriceDynamics2018}, but does not consider the credit risk of the individual firms. Conversely, the structural models of climate-related credit risk take the carbon price as given \citep{leguenedalCorporateDebtValue2025}. This chapter bridges the two strands. 

In this chapter, we extend the model of \citet{bourgeyBridgingSocioeconomicPathways2024,bourgeyEfficientSharedSocioeconomic2026} to an ETS. The firms are grouped into sector-region cells, each with its own cap-and-trade scheme. Each firm chooses its emission strategy taking the an anticipated carbon price as given, the realized carbon price clears the carbon market at each time, and the anticipated carbon price is required to be consistent with the expected realized one. The carbon market is then in a temporary equilibrium \citep{grandmontTemporaryGeneralEquilibrium1977}, in the spirit of the consistent expectations equilibrium \citep{hommesCONSISTENTEXPECTATIONSEQUILIBRIA1998}. Moreover, we introduce a downside demand shock driven by the Network for Greening the Financial System (NGFS) scenarios, and we model the physical risk by extreme events arriving according to a point process, whose losses are reduced by adaptation investment, following \citet{vandantzigEconomicDecisionProblems1956}, and scale with the damage function of the Dynamic Integrated Model of Climate and Economy (DICE) \citep{nordhausDICE2013RIntroduction2013,barragePoliciesProjectionsSocial2024}. 
We derive the optimal emission of the firm in feedback form, show that the equilibrium carbon price always exists, and propose a linearized case, in which the carbon price is deterministic, and the default probability is explicit. Moreover, we calibrate the model to the European power sector under the EU ETS, and find that the sign of the effect of a slow transition is set by the energy mix of the firms, which lowers the default probability of renewable producers and coal fleets, but raises that of some gas-fired integrated utilities.

The rest of the chapter is organized as follows. Section~\ref{sec:climate_risk_model} introduces the model framework for both transition and physical risks, and Section~\ref{sec:climate_risk_main_results} provides the optimal emission strategy of the firm, the equilibrium carbon price at the level of the sector-region cell, and the default probability. Section~\ref{sec:numerical_procedure_climate_risk} describes the numerical procedure. In Section~\ref{sec:numerical_experiments_climate_risk}, we compare our model with the model of \citet{bourgeyBridgingSocioeconomicPathways2024,bourgeyEfficientSharedSocioeconomic2026} numerically, as well as the linearized case with the general case, and calibrate it to the European power sector under the EU ETS in Section~\ref{sec:calibration_climate_risk}. The corresponding mathematical proofs are detailed in Section~\ref{sec:appendix_climate_risk}.

\section{Model framework}
\label{sec:climate_risk_model}

Building on the framework of \citet{bourgeyBridgingSocioeconomicPathways2024,bourgeyEfficientSharedSocioeconomic2026}, which bridges climate risk and the credit risk of the firm, we extend it in three directions: we replace the reward-penalty scheme with carbon pricing, introduce a downside demand shock to production, and account for adaptation investment when evaluating the cost of physical risk.
In our framework, we classify firms along two dimensions: geographical region and economic sector. Let $\Lc$ denote the set of regions with a typical region indexed by $\ell$, and $\Sc$ denote the set of sectors with a typical sector indexed by $s$. Given these classifications, we define the basic unit of analysis as a sector-region cell $g := (\ell, s) \in \Lc \times \Sc$. Let $\Ic$ denote the set of firms with a typical firm indexed by $i$. For simplicity, we assume that each firm operates in a single cell, and denote by $g^i$ the cell to which firm $i$ belongs. Moreover, we define $\Ic^g := \{i \in \Ic: g^i = g\}$ as the set of firms in cell $g$. Within this framework, we now examine how transition and physical risks affect the firm.

\subsection{Transition risk}
\label{sec:transition_risk}

Transition risk arises from the process of adjusting to a low-carbon economy, which directly affects the energy usage patterns of firms. The pressure of transition for the firm may induce it to limit its energy usage or shift toward cleaner but more expensive energy sources. All these mitigations may lead to a reduction in the firm's profit, as they may increase the energy cost or decrease the related production. 

\begin{figure}[H]
\centering
\hspace{-0.5cm}
\begin{subfigure}{0.45\linewidth}
\centering
\includegraphics[width=\linewidth]{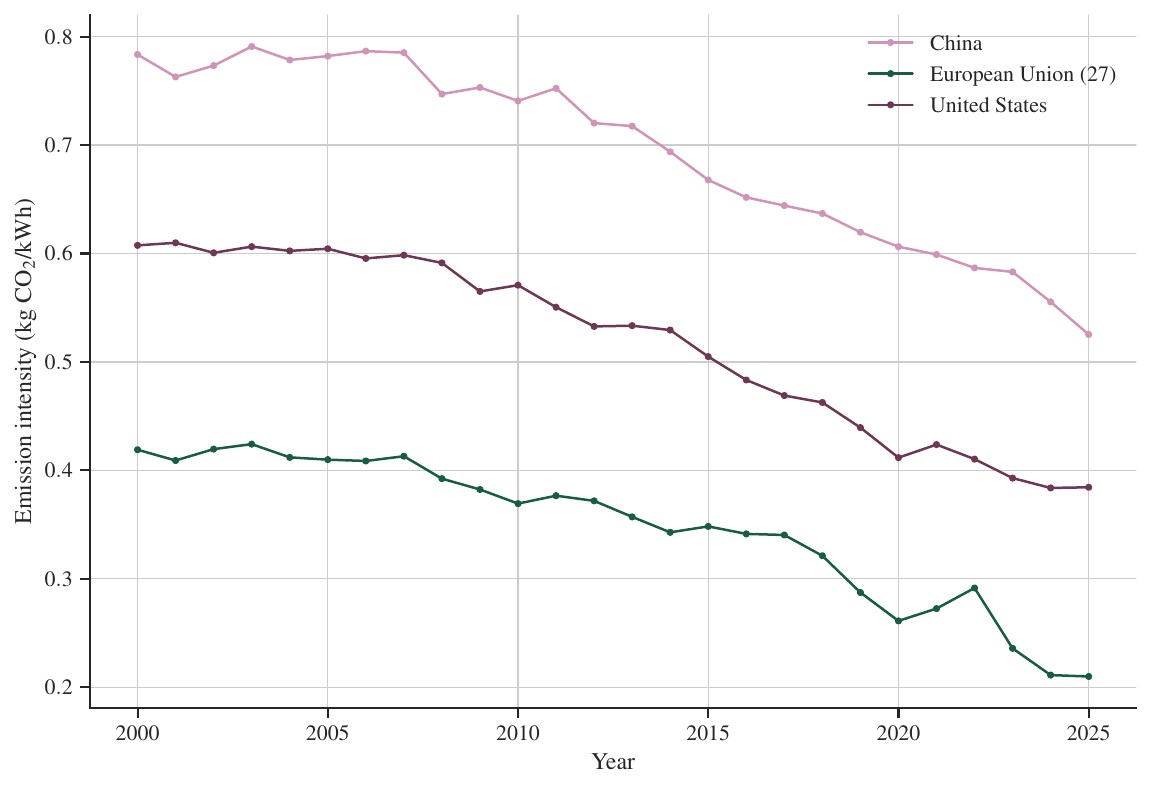}
\caption{Emission factors of electricity.}
\label{fig:carbon_intensity_electricity}
\end{subfigure}
\hfill
\begin{subfigure}{0.45\linewidth}
\centering
\includegraphics[width=\linewidth]{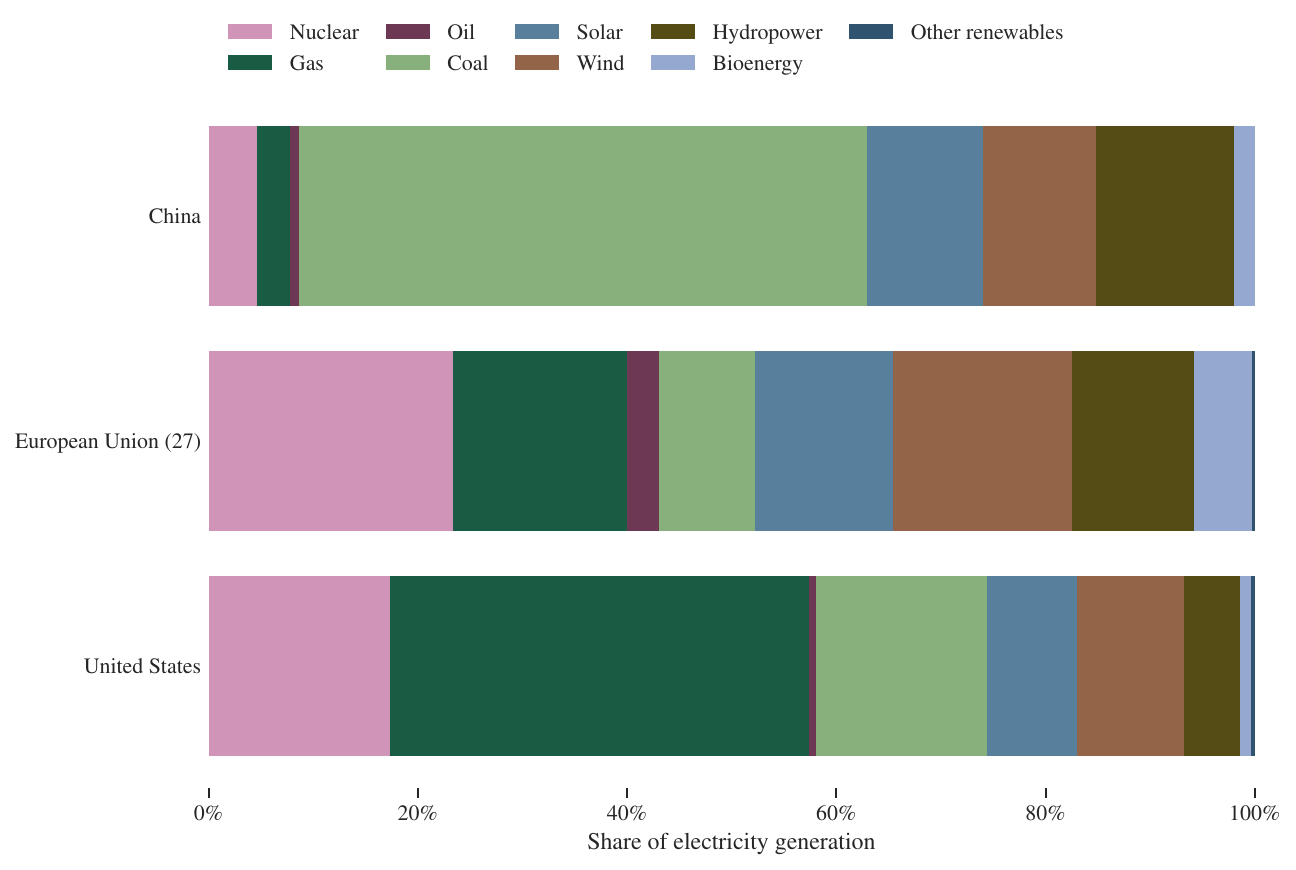}
\caption{Proportion of electricity generation by source in 2025.}
\label{fig:electricity_generation_proportion}
\end{subfigure}
\caption{Electricity emission factors and generation mix in China, the EU, and the US.}
\label{fig:electricity_emissions_generation}
\begin{minipage}{0.7\linewidth}
\footnotesize
\textit{Source: \citet{emberLifecycleCarbonIntensity2026}.}
\end{minipage}
\end{figure}

\begin{figure}[H]
\centering
\hspace{-0.5cm}
\begin{subfigure}[t]{0.3\linewidth}
\centering
\includegraphics[width=\linewidth]{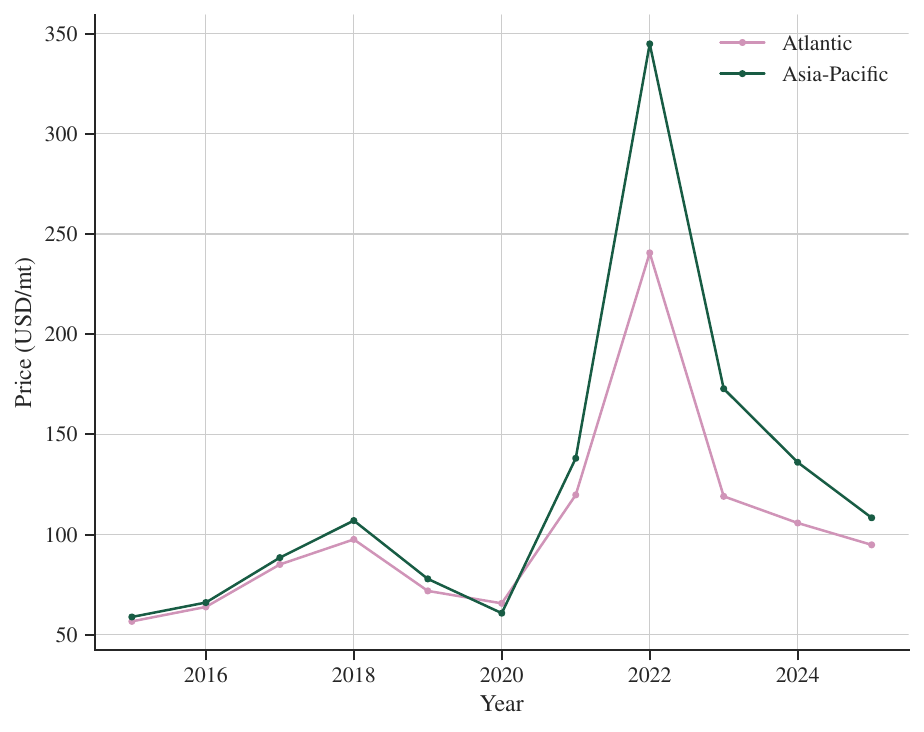}
\caption{Coal prices.}
\label{fig:coal_prices}
\end{subfigure}
\hfill
\begin{subfigure}[t]{0.3\linewidth}
\centering
\includegraphics[width=\linewidth]{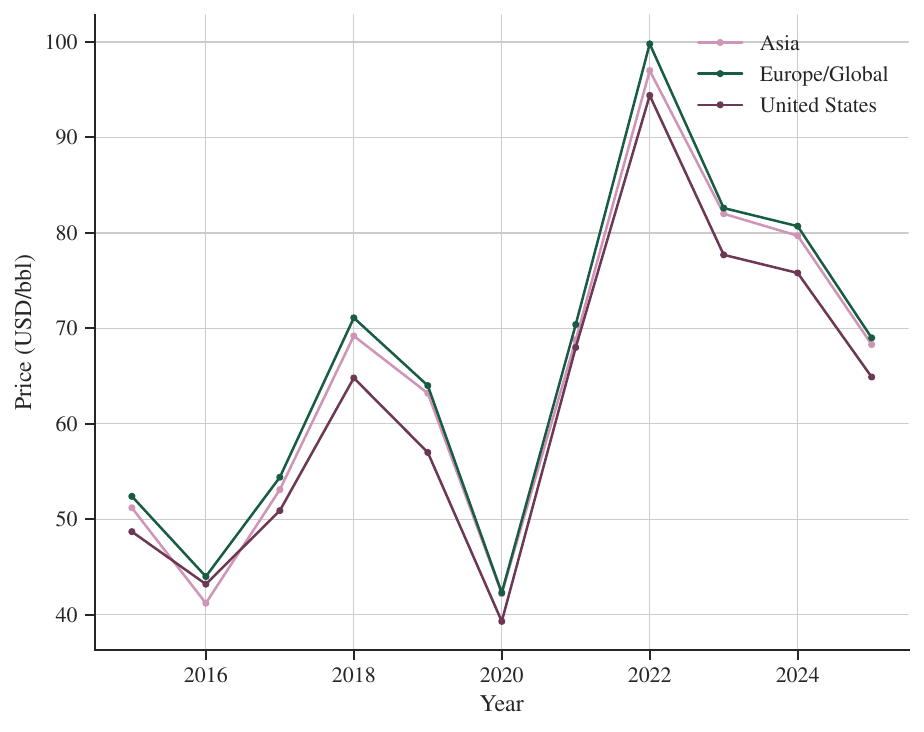}
\caption{Oil prices.}
\label{fig:oil_prices}
\end{subfigure}
\hfill
\begin{subfigure}[t]{0.3\linewidth}
\centering
\includegraphics[width=\linewidth]{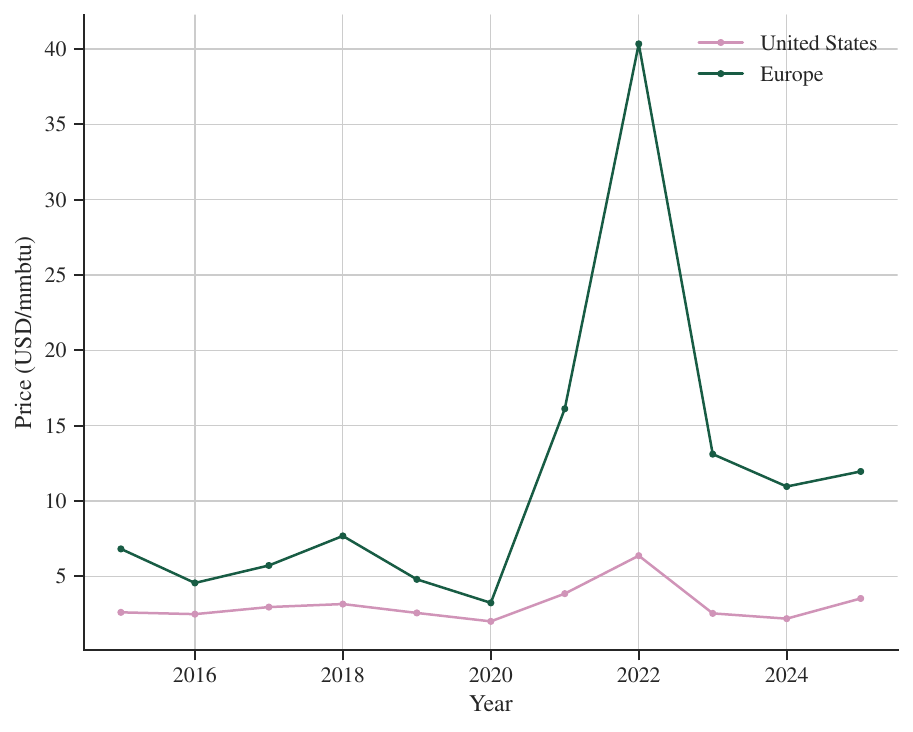}
\caption{Natural gas prices.}
\label{fig:natural_gas_prices}
\end{subfigure}
\caption{Energy prices in different regions between 2015 and 2025.}
\label{fig:energy_prices}
\begin{minipage}{0.7\linewidth}
\footnotesize
\textit{Source: World Bank Commodities Price Data (The Pink Sheet).}
\end{minipage}
\end{figure}

Let $\Ec$ denote the set of all available energy sources (e.g., coal, oil, natural gas, and renewable energy) with a typical energy source indexed by $e$, and $\Ec^i \subseteq \Ec$ denote the set available to firm $i$. To link with greenhouse gas (GHG) emissions, we introduce the emission intensity, denoted by $\theta_{e,t}^i > 0$, which quantifies the GHG emissions associated with consuming 1 kWh of energy source $e$ at time $t$ for firm $i$, measured in carbon dioxide equivalent (CO$_2$e)\footnote{CO$_2$e converts emissions of different GHGs into equivalent amounts of CO$_2$ based on their global warming potentials \citep{RN15}.}. Note that the emission intensity is assumed to be different across firms. As shown in Figure~\ref{fig:electricity_emissions_generation}, the emission intensity of electricity varies across regions, and this is mainly due to the different energy mix, as China and the United States (US), which rely more heavily on fossil fuels, have higher emission intensity. Energy prices are also assumed to vary across firms, as supported by Figure~\ref{fig:energy_prices}, where we can see that the prices of coal, oil, and natural gas are different in different regions. Let $\alpha_{e,t}^i$ denote the unit price of energy source $e$ faced by firm $i$ at time $t$, and $\beta_{e,t}^i$ denote the coefficient of the quadratic energy cost, which captures the increasing marginal cost of the energy consumption, such as the short-term capacity constraints. Thus, the energy cost of firm $i$ at time $t$ is defined as follows
\eqlnostar{eq:energy_cost}{
\Cc^i_\Ec \bigl(\bgam_t^i \bigr)
:= \sum_{e\in\Ec^i}
\Biggl(
\frac{\alpha_{e,t}^i\, \gamma_{e,t}^i}{\theta_{e,t}^i}
+ \beta_{e,t}^i \biggl(\frac{\gamma_{e,t}^i}{\theta_{e,t}^i} \biggr)^2
\Biggr)
= \balp_{\theta, t}^i \cdot \bgam_t^i
+ \bbet_{\theta, t}^i \cdot (\bgam^2)_t^i,
}
where $\gamma_{e,t}^i$ is the emission of firm $i$ from energy source $e$ at time $t$, $\bgam_t^i := (\gamma_{e,t}^i)_{e \in \Ec^i}$ is the corresponding vector, and
\eqstar{
(\bgam^2)_t^i := \Bigl(\bigl(\gamma_{e,t}^i \bigr)^2 \Bigr)_{e \in \Ec^i},
\quad
\balp_{\theta, t}^i := \biggl(\frac{\alpha_{e,t}^i}{\theta_{e,t}^i} \biggr)_{e \in \Ec^i}
=: \bigl( \alpha_{e, \theta, t}^i \bigr)_{e \in \Ec^i},
\quad
\bbet_{\theta, t}^i := \biggl(\frac{\beta_{e,t}^i}{(\theta_{e,t}^i)^2} \biggr)_{e \in \Ec^i}
=: \bigl( \beta_{e, \theta, t}^i \bigr)_{e \in \Ec^i}.
}

\begin{figure}[H]
\centering
\hspace{-0.5cm}
\includegraphics[width=0.7\linewidth]{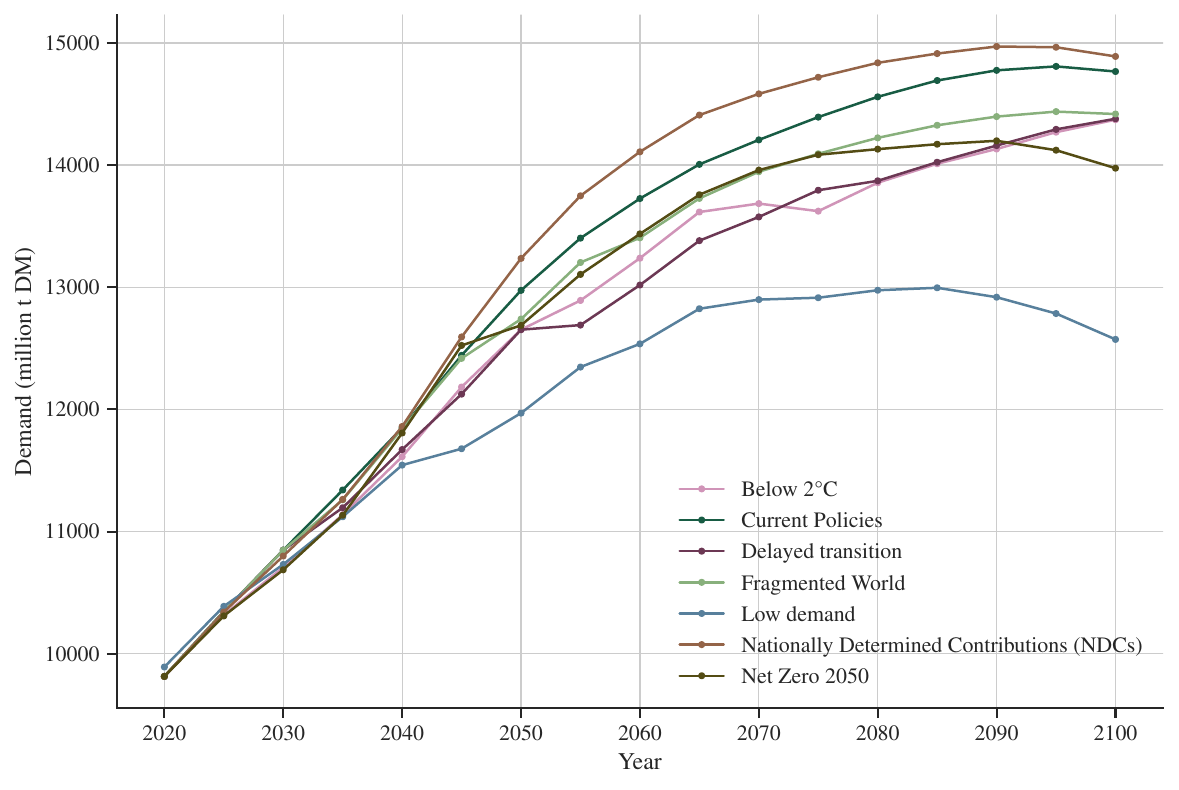}
\caption{Global agricultural demand under different NGFS scenarios.}
\label{fig:agricultural_demand_nfgs}
\begin{minipage}{0.7\linewidth}
\footnotesize
\textit{Source: NGFS Phase V Scenario Explorer (IIASA), GCAM 6.0.}
\end{minipage}
\end{figure}

Once the energy mix has been adjusted, the production of the firm will be affected, as well as the related revenue. Under the transition, the demand faced by the firm is also influenced. As presented in Figure~\ref{fig:agricultural_demand_nfgs}, the future agricultural demand differs across the Network for Greening the Financial System (NGFS) scenarios. In particular, the demand under Current Policies, which is usually regarded as the business-as-usual (BAU) scenario, is higher than that of other scenarios, with the Nationally Determined Contributions (NDCs) scenario being an exception. Similar long-term demand adjustments are observed in other sectors. Hence, the firm will face a demand shock, mostly downside, which will push it to further adjust its energy usage and production. Motivated by this observation, we model the downside demand shock faced by firms in cell $g$ as follows
\eqlnostar{eq:demand_shock}{
q_t^g := 1 + \min \biggl(0, \frac{D_t^{g, X} - D_t^{g, B}}{D_t^{g, B}} \biggr),
}
where $D_t^{g, X}$ and $D_t^{g, B}$ represent the demand of cell $g$ under any scenario $X$ and the NGFS BAU scenario, respectively.
Let us fix a filtered probability space $(\Omega, \Fc, \Pb)$ equipped with a filtration $\Fb = (\Fc_t)_{t \ge 0}$ satisfying the usual conditions. On this space, $(B^g, (B^i)_{i \in \Ic^g})$ are mutually independent standard Brownian motions, where $B^g$ is a cell-level systemic factor, and $(B^i)_{i \in \Ic^g}$ represent firm-specific idiosyncratic factors.
Instead of modelling the production, we model the normalized log-production of firm $i$ in cell $g$ at time $t$, defined as follows
\eqlnostar{eq:log_production}{
p_t^i := \log \biggl( \frac{P_t^i}{P_0^i} \biggr),
\qquad
p_0^i = 0,
}
where $P_t^i$ is the production of firm $i$ at time $t$. We assume that the normalized log-production $p_t^i$ solves the following stochastic differential equation (SDE)
\eqlnostar{eq:log_production_sde}{
\dd p_t^i &= \Bigl( a^i + b^i \log \bigl(q_t^g \bigr) - b^i p_t^i \Bigr) \dd t + \sum_{e \in \Ec^i} \frac{c_e^i \gamma_{e,t}^i}{\theta_{e,t}^i}\, \dd t + \sigma^i\, \dd W_t^i,
\qquad
i \in \Ic^g,
}
where
\eqlnostar{eq:combined_bm}{
\dd W_t^i &= \rho^i \dd B_t^g + \sqrt{1-(\rho^i)^2}\, \dd B_t^i,
}
with $\sigma^i > 0$, $\rho^i \in (-1, 1)$. Here, $a^i \in \Rb$ represents the drift coefficient, $b^i > 0$ is the mean-reversion speed, and $c_e^i \ge 0$ measures the sensitivity of firm $i$'s production to the consumption of energy source $e$. 
The production revenue of firm $i$ is then given by
\eqlnostar{eq:revenue_func}{
\pi^i_t := S_t^i P_t^i = S_t^i P_0^i \ee^{p_t^i},
}
where $S_t^i > 0$ denotes the deterministic selling price of firm $i$ at time $t$.

\begin{figure}[H]
\centering
\hspace{-0.5cm}
\includegraphics[width=0.7\linewidth]{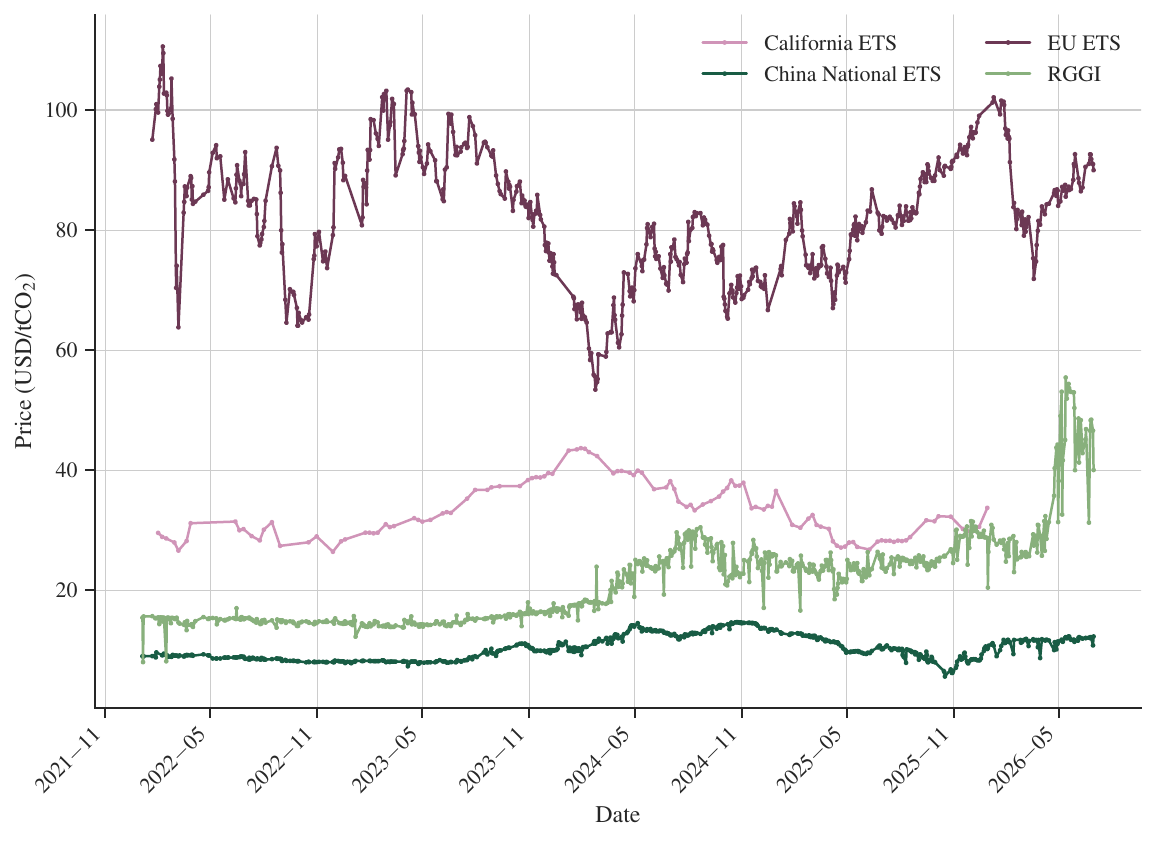}
\caption{Carbon prices under different ETSs.}
\label{fig:carbon_prices}
\begin{minipage}{0.7\linewidth}
\footnotesize
\textit{Source: International Carbon Action Partnership (ICAP).}
\end{minipage}
\end{figure}

\begin{figure}[H]
\centering
\hspace{-0.5cm}
\includegraphics[width=0.7\linewidth]{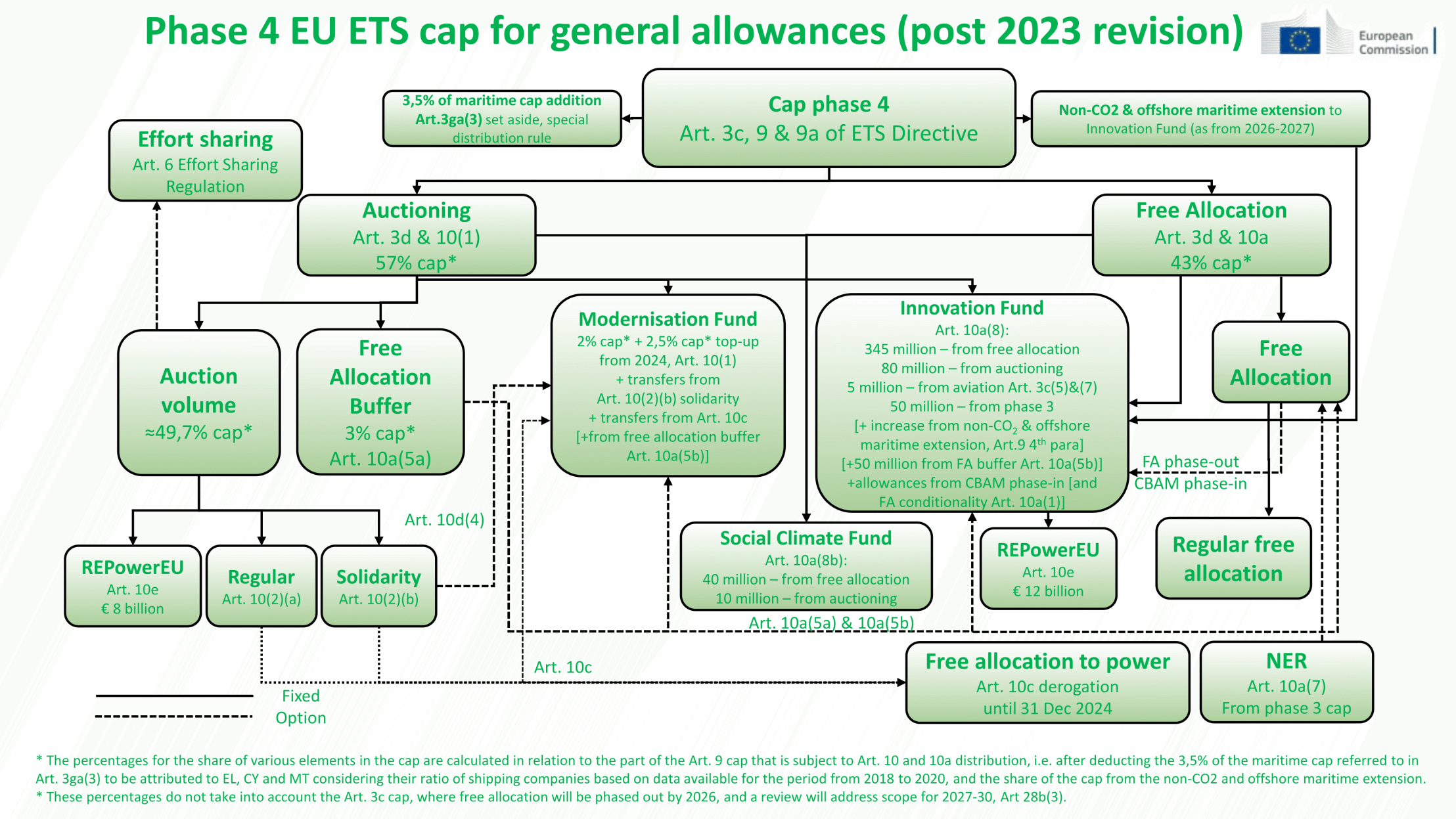}
\caption{Emission cap diagram under the EU ETS.}
\label{fig:eu_ets_cap}
\begin{minipage}{0.7\linewidth}
\footnotesize
\textit{Source: European Commission.}
\end{minipage}
\end{figure}

Besides, the firm faces additional costs from carbon pricing policies imposed by governments to meet the Paris Agreement goal of limiting global warming to 1.5$^\circ$C. In our model, we only consider the emission trading system (ETS) as the main instrument of carbon pricing. 
However, there exist different types of ETS, differing in how the emission constraint is set. The European Union Emission Trading System (EU ETS) is based on an absolute emission cap \citep{EuropeanCommission2015EUETSHandbook}, while China's national ETS adopts an intensity-based approach, under which the emission cap is linked to firms' output \citep{ministryofecologyandenvironmentofthepeoplesrepublicofchinaProgressReportChinas2025}. Moreover, the EU ETS covers electricity and heat generation, energy-intensive industries, aviation, and maritime transport, whereas China's National ETS initially covered only the power sector and was expanded in 2025 to include steel, cement, and aluminium smelting. The EU ETS and China's National ETS are both the only ETSs in their respective regions, while the US has several ETSs, such as the Regional Greenhouse Gas Initiative (RGGI) and the California Cap-and-Invest Program (Californian ETS, formerly the California Cap-and-Trade Program). These heterogeneities of ETSs are reflected in the carbon price, which is reported in Figure~\ref{fig:carbon_prices}. 
Given these heterogeneities, we assume that the ETS is operated separately at each sector-region cell, which abstracts from the trading of allowances across sectors within a region, and follows the cap-and-trade scheme, of which the emission cap is determined at the absolute level, and the allowances are distributed through auctioning and free allocation, as illustrated in Figure~\ref{fig:eu_ets_cap}. Although the free allowance allocation is determined by the product benchmark and firms' activity level under the EU ETS \citep{EuropeanCommission2015EUETSHandbook}, we treat it as exogenous and independent of the firm's current emission and production for simplicity. This approach is known as grandfathering, which was implemented in the early phases of the EU ETS \citep{EuropeanCommission2015EUETSHandbook}.

Let $\Upsilon_t^i$ denote the emission of firm $i$ at time $t$, given by
\eqlnostar{eq:firm_total_emission}{
\Upsilon_t^i := \sum_{e \in \Ec^i} \gamma_{e,t}^i 
= \mathbf{1}_{|\Ec^i|} \cdot \bgam_t^i,
}
where $\mathbf{1}_{|\Ec^i|}$ denotes the vector of ones of dimension $|\Ec^i|$. The aggregate emission of firms in cell $g$ at time $t$ is then defined as $\Gamma_t^g := \sum_{i \in \Ic^g} \Upsilon_t^i$. Let $\tGam_t^g$ denote the emission cap, and $\Ac_t^g$ the total free allowance allocation of cell $g$ at time $t$, satisfying $0 \le \Ac_t^g \le \tGam_t^g$. We do not consider the intertemporal allowance banking and bid-ask frictions, so that the firm's net allowance position is settled continuously at price $\Pc_t^g$. Thus, the carbon compliance cost of firm $i \in \Ic^g$ at time $t$ is given as follows
\eqlnostar{eq:carbon_compliance_cost}{
\Cc_\Cc^i \bigl(\Upsilon_t^i, \Pc_t^g \bigr) := \Pc_t^g \bigl(\Upsilon_t^i -  A^i_t \bigr),
}
where $A_t^i$ denotes the free allowance allocation of firm $i$ at time $t$, with $0 \le A_t^i \le \Ac_t^g$, and $\sum_{i \in \Ic^g} A_t^i = \Ac_t^g$. Here, we assume that the firm can purchase or sell allowances at the market price $\Pc_t^g$.

\subsection{Physical risk}
\label{sec:physical_risk}

\begin{figure}[H]
\centering
\hspace{-0.5cm}
\includegraphics[width=0.7\linewidth]{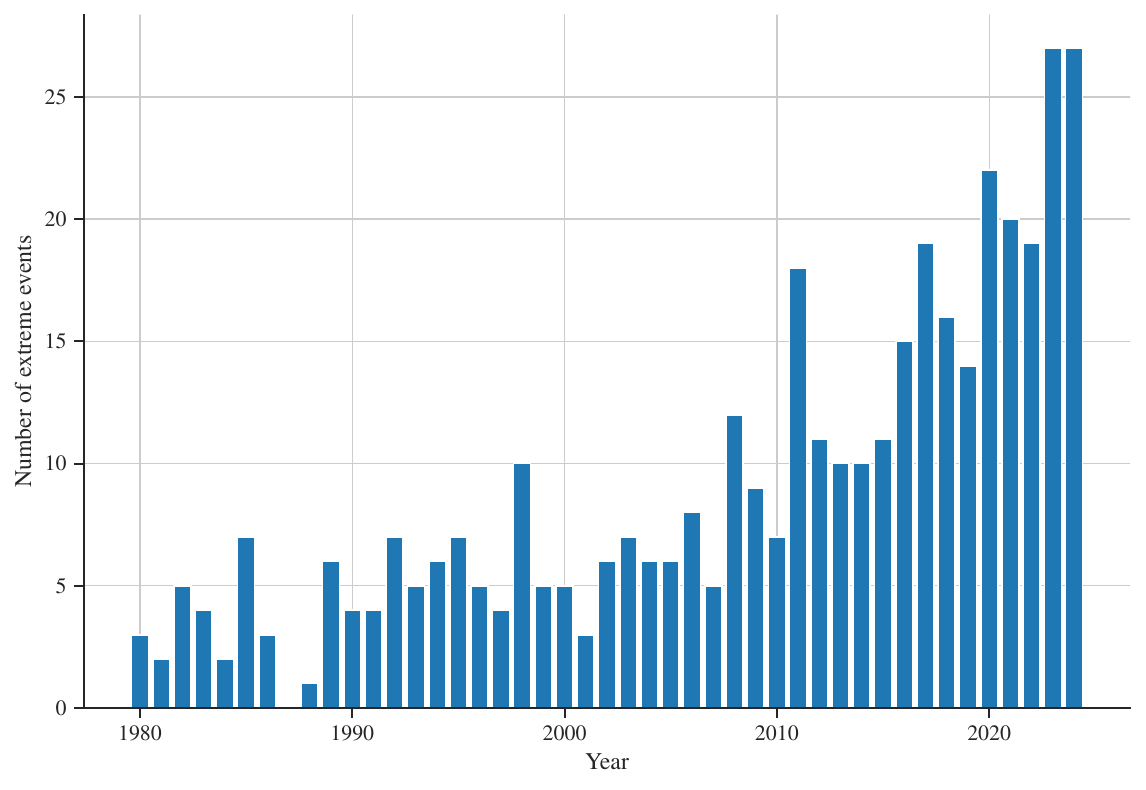}
\caption{Annual number of extreme events with billion-dollar losses in the US.}
\label{fig:extreme_events_us}
\begin{minipage}{0.7\linewidth}
\footnotesize
\textit{Source: \citet{NOAA2025BillionDollarDisasters}.}
\end{minipage}
\end{figure}

\begin{figure}[H]
\centering
\hspace{-0.5cm}
\includegraphics[width=0.7\linewidth]{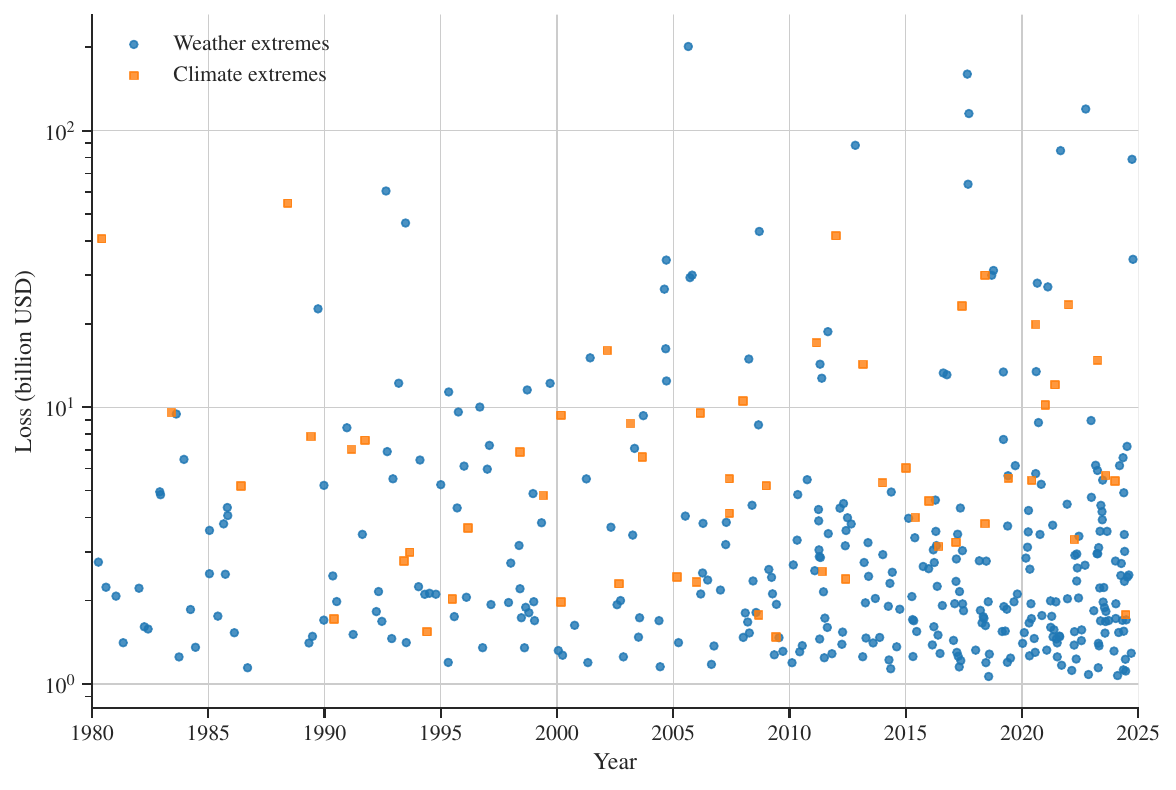}
\caption{Losses (CPI-adjusted) from extreme events in the US.}
\label{fig:physical_loss_us}
\begin{minipage}{0.7\linewidth}
\footnotesize
\textit{Source: \citet{NOAA2025BillionDollarDisasters}.}
\end{minipage}
\end{figure}

Although the impact of transition risk on the firm is beyond the three perspectives mentioned above, we put our focus only on these and leave the other aspects for future research. We now turn to the impact of physical risk, which brings direct losses from extreme events to the firm. Following \citet{RN1}, there are two categories of extreme events: extreme weather events and extreme climate events. Extreme weather events are rare at a particular place and time of year and typically last from hours to days, such as tornadoes and heavy precipitation. Extreme climate events arise when a pattern of extreme weather persists for some time, such as a season; examples include droughts and heatwaves.
Let $\Wc$ denote the set of all extreme events with a typical extreme event indexed by $w$. The exposure of firm $i$ to extreme events is determined by the geographical location of its offices and production facilities. Since each firm is assumed to operate in a single cell $g$, the exposure of firm $i \in \Ic^g$ is determined by its region, denoted by $\ell^i$. Moreover, we assume that all firms in the same region $\ell$ are exposed to the same occurrence of extreme events. Accordingly, we define a point process $N_w^\ell$ on $(\Omega, \Fc, \Pb)$ with a non-negative stochastic intensity $\lambda_w^\ell$, i.e., $N_{w,t}^\ell - \int_0^t \lambda_{w,s}^\ell \dd s$ is an $\Fb$-martingale. The intensity $\lambda_w^\ell$ reflects the frequency of the extreme event $w$, which has been observed to be increasing in the US over the past half-century (Figure~\ref{fig:extreme_events_us}).
When the extreme event $w$ occurs in region $\ell$ at time $t$, each firm $i$ with $\ell^i = \ell$ is assumed to suffer an immediate loss $Z_{w,t}^i \ge 0$, which is determined by firm $i$'s vulnerability and exposure to the extreme event $w$. We assume that $(Z_{w,t}^i)_{t \ge 0}$ is $\Fb$-predictable, and the cumulative loss of firm $i$ over any time interval $[t_0, t_1]$, $0 \le t_0 \le t_1$, before adaptation is given as follows
\eqstar{
\sum_{w \in \Wc} \int_{t_0}^{t_1} Z_{w,u}^i \dd N_{w,u}^{\ell^i}.
} 
This captures the irregular arrival times of extreme events and the randomness of the loss magnitude, which are observed behaviours of physical losses in Figure~\ref{fig:physical_loss_us}.

Following the classical model of \citet{vandantzigEconomicDecisionProblems1956} for flood protection, in which the expected flood loss decreases exponentially in the level of protection, while the cost of protection is linear in that level, we assume that the adaptation investment of firm $i$ reduces each loss $Z_{w,t}^i$ by a factor of $\ee^{-\xi^i K_t^i}$, where $\xi^i > 0$ is the adaptation efficiency of firm $i$, and $K_t^i \ge 0$ is the adaptation investment of firm $i$ observed at time $t$. For any valuation time $t$, the adaptation investment $K_t$ is assumed to be exogenous and held fixed at its current level throughout the valuation horizon. Thus, the expected physical loss (EPL) of firm $i$ evaluated at time $t$ is given as follows
\eqlnostar{eq:epl}{
\EPL^i (K_t^i) &:= \ee^{-\xi^i K_t^i}\, \Eb \biggl[ 
\sum_{w\in \Wc} \int_t^\infty \ee^{-r(u-t)}\, Z_{w, u}^i \dd N_{w,u}^{\ell^i} \Big| \Fc_t
\biggr] \nonumber\\
&= \ee^{- \xi^i K_t^i}\, \int_t^\infty \ee^{-r(u-t)}\, \Eb \biggl[
\sum_{w\in \Wc} Z_{w, u}^i \lambda_{w,u}^{\ell^i} \Big| \Fc_t 
\biggr] \dd u,
}
where $r > 0$ is the discount rate. The adaptation investment $K_t^i$ entails a flow cost consisting of annualized capital cost and maintenance cost, given by
\eqlnostar{eq:adaptation_cost}{
\Cc_\Kc^i (t) := (r + \kappa^i) K_t^i,
}
where $\kappa^i>0$ is the maintenance cost rate per unit of adaptation investment.

However, the conditional expected loss intensity in \eqref{eq:epl} cannot be directly estimated from the available data at the temporal and spatial resolution required for forward-looking calibration. Therefore, we assume that the conditional expected loss intensity is proportional to the damage function of the Dynamic Integrated Model of Climate and Economy (DICE) \citep{nordhausDICE2013RIntroduction2013}. Let $\Tc_\ATM$ denote the trajectory of the global mean atmospheric temperature (ATM) relative to the pre-industrial era, and $\Dc$ denote the damage function, with $\Dc (\Tc_\ATM) = \psi \Tc_\ATM^2$, and $\psi = 0.003467$, following the DICE-2023 calibration in \citet{barragePoliciesProjectionsSocial2024}. The approximation of the conditional expected loss intensity is then given as follows
\eqlnostar{eq:dice_ansatz}{
\Eb \biggl[
\sum_{w \in \Wc} Z_{w, u}^i \lambda_{w,u}^{\ell^i} \Big| \Fc_t \biggr]
\approx \chi^i \Dc \bigl(\Tc_\ATM (u) \bigr),
\qquad
u \ge t >0.
}
Assuming that the damage function $\Dc$ varies little over a one-year interval, we calibrate $\chi^i$ by matching the conditional expected loss intensity of firm $i$ at a reference time $t_{\mathrm{ref}}$ with its expected realized physical loss before adaptation over the preceding year. That is,
\eqstar{
\chi^i = \frac{\Eb \bigl[\PL_{t_{\mathrm{ref}}}^i \bigr]}{\Dc (\Tc_{\mathrm{ATM}} (t_{\mathrm{ref}}))},
}
where
\eqstar{
\PL^i_{t_\mathrm{ref}} :=
\sum_{w \in \Wc} \int_{t_\mathrm{ref} - 1}^{t_\mathrm{ref}} Z^i_{w,s} \dd N_{w,s}^{\ell^i}.
}
If $\Eb [\PL_{t_{\mathrm{ref}}}^i]$ is estimated from observed losses, which already reflect the adaptation investment $K_{t_\mathrm{ref}}^i$, the ratio above is multiplied by $\ee^{\xi^i K_{t_\mathrm{ref}}^i}$. The approximation of the EPL in \eqref{eq:epl} is then given as follows
\eqlnostar{eq:epl_approx}{
\widehat{\EPL}^i \bigl(t, K^i \bigr) := \ee^{- \xi^i K^i}\, \Eb \bigl[\PL_{t_{\mathrm{ref}}}^i \bigr]
\int_t^\infty \ee^{-r(u-t)}\, \frac{\Dc (\Tc_{\mathrm{ATM}} (u))}{\Dc (\Tc_{\mathrm{ATM}} (t_{\mathrm{ref}}))} \dd u.
}

\section{Main results}
\label{sec:climate_risk_main_results}

Within this framework, we now try to answer what exactly the firm will do under the pressure of the transition, and what is the contribution of all firms' actions to the carbon price. Before going forward, we assume that each firm takes the carbon price as given, and the price impact of an individual firm is negligible. Hence, the carbon price is determined by emission constraint in market equilibrium. 
Note that we do not discuss the adjustment of adaptation investment, keeping it fixed. We leave it for future research, as the adaptation investment is usually a long-term decision, and will not change much and has little impact in the short term.

\subsection{Optimal emission strategy}
\label{sec:optimal_emission_strategy}

Given the competitive price-taking assumption, firm $i \in \Ic^g$ relies on two carbon prices: the equilibrium carbon price $\Pc^g := (\Pc_t^g)_{t \ge 0}$ and the anticipated carbon price $\Pbar^g := (\Pbar_t^g)_{t \ge 0}$. The equilibrium carbon price $\Pc^g$ is a stochastic process determined by the market-clearing condition, and the anticipated carbon price $\Pbar^g$ is a non-negative deterministic function of time. We specify these two carbon prices in Section~\ref{sec:equilibrium_carbon_price}. 
We assume that each firm $i$ in cell $g$ adjusts its energy structure, namely, its emission strategy $\bgam^i := (\bgam_t^i)_{t \ge 0}$, to maximize its expected discounted net profit, taking the anticipated carbon price $\Pbar^g$ as given.
Mathematically, the objective function of firm $i$ is given by
\eqlnostar{eq:obj_func}{
\Jc^i \bigl(\bgam^i \bigr)
&:= \Eb \biggl[ \int_0^\infty \ee^{-rt}\, \Bigl(
\pi^i_t
- \Cc_\Ec^i \bigl( \bgam_t^i \bigr)
- \Cc_\Cc^i \bigl(\mathbf{1}_{|\Ec^i|} \cdot \bgam_t^i, \Pbar_t^g \bigr)
\Bigr) \dd t \biggr] \nonumber\\
&= \Eb \biggl[ \int_0^\infty \ee^{-rt}\, \Bigl(
S_t^i\, P_0^i\, \ee^{p_t^i}
- \balp_{\theta, t}^i \cdot \bgam_t^i
- \bbet_{\theta, t}^i \cdot (\bgam^2)_t^i
- \Pbar_t^g \bigl(\mathbf{1}_{|\Ec^i|} \cdot \bgam_t^i - A^i_t \bigr)
\Bigr) \dd t \biggr].
}
Since the EPL in \eqref{eq:epl} and the adaptation cost in \eqref{eq:adaptation_cost} are independent of the emission $\bgam^i$, they are omitted without affecting the decision of firm $i$. The optimization problem for firm $i$ is then given as follows
\eqlnostar{eq:optimization_problem}{
\argsup_{\bgam^i \in \Gc^i} \Jc^i \bigl(\bgam^i \bigr),
}
where the admissible control set $\Gc^i$ consists of all progressively measurable processes $\bgam^i$ satisfying $\bgam_t^i \in \Bc_t^i$, for $t \ge 0$, with
\eqlnostar{eq:domain_control}{
\Bc_t^i := \prod_{e \in \Ec^i} \bigl[0, \eta_{e, t}^{\mathrm{max}} \bigr],
}
where $t \mapsto \eta_{e, t}^{\mathrm{max}}$ is a non-negative and deterministic function for each $e \in \Ec^i$. Hence, $\Bc_t^i$ is a non-empty compact subset of $[0, + \infty)^{|\Ec^i|}$. 

\begin{assumption}
\label{assump:integrability}
The functions $t \mapsto \alpha_{e, \theta, t}^i$, $\beta_{e, \theta, t}^i$, $S_t^i$, $\log (q_t^g)$, $c_e^i/ \theta_{e,t}^i$, $A_t^i$, and $\eta_{e,t}^\mathrm{max}$, for any $e \in \Ec^i$, are deterministic, measurable, and bounded. Moreover, there exists $\delta \in (0, r)$ such that
\eqlnostar{eq:integrability}{
\int_0^\infty \ee^{-\delta t}\, \sup_{\bgam_t^i \in \Bc_t^i} \big|\bgam_t^i \big|^2 \dd t <\infty,
}
and, for every $t \ge 0$,
\eqlnostar{eq:integrability_carbon_price}{
\int_t^\infty \ee^{-\delta (u-t)} \big| \Pbar_u^g \big|^2 \dd u < \infty.
}
\end{assumption}

Under Assumption~\ref{assump:integrability}, the value function of firm $i$ is well-defined and given as follows
\eqlnostar{eq:value_func}{
\Vbar^i (t, p) := \sup_{\bgam^i \in \Gc^i} 
\Eb \biggl[ \int_t^\infty \ee^{-r(u-t)}\, \Bigl(
S_u^i P_0^i \ee^{p_u^i}
- \balp_{\theta, u}^i \cdot \bgam_u^i
- \bbet_{\theta, u}^i \cdot (\bgam^2)_u^i \nonumber\\
- \Pbar_u^g \bigl(\mathbf{1}_{|\Ec^i|} \cdot \bgam_u^i -  A^i_u \bigr)
\Bigr) \dd u \Big| p_t^i = p \biggr],
}
for $(t, p) \in [0, \infty) \times \Rb$.

\begin{assumption}
\label{assump:regularity_value_func}
The value function $\Vbar^i \in \Cc^{1,2} ([0, \infty) \times \Rb)$ in \eqref{eq:value_func} satisfies the Hamilton--Jacobi--Bellman (HJB) equation
\eqlnostar{eq:hjb_equation}{
\partial_t \Vbar^i (t, p)
+ \frac12 (\sigma^i)^2 \partial_p^2 \Vbar^i (t, p)
+ \sup_{\bgam \in \Bc_t^i} \Hc_t^i \bigl(p, \bgam, \Pbar_t^g, \partial_p \Vbar^i (t, p) \bigr)
= r \Vbar^i (t, p),
}
where the function $\Hc^i$ is defined as follows 
\eqlnostar{eq:hamiltonian}{
\Hc_t^i \bigl(p, \bgam, \Pc, z \bigr) &:=
S_t^i\, P_0^i\, \ee^p
- \balp_{\theta, t}^i \cdot \bgam
- \bbet_{\theta, t}^i \cdot \bgam^2
- \Pc \bigl(\mathbf{1}_{|\Ec^i|} \cdot \bgam - A^i_t \bigr) 
+ z \mu_t^i (p, \bgam).
}
Here, $\mu_t^i$ is the coefficient of the drift term in \eqref{eq:log_production_sde}, given by
\eqlnostar{eq:drift_log_production}{
\mu_t^i (p, \bgam) := a^i + b^i \log \bigl(q_t^g \bigr) - b^i p + \bc_{\theta, t}^i \cdot \bgam,
}
where
\eqstar{
\bc_{\theta, t}^i := \Bigl( \frac{c_e^i}{\theta_{e,t}^i} \Bigr)_{e \in \Ec^i}
=: \bigl( c_{e, \theta, t}^i \bigr)_{e \in \Ec^i}.
}
\end{assumption}

\begin{assumption}
\label{assump:log_production}
For every $(t,p) \in [0,\infty) \times \Rb$ and $\bgam^i \in \Gc^i$, the normalized log-production $p_t^i$ satisfies the condition of transversality, that is,
\eqlnostar{eq:transversality_condition}{
\lim_{T \rightarrow \infty} \Eb \Bigl[ 
\ee^{-r(T-t)}\, \Vbar^i \bigl(T, p^i_T \bigr) \big| p^i_t = p
\Bigr] = 0.
}
Moreover, for every $T > 0$,
\eqstar{
\Eb \Bigl[ \int_0^T\partial_p \Vbar^i \bigl(s, p^i_s \bigr)^2 \dd s \Bigr] < \infty.
}
\end{assumption}


\begin{theorem}
\label{thm:verification_of_optimal_emission}
Suppose that Assumptions~\ref{assump:integrability} -- \ref{assump:log_production} hold. If there exists a measurable selector
\eqlnostar{eq:pointwise_optimal_emission}{
\hbgam^i (t,p) \in \argsup_{\bgam \in \Bc_t^i} \Hc_t^i \bigl(p, \bgam, \Pbar_t^g, \partial_p \Vbar^i (t,p) \bigr),
}
such that the associated feedback process
\eqlnostar{eq:optimal_emission}{
\hbgam^i := \Bigl( \hbgam^i \bigl(t, p_t^i \bigr) \Bigr)_{t \ge 0},
}
belongs to $\Gc^i$, then $\hbgam^i$ solves the stochastic control problem in \eqref{eq:optimization_problem}.
\end{theorem}

\begin{proof}
Please refer to Appendix~\ref{sec:proof_of_thm_verification_of_optimal_emission}.
\end{proof}

\begin{remark}
\label{re:hamiltonian}
Indeed, $\Hc_t^i (p, \bgam, \Pbar_t^g, \partial_p \Vbar^i (t, p))$ in \eqref{eq:hjb_equation} is the Hamiltonian of the stochastic control problem \eqref{eq:optimization_problem}. In equilibrium, firm $i$ evaluates the Hamiltonian at the realized carbon price instead of the anticipated carbon price, while $\partial_p \Vbar^i$ remains determined by $\Pbar^g$.
\end{remark}

\begin{assumption}
\label{assump:positive_quadratic_cost}
For each energy source $e \in \Ec^i$, the quadratic cost coefficient satisfies $\beta_{e,t}^i > 0$, for $t \ge 0$. 
\end{assumption}

\begin{theorem}
\label{thm:existence_uniqueness}
Under Assumptions~\ref{assump:integrability} -- \ref{assump:positive_quadratic_cost}, the Hamiltonian $\Hc_t^i$ defined in \eqref{eq:hamiltonian}, for every $(t, p) \in [0, \infty) \times \Rb$, is continuous and strictly concave in $\bgam \in \Bc_t^i$. Thus, there exists a unique $\hbgam^i (t,p)$ maximizing $\Hc_t^i$ over $\Bc_t^i$, and the associated feedback process $\hbgam^i$ in \eqref{eq:optimal_emission} is the unique solution to the stochastic control problem in \eqref{eq:optimization_problem}, up to $\dd t \otimes \dd \Pb$-null sets.
\end{theorem}

\begin{proof}
Since $\beta_{e,t}^i > 0$ and $\theta_{e,t}^i > 0$, the Hessian of $\Hc_t^i (p, \bgam, \Pbar_t^g, \partial_p \Vbar^i (t, p))$ with respect to $\bgam$ is negative definite, for all $e \in \Ec^i$ and $t \ge 0$. Thus, $\Hc_t^i$ is strictly concave and continuous in $\bgam \in \Bc_t^i$. According to the Weierstrass extreme value theorem \citep[Theorem B.1.1]{phamContinuoustimeStochasticControl2009}, the non-emptiness, compactness and convexity of $\Bc_t^i$ ensure the existence of a maximizer $\hbgam^i (t, p)$ of $\Hc_t^i$ over $\Bc_t^i$, while the strict concavity of $\Hc_t^i$ ensures its uniqueness. 
The associated feedback process $\hbgam^i$ defined in \eqref{eq:optimal_emission} belongs to $\Gc^i$, since it takes values in $\Bc_t^i$ and the SDE \eqref{eq:log_production_sde} with the feedback $\hbgam^i (t, p)$ admits a pathwise unique strong solution, its drift being of linear growth and locally Lipschitz continuous in $p$ by Assumption~\ref{assump:regularity_value_func} and the Lipschitz continuity of the projection onto $\Bc_t^i$, and Theorem~\ref{thm:verification_of_optimal_emission} implies that it solves the stochastic control problem in \eqref{eq:optimization_problem}.
Moreover, if $\bgam^i \in \Gc^i$ also solves \eqref{eq:optimization_problem}, the proof of Theorem~\ref{thm:verification_of_optimal_emission} gives $\Eb [\int_0^\infty \ee^{-rt} \Delta \Hc_t^i \dd t] = 0$, so that $\Delta \Hc_t^i = 0$, $\dd t \otimes \dd \Pb$-a.e. By the uniqueness of the maximizer, $\bgam_t^i = \hbgam^i (t, p_t^i)$, $\dd t \otimes \dd \Pb$-a.e., and hence $p^i$ solves the SDE \eqref{eq:log_production_sde} with the feedback process $\hbgam^i (t, p)$. By pathwise uniqueness, $p^i = \hp^i$, and thus $\bgam^i = \hbgam^i$, $\dd t \otimes \dd \Pb$-a.e. 
\end{proof}

\begin{corollary}
\label{coro:explicit_optimal_emission}
Suppose that Assumptions~\ref{assump:integrability} -- \ref{assump:positive_quadratic_cost} hold. For each energy source $e \in \Ec^i$, the interior first-order condition (FOC) for the pointwise maximization of the Hamiltonian in \eqref{eq:hamiltonian} is given by
\eqlnostar{eq:foc}{
c_{e, \theta, t}^i \partial_p \Vbar^i (t, p) - \alpha_{e, \theta, t}^i - 2 \beta_{e, \theta, t}^i \gamma_e - \Pbar_t^g  = 0.
}
Therefore, the unconstrained maximizer is derived as follows
\eqlnostar{eq:unconstrained_max_emission}{
\hgam_e^{i,\mathrm{int}} (t, p) = \frac{c_{e, \theta, t}^i \partial_p \Vbar^i (t, p) - \alpha_{e, \theta, t}^i - \Pbar_t^g}{2 \beta_{e, \theta, t}^i}.
}
The unique optimal emission strategy is then given by
\eqlnostar{eq:explicit_optimal_emission_general}{
\hbgam^i (t, p) = \bigl( \hgam_e^i (t, p) \bigr)_{e \in \Ec^i}, 
\qquad 
\hgam_e^i (t, p) = \min \Bigl( \max \bigl( 0, \hgam_e^{i, \mathrm{int}} (t, p) \bigr), \eta_{e, t}^{\mathrm{max}} \Bigr).
}
\end{corollary}

\begin{proof}
By Theorem~\ref{thm:existence_uniqueness}, the Hamiltonian in \eqref{eq:hamiltonian} is continuously differentiable and strictly concave in $\bgam \in \Bc_t^i$. Thus, the unique maximizer of $\Hc_t^i$ over $\Bc_t^i$ is either an interior point satisfying the FOC, or a boundary point of $\Bc_t^i$. Taking the first-order partial derivative of $\Hc^i_t$ with respect to $\gamma_e^i$, for any $e \in \Ec^i$, and setting it equal to zero gives the FOC in \eqref{eq:foc}. The corresponding unconstrained interior maximizer of $\Hc_t^i$ is then given by \eqref{eq:unconstrained_max_emission}.
Since the Hamiltonian is separable across energy sources with respect to $\bgam$, and $\Bc_t^i$ is a product of intervals, the maximization problem can be solved componentwise by projecting the unconstrained maximizer onto the corresponding admissible interval. This yields the expression of optimal emission in \eqref{eq:explicit_optimal_emission_general}.
\end{proof}

\subsection{Equilibrium carbon price}
\label{sec:equilibrium_carbon_price}

In the market equilibrium, the carbon price is determined by the market-clearing condition, which requires that the aggregate emission of firms in cell $g$ does not exceed the emission cap $\tGam_t^g$, for $t \ge 0$. By Remark~\ref{re:hamiltonian}, firm $i \in \Ic^g$ now evaluates the Hamiltonian at the realized carbon price $\Pc^g$ instead of the anticipated carbon price $\Pbar^g$. Thus, the optimal emission of firm $i$ is obtained by replacing $\Pbar^g$ with $\Pc^g$ in \eqref{eq:explicit_optimal_emission_general}. Specifically, the optimal emission of firm $i$ is given as follows
\eqlnostar{eq:optimal_emission_func}{
\hbgami (t, p^i, \Pc^g) := (\hgamei(t, p^i, \Pc^g))_{e \in \Ec^i},
}
where the optimal emission of each energy source $e \in \Ec^i$ is given by
\eqlnostar{eq:optimal_emission_func_per_source}{
\hgamei(t, p^i, \Pc^g) := \min \Bigl( 
\max \bigl( 0, \hgameii(t, p^i, \Pc^g) \bigr), 
\eta_{e, t}^{\mathrm{max}} \Bigr),
}
with the unconstrained maximizer given by
\eqlnostar{eq:unconstrained_max_emission_func}{
\hgameii (t, p^i, \Pc^g) := \frac{c_{e, \theta, t}^i \partial_p \Vbar^i (t, p^i) - \alpha_{e, \theta, t}^i - \Pc^g}{2 \beta_{e, \theta, t}^i}.
}
The total optimal emission of firm $i$ is then given by
\eqlnostar{eq:total_optimal_emission_func}{
\widehat{\Upsilon}^{i, *} (t, p^i, \Pc^g) := \mathbf{1}_{|\Ec^i|} \cdot \hbgami (t, p^i, \Pc^g),
}
and the aggregate optimal emission of firms in cell $g$ is given by
\eqlnostar{eq:cell_optimal_emission}{
\hGam^g (t, \bp, \Pc^g) := \sum_{i \in \Ic^g} \widehat{\Upsilon}^{i, *} (t, [\bp]_i, \Pc^g),
}
where $\bp := (p^i)_{i \in \Ic^g} \in \Rb^{|\Ic^g|}$, and $[\bp]_i := p^i$. For any time $t \ge 0$, the aggregate optimal emission $\hGam^g$ is capped by the emission cap $\tGam_t^g$, and the resulting carbon price $\Pc_t^g$ is non-negative. Thus, the equilibrium carbon price $\Pc^{g, *}$ is determined by the following market-clearing conditions
\eqlnostar{eq:clearing_condition}{
\Pc^{g, *} (t, \bp) \ge 0, 
\quad
\tGam_t^g - \hGam^g \bigl(t, \bp, \Pc^{g, *} (t, \bp) \bigr) \ge 0, 
\quad
\Pc^{g, *} \Bigl( \tGam_t^g - \hGam^g \bigl(t, \bp, \Pc^{g, *} (t, \bp) \bigr) \Bigr) = 0.
}

\begin{theorem}
\label{thm:existence_uniqueness_carbon_price}
Suppose that Assumptions~\ref{assump:regularity_value_func} and \ref{assump:positive_quadratic_cost} hold for each firm $i \in \Ic^g$. Then, the aggregate optimal emission $\hGam^g$ is continuous and non-increasing in $\Pc^g \in [0, \infty)$, and 
\eqstar{
\lim_{\Pc^g \rightarrow \infty} \hGam^g (t, \bp, \Pc^g) = 0,
}
for any $(t, \bp) \in [0, \infty) \times \Rb^{|\Ic^g|}$. Thus, the equilibrium carbon price $\Pc^{g, *}$ exists and is determined differently by the following three cases.
\begin{enumerate}[label=(\roman*)]
\item If $\hGam^g (t, \bp, 0) < \tGam_t^g$, the equilibrium carbon price is uniquely given by $\Pc^{g, *} (t, \bp) = 0$, and the cap is slack. 
\item If $0 < \tGam_t^g < \hGam^g (t, \bp, 0)$, there exists at least one equilibrium carbon price $\Pc^{g, *} (t, \bp) > 0$, such that $\hGam^g (t, \bp, \Pc^{g, *} (t, \bp)) = \tGam_t^g$, and hence the cap binds. 
\item If $\tGam_t^g = 0 < \hGam^g (t, \bp, 0)$, the set of equilibrium carbon prices is $[\Pc^{g,c} (t, \bp), \infty)$, at which all firms in cell $g$ stop emitting. Here, $\Pc^{g,c} (t, \bp)$ represents the choke price, the lowest price at which no firm finds it profitable to emit, defined as follows
\eqstar{
\Pc^{g,c} (t, \bp) := \max_{i \in \Ic^g, e \in \Ec_t^i} \bigl(
c_{e, \theta, t}^i \partial_p \Vbar^i (t, p^i) - \alpha_{e, \theta, t}^i 
\bigr) > 0,
}
where $\Ec_t^i := \{e \in \Ec^i : \eta_{e,t}^{\mathrm{max}} > 0 \}$ denotes the set of energy sources that firm $i$ can use at time $t$.
\end{enumerate}
Moreover, if there exists at least one firm-energy pair $(i, e)$ for the unconstrained maximizer in \eqref{eq:unconstrained_max_emission_func} at the equilibrium carbon price $\Pc^{g, *} (t, \bp)$, with $\tGam_t^g > 0$, satisfying $\hgameii (t, p^i, \Pc^{g, *} (t, \bp)) \in (0, \eta_{e,t}^{\mathrm{max}})$, then $\Pc^{g, *} (t, \bp)$ is unique.
\end{theorem}

\begin{proof}
Please refer to Appendix~\ref{sec:proof_of_thm_existence_uniqueness_carbon_price}.
\end{proof}

In the boundary case, where $\hGam^g (t, \bp, 0) = \tGam_t^g$, the equilibrium carbon price is $\Pc^{g, *} (t, \bp) = 0$, and the cap is binding. However, without an additional strict monotonicity condition of the aggregate optimal emission $\hGam^g$, the uniqueness of $\Pc^{g, *} (t, \bp)$ cannot be guaranteed in this case. Similarly, the equilibrium carbon price is not unique when the emission cap is zero. Therefore, we select the minimum of the set of equilibrium carbon prices as the unique equilibrium carbon price, defined as follows
\eqlnostar{eq:min_equilibrium_carbon_price}{
\hP^{g, *} (t,\bp) := \min \{ \Pc \ge 0: \Pc\ \text{satisfies \eqref{eq:clearing_condition}} \}.
}
The smallest equilibrium carbon price coincides with the unique equilibrium carbon price in Theorem~\ref{thm:existence_uniqueness_carbon_price}, and equals the choke price $\Pc^{g,c} (t, \bp)$ when the emission cap is zero. 

\begin{corollary}
\label{coro:explicit_carbon_price}
Suppose that the carbon market is in the binding regime, and that the unconstrained maximizer $\hgameii$ in \eqref{eq:unconstrained_max_emission_func} satisfies $0 < \hgameii < \eta_{e,t}^{\mathrm{max}}$, for each firm $i \in \Ic^g$ and each energy source $e \in \Ec^i$. Under the assumptions of Theorem~\ref{thm:existence_uniqueness_carbon_price}, the equilibrium carbon price $\Pc_t^{g, *}$ is unique and is given explicitly as follows
\eqlnostar{eq:equilibrium_carbon_price}{
\Pc^{g, *} (t, \bp) = \frac{M^g (t, \bp) - \tGam_t^g}{W_t^g},
}
where
\eqlnostar{eq:aggregate_emission_coefficients_equilibrium}{
M^g (t, \bp) := \sum_{i \in \Ic^g} \sum_{e \in \Ec^i} \frac{\varphi_{e, \theta, t}^i ([\bp]_i)}{2 \beta_{e, \theta, t}^i},
\qquad
W_t^g := \sum_{i \in \Ic^g} \sum_{e \in \Ec^i} \frac{1}{2 \beta_{e, \theta, t}^i},
}
with
\eqlnostar{eq:marginal_emission_value_without_carbon_price}{
\varphi_{e, \theta, t}^i ([\bp]_i) := c_{e, \theta, t}^i \partial_p \Vbar^i (t, [\bp]_i) - \alpha_{e, \theta, t}^i,
}
denoting the marginal value of emissions from energy source $e \in \Ec^i$ for firm $i$, excluding the marginal cost induced by the carbon price.
\end{corollary}

\begin{proof}
Since the unconstrained maximizer $\hgameii$ in \eqref{eq:unconstrained_max_emission_func} satisfies $0 < \hgameii < \eta_{e,t}^{\mathrm{max}}$, for each firm $i \in \Ic^g$ and each energy source $e \in \Ec^i$, the corresponding optimal emission in \eqref{eq:optimal_emission_func_per_source} at the equilibrium carbon price $\Pc^{g, *} (t, \bp)$ is given as follows
\eqstar{
\hgamei  \bigl(t, [\bp]_i, \Pc^{g, *} (t, \bp) \bigr) = \frac{c_{e, \theta, t}^i \partial_p \Vbar^i (t, [\bp]_i) - \alpha_{e, \theta, t}^i - \Pc^{g, *} (t, \bp)}{2 \beta_{e, \theta, t}^i} 
= \frac{1}{2 \beta_{e, \theta, t}^i} \bigl(\varphi_{e, \theta, t}^i ([\bp]_i) - \Pc^{g, *} (t, \bp) \bigr).
}
The aggregate emission in cell $g$ at $\Pc^{g, *} (t, \bp)$ is then given by
\eqstar{
\hGam^g \bigl(t, \bp, \Pc^{g, *} (t, \bp) \bigr)
= \sum_{i \in \Ic^g} \sum_{e \in \Ec^i} \hgamei \bigl(t, [\bp]_i, \Pc^{g, *} (t, \bp) \bigr)
= M^g (t, \bp) - W_t^g\, \Pc^{g, *} (t, \bp).
}
In the binding regime, the market-clearing condition in \eqref{eq:clearing_condition} gives $\hGam^g (t, \bp, \Pc^{g, *} (t, \bp)) = \tGam_t^g$, which yields the explicit formula in \eqref{eq:equilibrium_carbon_price}. Additionally, $\Pc^{g, *}$ is unique, as $\hgameii$ is interior for each pair $(i, e)$.
\end{proof}

Given the competitive price-taking assumption, the stochastic control problem in \eqref{eq:optimization_problem} of firm $i \in \Ic^g$ is solved by replacing the equilibrium carbon price $\Pc^{g, *}$ with the anticipated carbon price $\Pbar^g$ in the value function $\Vbar^i$ \eqref{eq:value_func}. Thus, the equilibrium carbon price $\Pc^g (t, \bp)$ depends on $\Pbar^g$, so does the smallest equilibrium carbon price $\hP^{g, *} (t, \bp)$ in \eqref{eq:min_equilibrium_carbon_price}.  We write $\Pc^g (t, \bp; \Pbar^g)$ and $\hP^{g, *} (t, \bp; \Pbar^g)$ to emphasize this dependence. 
In the market equilibrium, the anticipated carbon price is assumed to be consistent with the expectation of the equilibrium carbon price, that is,
\eqlnostar{eq:consistency_condition}{
\Pbar_t^g = \Eb \Bigl[ \hP^{g, *} \bigl(t, \hbp_t; \Pbar^g \bigr) \Bigr], 
\qquad t \ge 0.
}
Here, $\hbp_t := (\hp_t^i)_{i \in \Ic^g}$, and $\hp_t^i$ is the solution to the system of SDEs given as follows
\eqlnostar{eq:optimal_log_production_sde}{
\dd \hp_t^i &= 
\mu_t^i \Bigl(\hp_t^i, \hbgami \bigl( t, \hp_t^i, \hP^{g, *} (t, \hbp_t; \Pbar^g) \bigr) \Bigr) \dd t 
+ \sigma^i\, \dd W_t^i,
\qquad 
\hp_0^i = p_0^i.
}
The carbon market is then in the temporary equilibrium \citep{grandmontTemporaryGeneralEquilibrium1977}, under which the carbon price clears the market at each time $t$, given the current state and the firms' expectations about future carbon prices. Although the anticipated carbon price $\Pbar^g$ is required to be consistent with the expected equilibrium carbon price $\hP^{g, *}$ at each point in time, it is weaker than rational expectations \citep{muthRationalExpectationsTheory1961}. Our approach is in the spirit of the consistent expectations equilibrium \citep{hommesCONSISTENTEXPECTATIONSEQUILIBRIA1998}, where the expectations of the market participants are only required to be consistent with the average behaviour of the market participants under the realized price. Note that the emission $\hbgami ( t, \hp_t^i, \hP^{g, *} (t, \hbp_t; \Pbar^g))$ does not solve the stochastic control problem \eqref{eq:optimization_problem} in general. It maximizes the Hamiltonian in \eqref{eq:hjb_equation} at the realized carbon price, while the continuation value is computed under the anticipated carbon price. Hence, firms only adjust their current emissions to the realized carbon price, but do not account for the randomness of the future carbon price. The emission $\hbgami ( t, \hp_t^i, \hP^{g, *} (t, \hbp_t; \Pbar^g))$ coincides with the solution of \eqref{eq:optimization_problem} if and only if the equilibrium carbon price is deterministic, that is, $\hP^{g, *} (t, \hbp_t; \Pbar^g) = \Pbar_t^g$. With a slight abuse of notation, we still refer to $\hgam^{i, *}$ as the optimal emission of firm $i$. 

\subsection{Default probability}
\label{sec:default_probability}

With the optimal emission $\hbgami$ in \eqref{eq:optimal_emission_func} of firm $i \in \Ic^g$, and the equilibrium carbon price $\hP^{g, *}$ in \eqref{eq:min_equilibrium_carbon_price} in the cell $g$, the optimal firm value is given as follows
\eqlnostar{eq:optimal_firm_value}{
\hV_t^i &:=
\Eb \biggl[
\int_t^\infty \ee^{-r (u-t)}\, \Bigl(
S_u^i P_0^i \ee^{\hp_u^i}
- \balp_{\theta, u}^i \cdot \widehat{\bgam}^{i, *} \bigl(u, \hp_u^i, \hP^{g, *} (u, \hbp_u; \Pbar^g) \bigr) \nonumber \\
&\phantom{= \Eb \biggl[\int_t^\infty \ee^{-r (u-t)}\, \Bigl(}
- \bbet_{\theta, u}^i \cdot \Bigl(\widehat{\bgam}^2 \bigl(u, \hp_u^i, \hP^{g, *} (u, \hbp_u; \Pbar^g) \bigr) \Bigr)^{i, *} \nonumber \\
&\phantom{= \Eb \biggl[\int_t^\infty \ee^{-r (u-t)}\, \Bigl(}
- \hP^{g, *} (u, \hbp_u; \Pbar^g)\Bigl(\widehat{\Upsilon}^{i, *} \bigl(u, \hp_u^i, \hP^{g, *} (u, \hbp_u; \Pbar^g) \bigr) -  A^i_u \Bigr)
\Bigr) \dd u \Big| \Fc_t
\biggr]\nonumber\\
&\quad- \EPL^i (t, K^i)
- \frac{(r + \kappa^i) K^i}{r}.
}
In structural models, such as the Merton model \citep{mertonPricingCorporateDebt1974} and its extensions \citep{blackValuingCorporateSecurities1976,longstaffSimpleApproachValuing1995}, default occurs when the firm's value is insufficient to meet its debt obligations. We assume that there exists a deterministic default threshold $L^i (t)$ for each firm $i \in \Ic^g$, representing the minimum firm value to cover its debt reimbursement, labour costs, and other operational expenses at time $t$. When the optimal firm value $\hV_t^i$ falls below $L^i (t)$, the firm is considered to be in default. The default probability of firm $i$ is then defined as follows
\eqlnostar{eq:default_probability}{
\DP_t^i := \Pb \Bigl(\hV_t^i \le L^i (t) \Bigr).
}
Note that $\DP_t^i$ is not the probability of a first default before time $t$.

\section{Numerical procedure}
\label{sec:numerical_procedure_climate_risk}

Although the formula for the optimal emission $\hbgam^i$ in \eqref{eq:optimal_emission_func} is explicit, it depends on the marginal value of log-production $\partial_p \Vbar^i$. Suppose that $\partial_p \Vbar^i$ belongs to $\Cc^{1,2} ([0, \infty) \times \Rb)$. Differentiating the HJB equation \eqref{eq:hjb_equation} with respect to the normalized log-production $p$ yields the following partial differential equation (PDE) for $\partial_p \Vbar^i$
\eqlnostar{eq:marginal_value_pde}{
\partial_t \partial_p \Vbar^i (t, p)
+ \frac12 (\sigma^i)^2 \partial_p^3 \Vbar^i (t, p)
+ \mu_t^i (p, \hbgam^i (t,p) ) \partial_p^2 \Vbar^i (t, p)
+ S_t^i P_0^i \ee^p
= (r + b^i) \partial_p \Vbar^i (t, p).
}
Since the term $\mu_t^i$ depends on $\partial_p \Vbar^i$ itself through the optimal emission $\hbgam^i$, this PDE is quasi-linear, and has no closed-form solution in general. We hence solve this PDE numerically, backward in time by an implicit finite difference scheme with upwind differences for the convection term \citep{courantSolutionNonlinearHyperbolic1952}, and handle its non-linearity by a Picard iteration in each time step. Given the anticipated carbon price $\Pbar^g$, the PDE in \eqref{eq:marginal_value_pde} involves only the normalized log-production of firm $i$, so that it is a one-dimensional PDE for each firm $i$ in the cell $g$.
Since the equilibrium carbon price depends on the state of the normalized log-production of all firms in cell $g$, we have to compute it at each time node of each simulation path for the optimal normalized log-production. We then compute the equilibrium carbon price by a safeguarded Newton iteration, with bisection steps as safeguard.
With the optimal emission $\hbgam^i$ and the equilibrium carbon price $\hP^{g, *}$, we simulate the optimal normalized log-production $\hp^i$ in \eqref{eq:optimal_log_production_sde} by the exponential Euler-Maruyama scheme, and then compute the optimal firm value $\hV^i$ in \eqref{eq:optimal_firm_value} and the default probability $\DP^i$ in \eqref{eq:default_probability} by the least-squares Monte Carlo (LSMC) method \citep{longstaffValuingAmericanOptions2001}.

Let $0 = t_0 < t_1 < \cdots < t_J = T$ be a partition of the time interval $[0, T]$, and $\Delta_j := t_{j+1} - t_j$, for $j \in \{0, 1, \dots, J-1\}$. The equilibrium carbon price is computed at each time node $t_j$ and held constant on $[t_j, t_{j+1})$. At iteration $k \in \Nb$, given the anticipated carbon price $\Pbar^{g, (k)}$, we first solve the PDE in \eqref{eq:marginal_value_pde} for each firm $i \in \Ic^g$, then simulate $N$ paths of the optimal normalized log-production $(\hbp^{(n)})_{1 \le n \le N}$, and replace the consistency condition under the temporary equilibrium in \eqref{eq:consistency_condition} by the sample average over the $N$ paths, that is,
\eqlnostar{eq:fixed_point_iteration}{
\Pbar_{t_j}^{g, (k+1)} = (1 - \vartheta) \Pbar_{t_j}^{g, (k)} 
+ \frac{\vartheta}{N} \sum_{n=1}^N \widehat{\Pc}^{g, *} \bigl(t_j, \hbp_{t_j}^{(n)}; \Pbar^{g, (k)} \bigr), 
}
with a weight $\vartheta \in (0,1]$. We use the same random numbers in each iteration, so that the sample averages change across iterations only through $\Pbar^{g, (k)}$. The iteration stops when the relative change in the supremum norm of  $\Pbar^g$ is below a tolerance $\epsilon > 0$. After convergence, we run a final simulation with a larger number of paths $N' > N$. For each of the $N'$ paths, we simulate $R$ independent copies of the normalized log-production of firm $i$. At each time node $t_j$, the discounted cash flows of firm $i$ from $t_j$ to $T$, computed by the trapezoidal rule on each path and each copy, are regressed on the basis functions 1, $\hp_{t_j}^i$, $\ee^{\hp_{t_j}^i}$, $\ee^{\hp_{t_j}^i/2}$, $\hP_{t_j}^{g, *}$, and $\hP_{t_j}^{g, *} \ee^{\hp_{t_j}^i}$ across the $N' R$ samples, and the optimal firm value $\hV_{t_j}^{i, (n)}$ on the $n$th sample is given by the fitted value.
The default probability $\DP^i$ in \eqref{eq:default_probability} is then estimated by the sample average of the indicator function over the $N' R$ samples, that is, 
\eqstar{
\Pb \Bigl( \hV_{t_j}^i \le L^i (t_j) \Bigr) \approx \frac{1}{N' R} \sum_{n=1}^{N' R} \ind_{\{\hV_{t_j}^{i, (n)} \le L^i (t_j)\}}.
} 
The numerical procedure is summarized in Algorithm~\ref{alg:equilibrium_carbon_price}.

\begin{algorithm}[H]
\caption{Equilibrium carbon price in cell $g$}
\label{alg:equilibrium_carbon_price}
\begin{algorithmic}[1]
\State Set $\Pbar^{g, (0)} = (0, 0, \ldots, 0)$.
\For{$k = 0, 1, 2, \ldots$}
\State Solve the PDE \eqref{eq:marginal_value_pde} with $\Pbar^{g, (k)}$ for each firm $i \in \Ic^g$ by the upwind finite difference scheme with Picard iteration in each time step $0 = t_0 < \cdots < t_J=T$.
\State Set $\hp_{t_0}^{i, (n)} = 0$, for each firm $i \in \Ic^g$ and each path $1 \le n \le N$.
\For{$j = 0, 1, \ldots, J-1$}
\State Compute $\partial_p \Vbar^i (t_j, \hp_{t_j}^i)$ on all $N$ paths.
\State Compute $\Pc_{t_j}^{g, *}$ on each path by the safeguarded Newton iteration.
\State Compute $\hp_{t_{j+1}}^i$ by the exponential Euler-Maruyama scheme on each path.
\EndFor
\State Compute $\Pbar^{g, (k+1)}$ by the fixed point iteration in \eqref{eq:fixed_point_iteration}.
\If{the relative change of $\Pbar^g$ in the supremum norm is below $\epsilon$}
\State \textbf{stop}
\EndIf
\EndFor
\State Repeat the inner loop over $j$ for $N'$ paths at the final $\Pbar^g$. 
\end{algorithmic}
\end{algorithm}

\section{Numerical experiments}
\label{sec:numerical_experiments_climate_risk}

In this section, we first compare our model with the \textit{reference model} introduced in \citet{bourgeyBridgingSocioeconomicPathways2024,bourgeyEfficientSharedSocioeconomic2026} and illustrate the difference between the two models in terms of the optimal emission and default probability. Since the carbon price is not involved in the \textit{reference model}, and its emission decision does not depend on the production state, we introduce a special case of our model, in which the equilibrium carbon price is deterministic and the optimal emission does not depend on the state of the firms in the cell. Afterwards, we discuss the difference between the special case and the general case of our model, further showing the exact impact of carbon pricing on the firms' credit risk.

\subsection{Linearized case}
\label{sec:linearized_case}

Let $\pbar_t^i$ be a deterministic reference path of the normalized log-production of firm $i$. Around $\pbar_t^i$, the production revenue $\pi^i_t$ in \eqref{eq:revenue_func} can be approximated by a first-order Taylor expansion as follows
\eqlnostar{eq:revenue_approx}{
\pi^i_t = S_t^i P_0^i \ee^{p_t^i} 
\approx S_t^i P_0^i \ee^{\pbar_t^i} + S_t^i P_0^i \ee^{\pbar_t^i}\, \bigl(p_t^i - \pbar_t^i \bigr).
}
The marginal value of log-production is then approximated by
\eqlnostar{eq:marginal_value_approx}{
\partial_p \Vbar^i (t, p) \approx \int_t^\infty \ee^{-(r+b^i) (u-t)}\, S_u^i P_0^i \ee^{\pbar_u^i}\, \dd u.
}
Thus, up to terms independent of the emission $\bgam^i$, maximizing the Hamiltonian $\Hc_t^i (p, \bgam, \Pbar_t^g, \partial_p \Vbar^i (t, p))$ in \eqref{eq:hjb_equation} is equivalent to maximizing the following reduced Hamiltonian
\eqstar{
\Hbar_t^i \bigl(\bgam, \Pbar_t^g \bigr) &:=
\Bigl(\int_t^\infty \ee^{-(r+b^i) (u-t)}\, S_u^i\, P_0^i\, \ee^{\pbar_u^i}\, \dd u \Bigr)
\bc_{\theta, t}^i \cdot \bgam
- \balp_{\theta, t}^i \cdot \bgam
- \bbet_{\theta, t}^i \cdot \bgam^2
- \Pbar_t^g \bigl(\mathbf{1}_{|\Ec^i|} \cdot \bgam - A^i_t \bigr).
}
Consequently, the unconstrained maximizer in \eqref{eq:unconstrained_max_emission} is approximated explicitly as
\eqlnostar{eq:unconstrained_max_emission_approx}{
\hgameii (t, p) \approx \frac{c_{e, \theta, t}^i \int_t^\infty \ee^{-(r+b^i) (u-t)}\, S_u^i P_0^i \ee^{\pbar_u^i}\, \dd u - \alpha_{e, \theta, t}^i - \Pbar_t^g}{2 \beta_{e, \theta, t}^i},
}
for any energy source $e \in \Ec^i$. Moreover, if the marginal revenue with respect to $p$ along the reference path is constant, that is, if there exists a constant $\overline{\Pi}^i \ge 0$ such that
\eqlnostar{eq:constant_marginal_revenue}{
S_t^i P_0^i \ee^{\pbar_t^i} = \overline{\Pi}^i,
}
then \eqref{eq:marginal_value_approx} further simplifies to a constant, given by
\eqlnostar{eq:marginal_value_constant}{
\partial_p \Vbar^i (t, p) \approx \frac{\overline{\Pi}^i}{r+b^i},
}
and \eqref{eq:unconstrained_max_emission_approx} admits the following closed-form approximation
\eqlnostar{eq:unconstrained_max_emission_reduced}{
\hgameii (t, p) \approx \frac{\frac{\overline{\Pi}^i c_{e, \theta, t}^i}{r+b^i} - \alpha_{e, \theta, t}^i - \Pbar_t^g}{2 \beta_{e, \theta, t}^i}.
}

\begin{proposition}
\label{prop:linearized_case}
Suppose each firm $i \in \Ic^g$ evaluates its production revenue by the first-order expansion in \eqref{eq:revenue_approx} around a deterministic reference path $\pbar^i$. If the equilibrium carbon price in \eqref{eq:min_equilibrium_carbon_price} satisfies \eqref{eq:integrability_carbon_price} in Assumption~\ref{assump:integrability}, the equilibrium carbon price is equal to the anticipated carbon price, which indicates all firms in cell $g$ have perfect foresight of the future carbon price.
\end{proposition}

\begin{proof}
Under the assumption, the profit function of firm $i$ within the expectation of \eqref{eq:obj_func} is affine in the normalized log-production $p^i$, and the cost terms do not depend on $p^i$. For any $\bgam^i \in \Gc^i$ and $u \ge t$, \eqref{eq:log_production_sde} gives 
\eqstar{
p_u^i
&= \ee^{-b^i (u-t)}\, p_t^i
+ \int_t^u \ee^{-b^i (u-s)}\, \Bigl(
a^i + b^i \log \bigl(q_s^g \bigr) + \bc_{\theta, s}^i \cdot \bgam_s^i 
\Bigr) \dd s 
+ \sigma^i \int_t^u \ee^{-b^i (u-s)}\, \dd W_s^i,
}
where only the first term depends on $p^i$. Thus, the value function $\Vbar^i$ in \eqref{eq:value_func} equals $p^i$ multiplied by the right-hand side of \eqref{eq:marginal_value_approx} plus terms independent of $p^i$. Since the first term does not depend on $\bgam^i$, taking the supremum over $\bgam^i$ preserves this structure, so that $\partial_p \Vbar^i$ equals the right-hand side of \eqref{eq:marginal_value_approx}, which involves no carbon price. It follows that the optimal emission $\hbgami$ in \eqref{eq:optimal_emission_func} depends neither on $p^i$ nor on $\Pbar^g$. Hence, the consistency condition in \eqref{eq:consistency_condition} is simplified as $\hP^{g, *} = \Pbar^g$, and the equilibrium carbon price is deterministic.
\end{proof}

In the special case of Proposition~\ref{prop:linearized_case} (referred to as the \textit{linearized case}), if the cap binds and all unconstrained maximizer in \eqref{eq:unconstrained_max_emission_approx} are interior, Corollary~\ref{coro:explicit_carbon_price} gives the equilibrium carbon price explicitly as follows
\eqlnostar{eq:equilibrium_carbon_price_approx}{
\Pc_t^{g, *} = \frac{
\sum_{i \in \Ic^g} \sum_{e \in \Ec^i}
\frac{1}{2 \beta_{e, \theta, t}^i} \biggl(
c_{e, \theta, t}^i \int_t^\infty \ee^{-(r+b^i) (u-t)}\, S_u^i P_0^i \ee^{\pbar_u^i}\, \dd u 
- \alpha_{e, \theta, t}^i
\biggr)
- \tGam_t^g
}{W_t^g}.
}
Furthermore, if \eqref{eq:constant_marginal_revenue} holds, the equilibrium carbon price admits the following closed-form approximation
\eqlnostar{eq:equilibrium_carbon_price_reduced}{
\Pc_t^{g, *} \approx \frac{
\sum_{i \in \Ic^g} \sum_{e \in \Ec^i}
\frac{1}{2 \beta_{e, \theta, t}^i} \biggl(
\frac{\overline{\Pi}^i c_{e, \theta, t}^i}{r+b^i} - \alpha_{e, \theta, t}^i
\biggr)
- \tGam_t^g
}{W_t^g}.
}

Given the open-loop optimal emission and deterministic equilibrium carbon price, the optimal normalized log-production $\hp^i$ of firm $i$ in \eqref{eq:optimal_log_production_sde} is a linear SDE with deterministic coefficients, and hence admits a Gaussian representation. For notational simplicity, we denote by $\hbgami (t, \Pc_t^{g, *})$ the optimal emission of firm $i$ at time $t$, obtained by replacing $\Pbar^g$ with $\Pc^{g, *}$ in \eqref{eq:unconstrained_max_emission_approx} for each unconstrained maximizer $\hgameii$. Solving the SDE in \eqref{eq:optimal_log_production_sde}, the optimal normalized log-production $\hp_u^i$ is given by
\eqlnostar{eq:optimal_log_production_approx}{
\hp_u^i
&= \ee^{-b^i (u-t)}\, \hp_t^i
+ \int_t^u \ee^{-b^i (u-s)}\, \Bigl( a^i + b^i \log \bigl(q_s^g \bigr)
+ \bc_{\theta, s}^i \cdot \hbgam^{i, *} \bigl(s, \Pc_s^{g, *} \bigr) \Bigr) \dd s \nonumber \\
&\quad+ \sigma^i \int_t^u \ee^{-b^i (u-s)}\, \Bigl(\rho^i \dd B_s^g + \sqrt{1-(\rho^i)^2} \dd B_s^i \Bigr),
}
for all $u \ge t$. Conditionally on $\Fc_t$, the vector $(\hp^i)_{i\in \Ic^g}$ is an $|\Ic^g|$-dimensional Gaussian vector with conditional mean
\eqlnostar{eq:optimal_log_production_approx_mean}{
\Eb \bigl[\hp_u^i \big| \Fc_t \bigr]
&= \ee^{-b^i (u-t)}\, \hp_t^i
+ \int_t^u \ee^{-b^i (u-s)}\, \Bigl( a^i + b^i \log \bigl(q_s^g \bigr) + \bc_{\theta, s}^i \cdot \hbgam^{i, *} \bigl(s, \Pc_s^{g, *} \bigr) \Bigr) \dd s \nonumber\\
&=: \ee^{-b^i (u-t)}\, \hp_t^i
+ m^i (u,t),
}
and conditional covariance
\eqlnostar{eq:optimal_log_production_approx_covariance}{
\cov \bigl(\hp_u^i, \hp_u^j \big| \Fc_t \bigr) = \sigma^i \sigma^j \bigl(\rho^i \rho^j \ind_{i \neq j} 
+ \ind_{i=j} \bigr) \frac{1 - \ee^{-(b^i + b^j) (u-t)}}{b^i + b^j}
=: \Sigma^{i,j} (u-t),
}
for every $i,j \in \Ic^g$. In particular, $\Sigma^{i, i}$ is the conditional variance of $\hp_u^i$. Thus, $\ee^{\hp_u^i}$ is log-normal, with conditional expectation
\eqstar{
\Eb \Bigl[\ee^{\hp_u^i} \big| \Fc_t \Bigr] = \exp \Bigl(
\ee^{-b^i (u-t)}\, \hp_t^i + m^i (u,t) + \frac{1}{2} \Sigma^{i,i} (u-t)
\Bigr).
}
Substituting this expression into \eqref{eq:optimal_firm_value}, we obtain
\eqlnostar{eq:optimal_firm_value_approx}{
\widehat{\Vc}_t^i = h^i \bigl(t, \hp_t^i \bigr),
}
where $h^i$ is the deterministic function defined, for $t \ge 0$ and $x \in \Rb$, as follows
\eqlnostar{eq:optimal_firm_value_approx_func}{
h^i (t, x) &:=
\int_t^\infty \ee^{-r (u-t)}\, S_u^i P_0^i \exp \Bigl(
\ee^{-b^i (u-t)}\, x + m^i (u,t) + \frac{1}{2} \Sigma^{i,i} (u-t)
\Bigr) \dd u \nonumber\\
&\quad- \int_t^\infty \ee^{-r (u-t)}\, \Bigl(
\balp_{\theta, u}^i \cdot \hbgam^{i, *} \bigl(u, \Pc_u^{g, *} \bigr)
+ \bbet_{\theta, u}^i \cdot \Bigl(\hbgam^2 \bigl(u, \Pc_u^{g, *} \bigr) \Bigr)^{i, *}
+ \Pc_u^{g, *} \bigl(\widehat{\Upsilon}^{i, *} \bigl(u, \Pc_u^{g, *} \bigr) -  A^i_u \bigr)
\Bigr) \dd u \nonumber \\
&\quad- \EPL^i \bigl(t, K^i \bigr)
- \frac{(r + \kappa^i) K^i}{r}.
}
Since only the first term in \eqref{eq:optimal_firm_value_approx_func} depends on $x$, and its integrand is strictly increasing in $x$, $h^i (t, \cdot)$ is continuous and strictly increasing. Thus, for the default threshold $L^i (t)$ in the range of $h^i (t, \cdot)$, the default probability in \eqref{eq:default_probability} is given explicitly as follows
\eqlnostar{eq:default_probability_approx}{
\DP_t^i
= \Pb \Bigl(\hp_t^i \le \bigl(h^i (t, \cdot) \bigr)^{-1} \bigl(L^i (t) \bigr) \Bigr)
= \Phi \biggl(\frac{\bigl(h^i (t, \cdot) \bigr)^{-1} \bigl(L^i (t)\bigr) - \ee^{-b^i t}\, \hp_0^i - m^i (t,0)}{\sqrt{\Sigma^{i,i} (t)}} \biggr),
}
where $\Phi$ is the cumulative distribution function of the standard Gaussian random variable.
In the numerical experiments, the default threshold $L^i (t)$ is calibrated to the case without carbon pricing, of which the firm value is denoted as $\hV^{i, \mathrm{ref}}$. It is obtained by setting $\hP^{g, *} = 0$ in \eqref{eq:optimal_firm_value_approx}. Assuming a constant default intensity $\zeta_{\mathrm{ref}}$, the default threshold $L^i (t)$ is then determined by
\eqstar{
\Pb \bigl( \hV_t^{i, \mathrm{ref}} < L^i (t) \bigr) = 1 - \ee^{-\zeta_{\mathrm{ref}} t}.
}
In the linearized case, inverting \eqref{eq:default_probability_approx}, we obtain the explicit expression of $L^i (t)$ as follows
\eqlnostar{eq:default_threshold_approx}{
L^i (t) = h^{i, \mathrm{ref}} \Bigl(t, \Phi^{-1} \bigl(1- \ee^{-\zeta_{\mathrm{ref}} t} \bigr) \sqrt{\Sigma^{i,i} (t)} + \ee^{-b^i t}\, \hp_0^i + m^{i, \mathrm{ref}} (t,0) \Bigr),
}
where $h^{i, \mathrm{ref}}$ and $m^{i, \mathrm{ref}}$ are defined similarly to \eqref{eq:optimal_firm_value_approx_func} and \eqref{eq:optimal_log_production_approx_mean}, but with a zero carbon price and without physical risk, i.e., $\Pc^{g, *} = 0$, $\EPL^i = 0$, and $\Cc_\Kc^i = 0$. 

\subsection{Comparison with the \textit{reference model}}
\label{sec:comparison_ref_model}

Compared with the \textit{reference model} in \citet{bourgeyBridgingSocioeconomicPathways2024,bourgeyEfficientSharedSocioeconomic2026}, there are three main differences in our model. First, we introduce an ETS, instead of a reward-penalty scheme, where the carbon price is determined endogenously, so that the policy cost faced by a firm depends on the emission decisions of all firms in the same cell. Moreover, we introduce a scenario-dependent downside demand shock into the production dynamics, and we consider the firm-level adaptation against the extreme events. Since the last two features are absent from the \textit{reference model}, the comparison focuses on the first feature to examine how differently the two policy mechanisms translate the same SSP scenario into the optimal emission and default probability.

\begin{figure}[H]
\centering
\hspace{-0.5cm}
\includegraphics[width=0.7\linewidth]{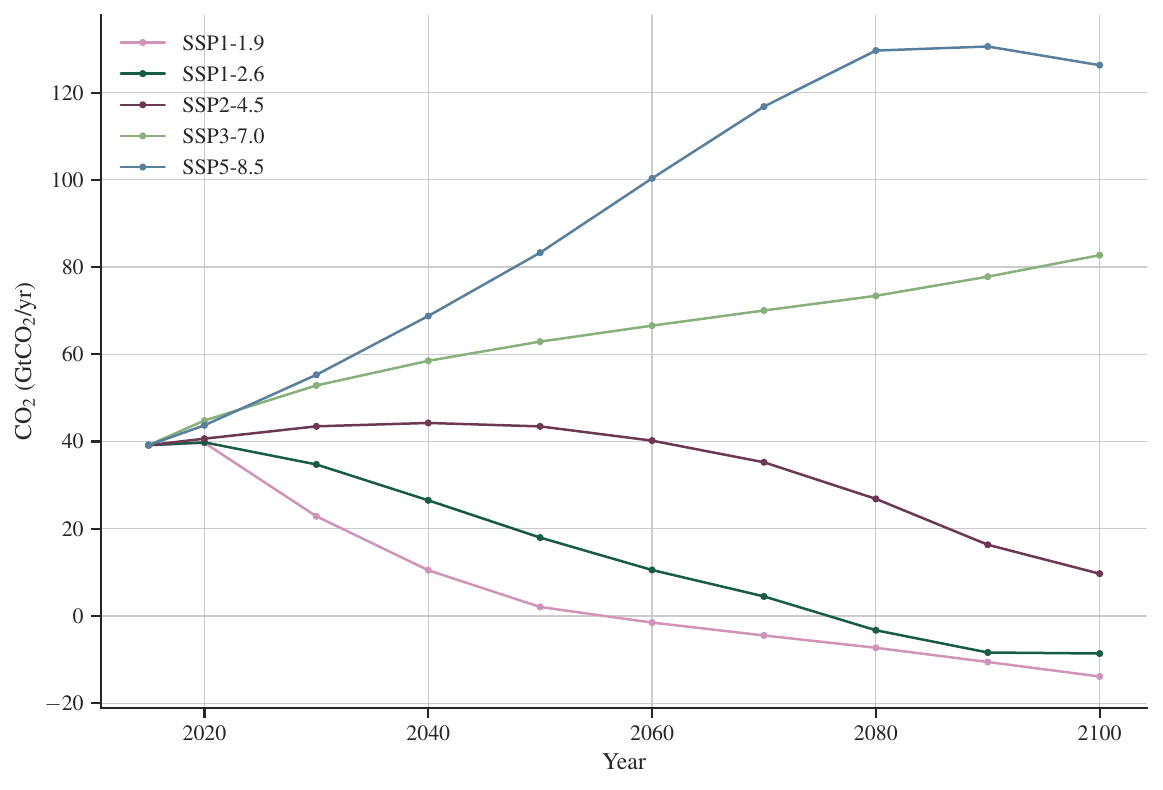}
\caption{Global CO$_2$ emissions under different SSP scenarios.}
\label{fig:emissions_ssp}
\begin{minipage}{0.7\linewidth}
\footnotesize
\textit{Source: IIASA SSP Scenario Database (CMIP6).}
\end{minipage}
\end{figure}

Specifically, we compare the linearized case introduced in Section~\ref{sec:linearized_case} with the explicit example of the \textit{reference model} described in \citet[Sections 2.5, 4 and 5.1]{bourgeyEfficientSharedSocioeconomic2026}. The emission decision of the \textit{reference model} relies on a constant marginal value of log-production $1 / (r+b)$, which corresponds to the linearized case of our model with a constant marginal revenue such that $\overline{\Pi} = 1$. That is, the reference path of the normalized log-production is given by $\pbar_t = 0$ for all $t \ge 0$.
The parameter settings used in the comparison are listed in Table~\ref{tab:comparison_parameter_settings}, with the common parameters from the implementation of the \textit{reference model} available in the accompanying GitHub repository\footnote{Available at: \url{https://github.com/fbourgey/climate-credit-risk/tree/main}.}. We use five illustrative SSP scenarios of IPCC AR6 \citep{RN15}, namely SSP1-1.9, SSP1-2.6, SSP2-4.5, SSP3-7.0, and SSP5-8.5. As shown in Figure~\ref{fig:emissions_ssp}, SSP1-1.9 and SSP1-2.6 represent futures with very low and low GHG emissions, respectively, SSP2-4.5 represents an intermediate-emission future, SSP3-7.0 and SSP5-8.5 represent a high- and very high-emission future, respectively. Each SSP scenario label encodes not only the GHG emission level, but also the underlying socioeconomic pathway and the radiative forcing level; we refer to \citet{gmd-9-3461-2016, oneillRoadsAheadNarratives2017} for more details. Since the numerical experiment of the \textit{reference model} includes SSP4-6.0 \citep[Figure 4]{bourgeyEfficientSharedSocioeconomic2026}, we also consider this scenario in our comparison, which is also an intermediate case like SSP2-4.5. The emission path of each SSP scenario enters into the \textit{reference model} as the benchmark for determining reward and penalty, while in our model, it is used as the emission cap for the ETS. 

\begin{table}[H]
\centering
\caption{Parameter settings used in the comparison with the \textit{reference model}.}
\label{tab:comparison_parameter_settings}
\small
\begin{tabular}{lcc}
\toprule
Parameters & \textit{Reference model} & Our model \\
\midrule
Sector                            & Industry & same \\
Initial production                & $P_0 = 1$ & $P_0 = 1$ \\
Revenue multiplier                & $1$ & $S = 1$ \\
Reference normalized log-production
& not applicable
& $\bar{p} = 0$ \\
$\alpha$                          & $(0, 0, 0)$ & same \\
$\beta$                           & $(0.1, 0.5, 0.8)$ & same \\
$\theta$                          & $(1, 1, 1)$ & same \\
$r, \sigma, a, b$                 & $0.02,\ 0.1,\ 0.1,\ 0.1$ & same \\
$c$                               & $(0.01, 0.01, 0.01)$ & same \\
Policy instrument
& $w_1 = 0.05,\ w_2 = 0.02$
& $\mathcal{P}_t^{g, *}$ \\
Reference default intensity       & $\zeta_{\mathrm{ref}} = 0.01$ & same \\
Free allowance allocation         & not applicable & $A^i_t = 0$ \\
Downside demand shock             & not applicable & $q_t^g = 1$ \\
Physical risk and adaptation      & excluded & excluded \\
\bottomrule
\end{tabular}
\end{table}

\begin{figure}[H]
\centering
\begin{subfigure}[t]{0.3\linewidth}
\centering
\includegraphics[width=\linewidth]
{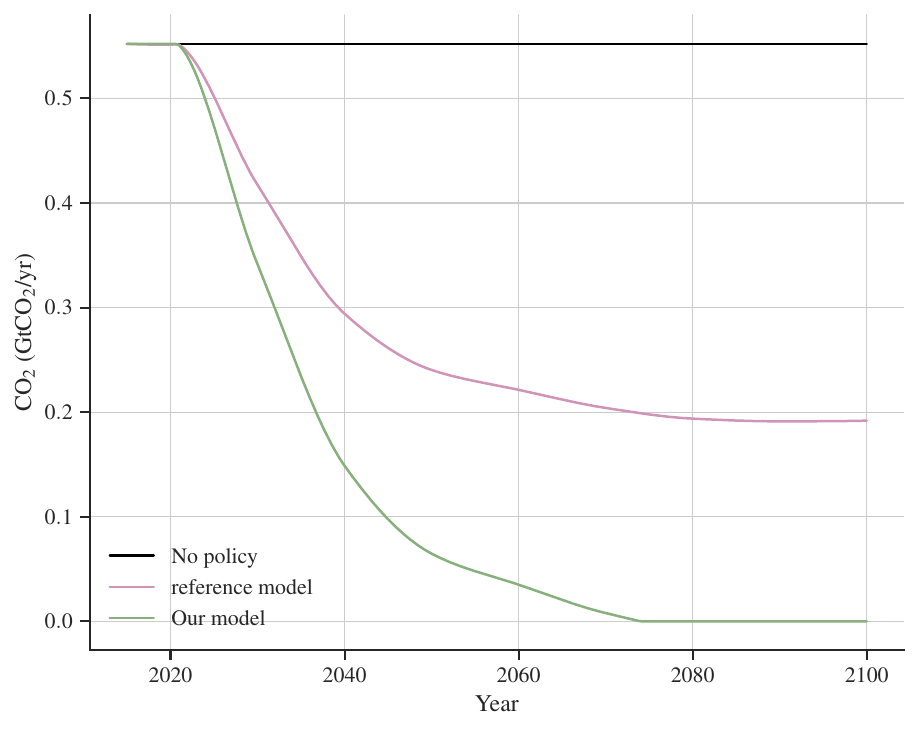}
\caption{SSP1-1.9.}
\label{fig:comparison_emissions_ssp119}
\end{subfigure}
\hfill
\begin{subfigure}[t]{0.3\linewidth}
\centering
\includegraphics[width=\linewidth]
{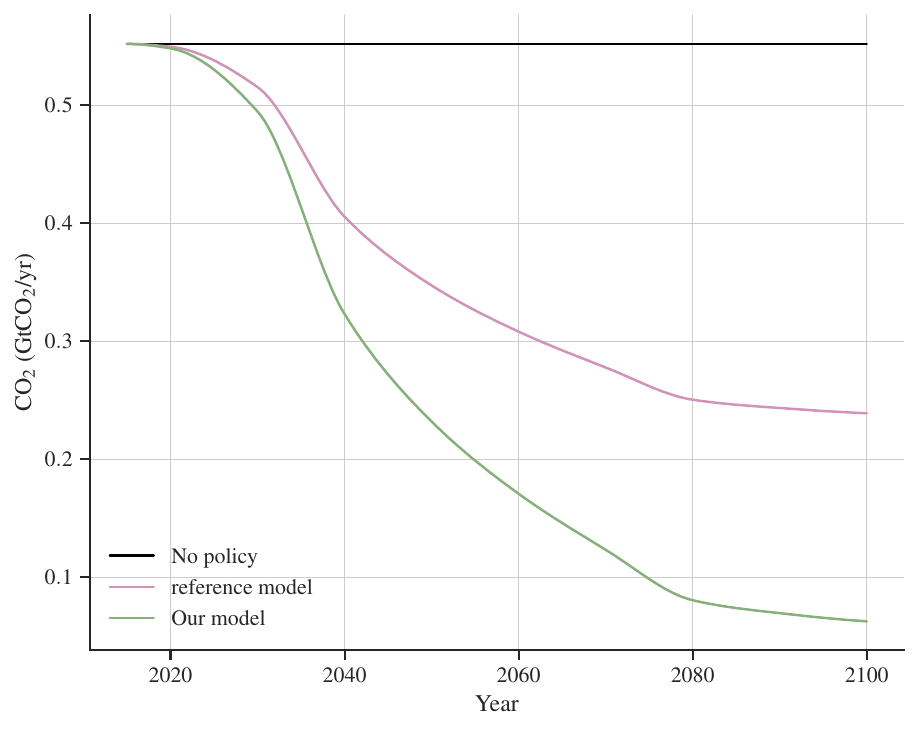}
\caption{SSP1-2.6.}
\label{fig:comparison_emissions_ssp126}
\end{subfigure}
\hfill
\begin{subfigure}[t]{0.3\linewidth}
\centering
\includegraphics[width=\linewidth]
{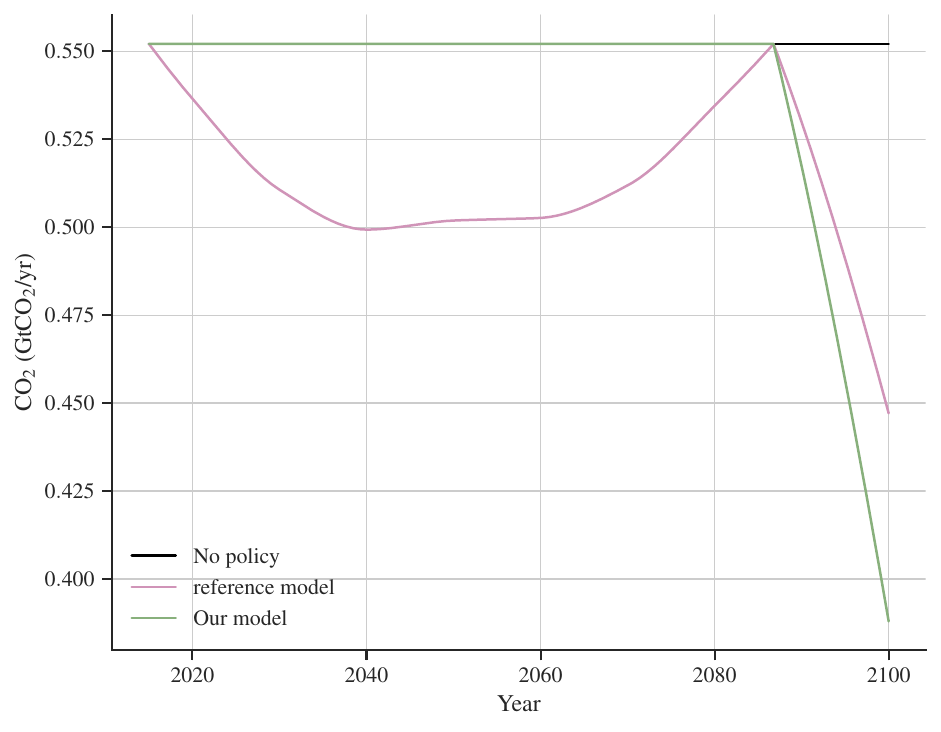}
\caption{SSP2-4.5.}
\label{fig:comparison_emissions_ssp245}
\end{subfigure}
\par\vspace{0.5cm}
\begin{subfigure}[t]{0.3\linewidth}
\centering
\includegraphics[width=\linewidth]
{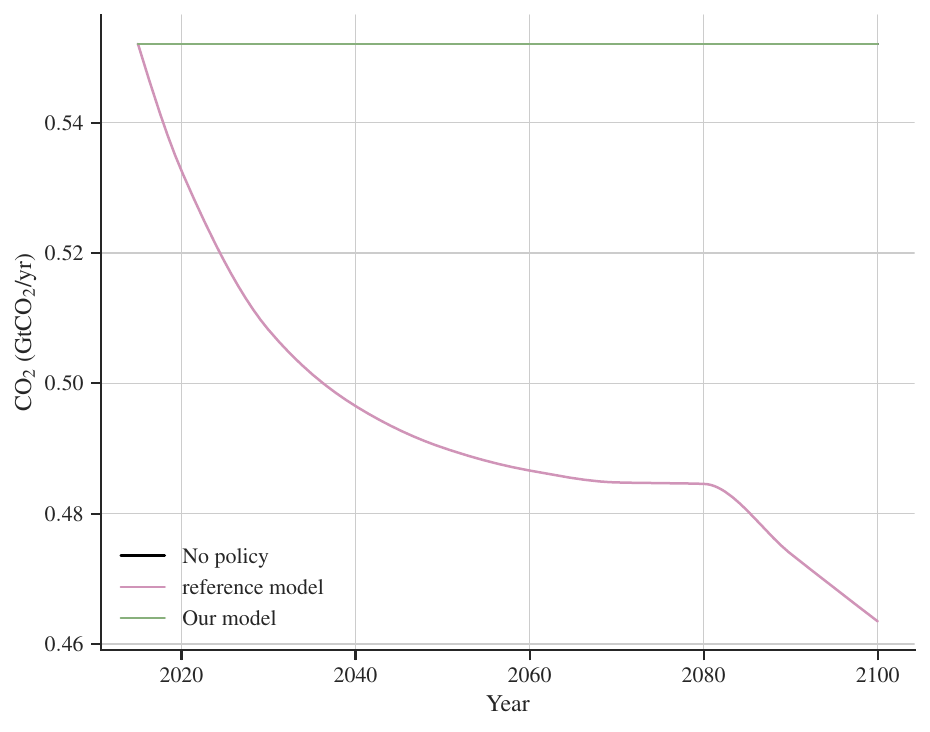}
\caption{SSP3-7.0.}
\label{fig:comparison_emissions_ssp370}
\end{subfigure}
\hfill
\begin{subfigure}[t]{0.3\linewidth}
\centering
\includegraphics[width=\linewidth]
{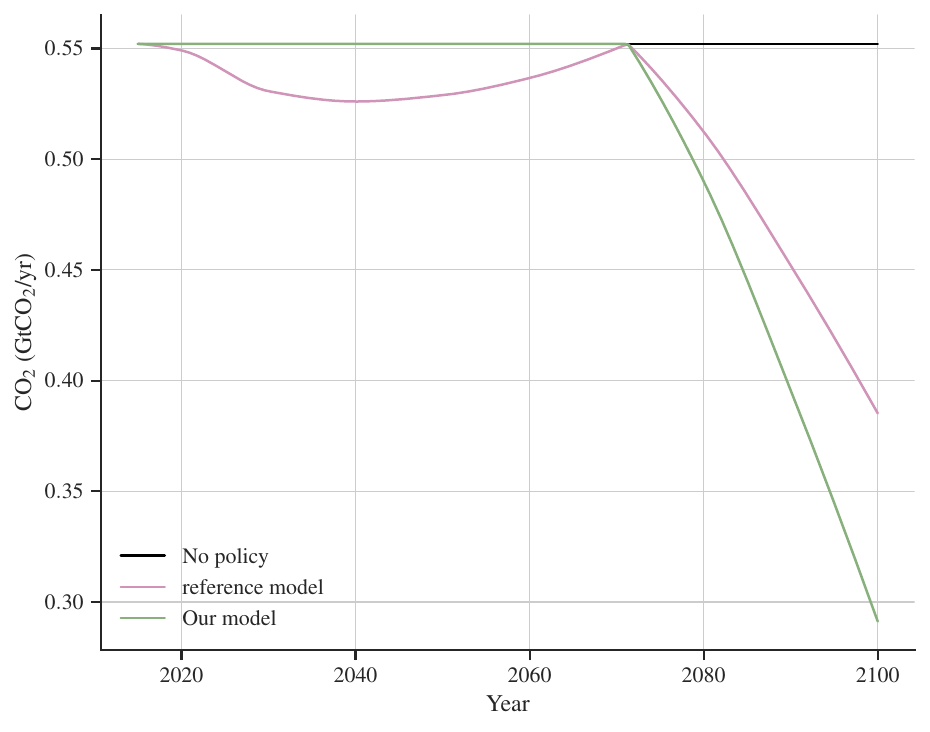}
\caption{SSP4-6.0.}
\label{fig:comparison_emissions_ssp460}
\end{subfigure}
\hfill
\begin{subfigure}[t]{0.3\linewidth}
\centering
\includegraphics[width=\linewidth]
{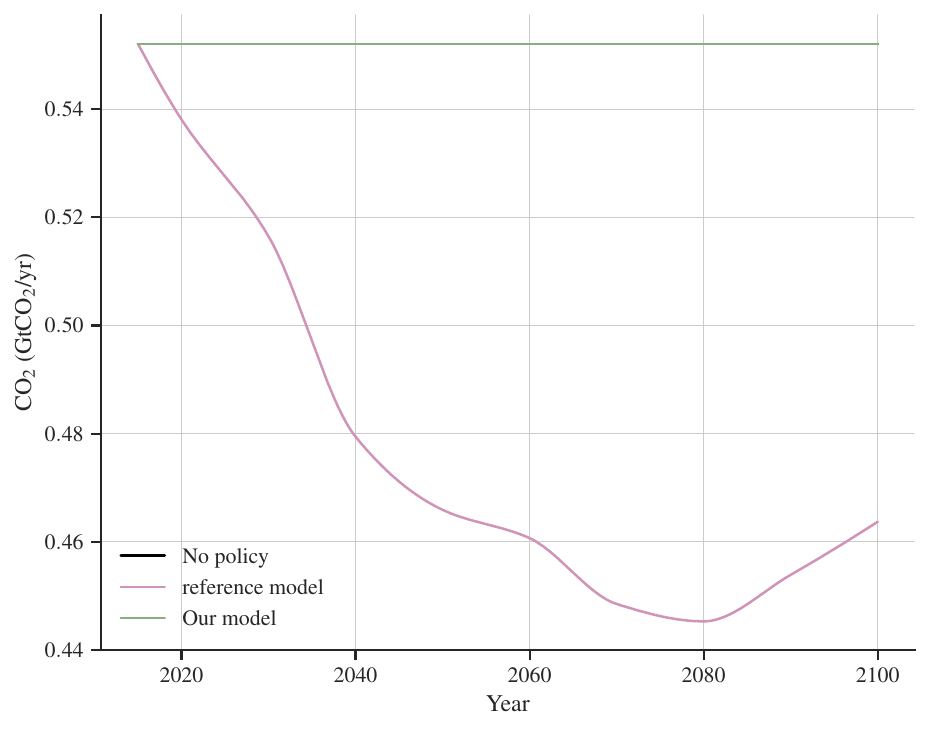}
\caption{SSP5-8.5.}
\label{fig:comparison_emissions_ssp585}
\end{subfigure}
\caption{Optimal emission in our model and in the \textit{reference model} under different SSP scenarios.}
\label{fig:comparison_emissions}
\end{figure}

Let us scale each SSP emission pathway by the factor $\Gamma_0^g / \Gamma_0^{g, \SSP}$, where $\Gamma_0^{g, \SSP}$ is the initial SSP emission in cell $g$, so that the scaled pathway starts from the aggregate emission $\Gamma_0^g$ implied by the parameters in Table~\ref{tab:comparison_parameter_settings}. Negative values of the scaled emission pathway are set to zero. The two models respond to the scenario in different ways. In our model, the emission cap matters only once it binds. As long as the scaled emission stays above $\Gamma_0^g$, the equilibrium carbon price is zero and the firm emits at its initial level. Once the scaled emission falls below $\Gamma_0^g$, the carbon price rises to the level needed for emissions to follow the cap exactly. In the \textit{reference model}, the firm is instead rewarded for emitting below the benchmark and penalized for emitting above it, with fixed weights. The corresponding optimal emission therefore decreases even when the benchmark lies above its initial emission, as in SSP3-7.0 and SSP5-8.5 (Figures~\ref{fig:comparison_emissions_ssp370} and~\ref{fig:comparison_emissions_ssp585}). However, it follows the declining emission benchmark only partially. For instance, its optimal emission in SSP1-1.9 stays well above zero, as shown in Figure~\ref{fig:comparison_emissions_ssp119}.

\begin{figure}[H]
\centering
\begin{subfigure}[t]{0.3\linewidth}
\centering
\includegraphics[width=\linewidth]
{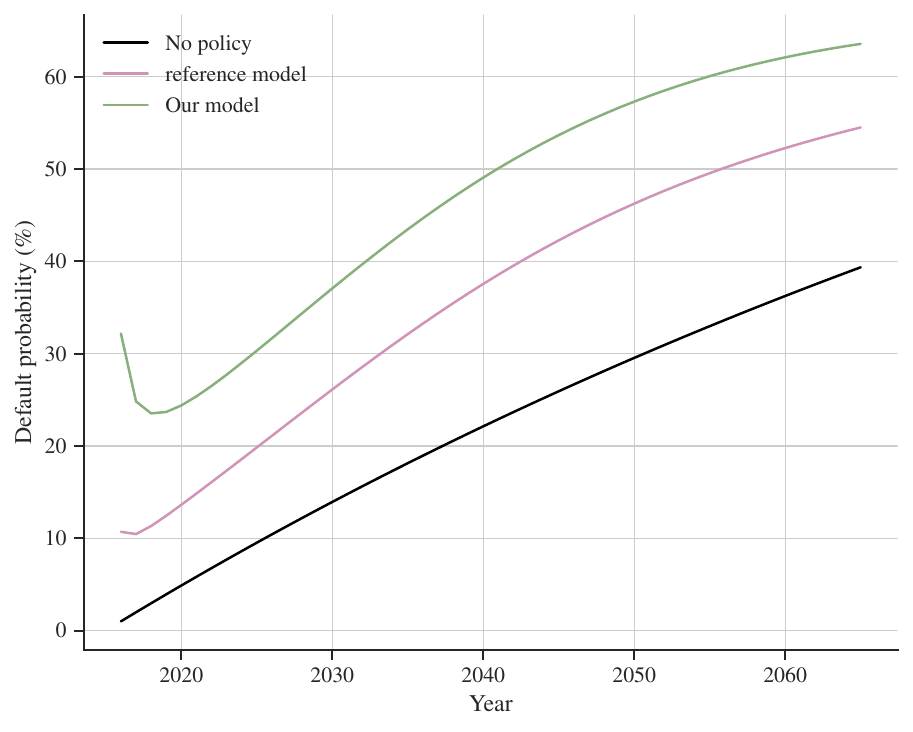}
\caption{SSP1-1.9.}
\label{fig:comparison_default_probability_ssp119}
\end{subfigure}
\hfill
\begin{subfigure}[t]{0.3\linewidth}
\centering
\includegraphics[width=\linewidth]
{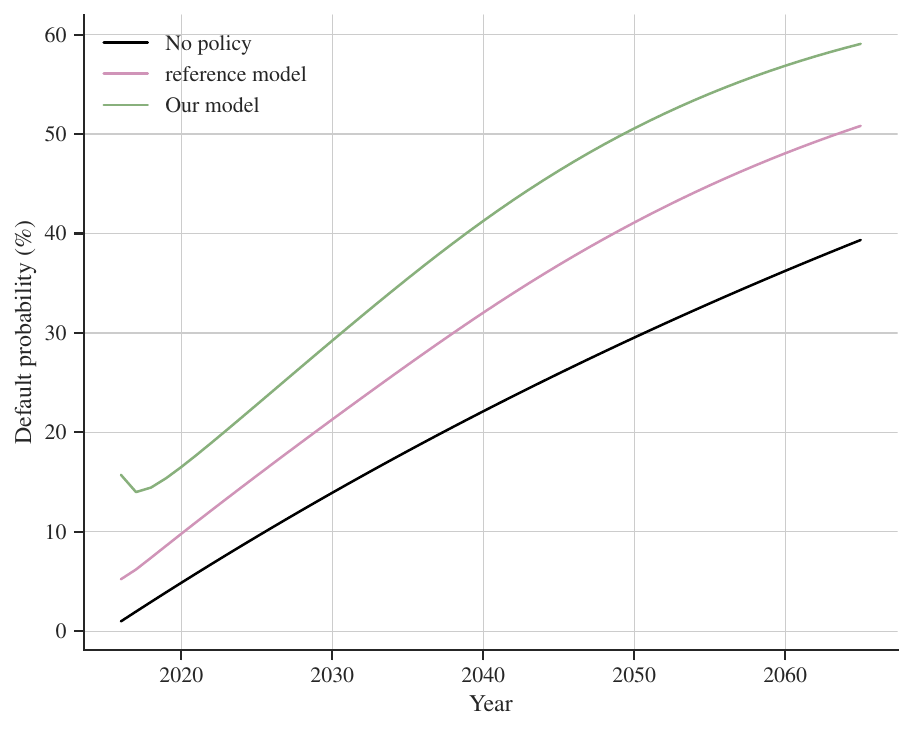}
\caption{SSP1-2.6.}
\label{fig:comparison_default_probability_ssp126}
\end{subfigure}
\hfill
\begin{subfigure}[t]{0.3\linewidth}
\centering
\includegraphics[width=\linewidth]
{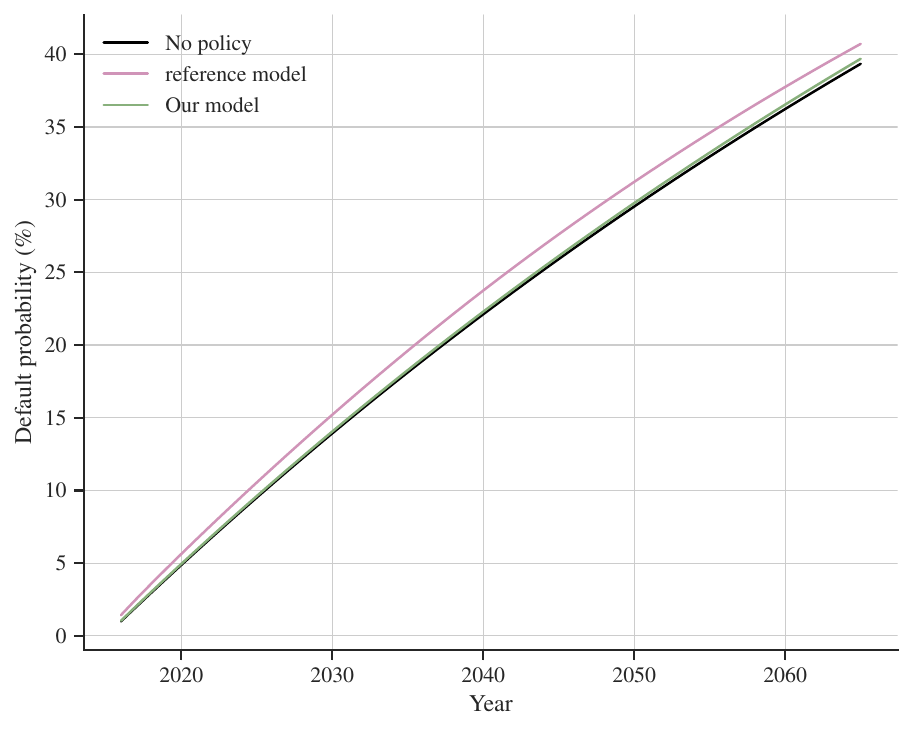}
\caption{SSP2-4.5.}
\label{fig:comparison_default_probability_ssp245}
\end{subfigure}
\par\vspace{0.5cm}
\begin{subfigure}[t]{0.3\linewidth}
\centering
\includegraphics[width=\linewidth]
{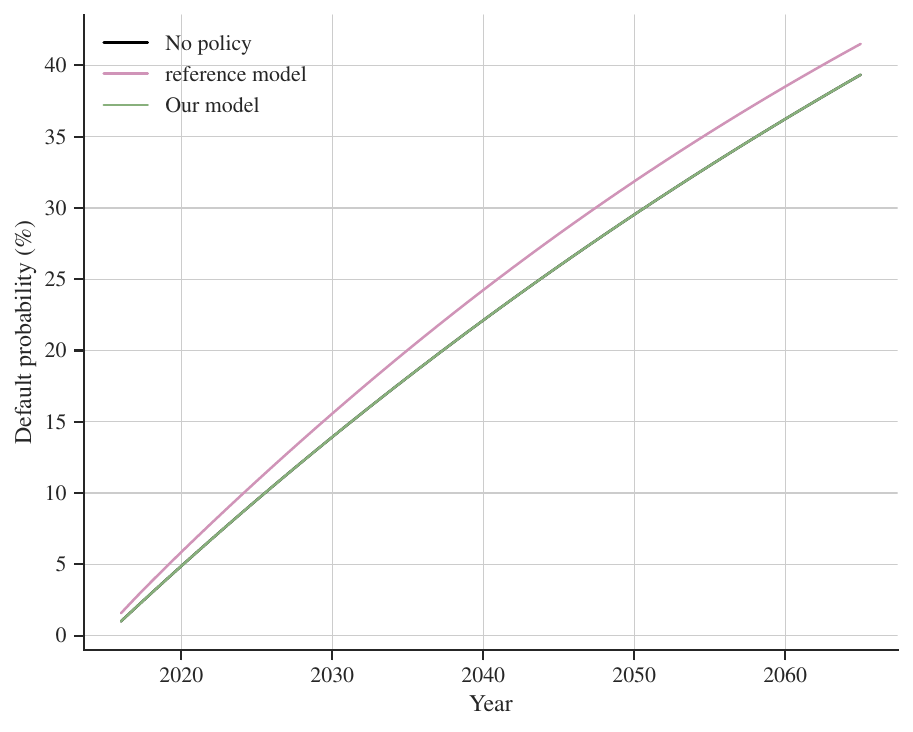}
\caption{SSP3-7.0.}
\label{fig:comparison_default_probability_ssp370}
\end{subfigure}
\hfill
\begin{subfigure}[t]{0.3\linewidth}
\centering
\includegraphics[width=\linewidth]
{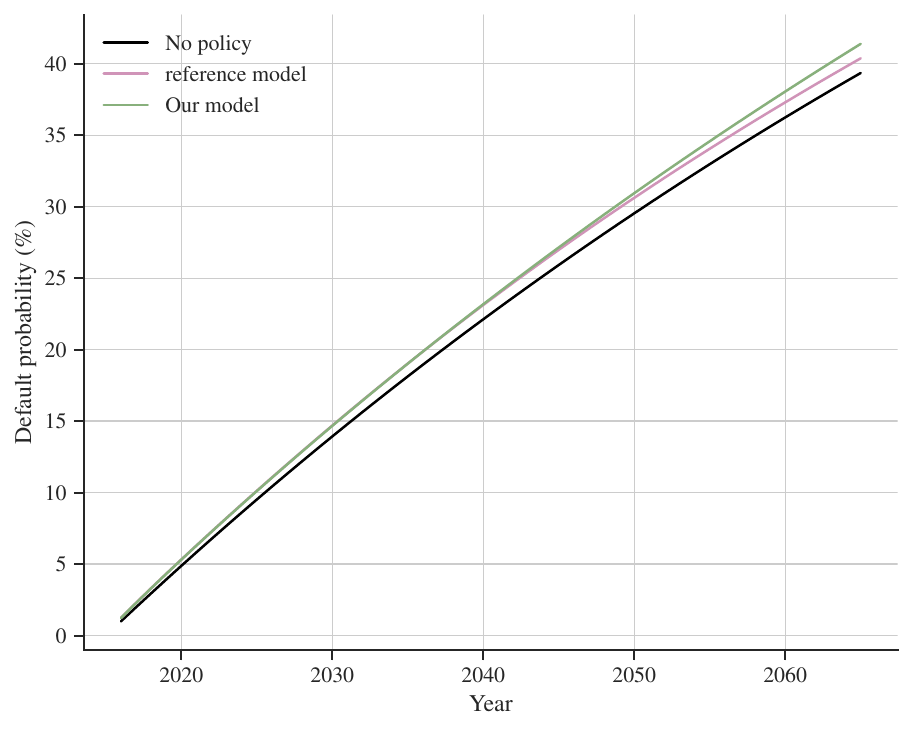}
\caption{SSP4-6.0.}
\label{fig:comparison_default_probability_ssp460}
\end{subfigure}
\hfill
\begin{subfigure}[t]{0.3\linewidth}
\centering
\includegraphics[width=\linewidth]
{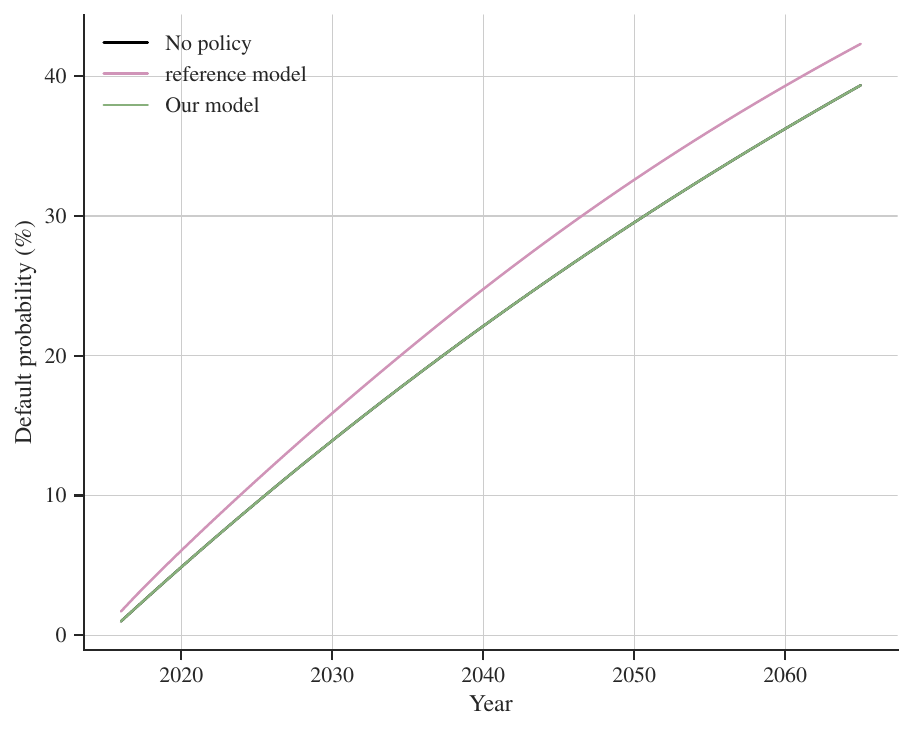}
\caption{SSP5-8.5.}
\label{fig:comparison_default_probability_ssp585}
\end{subfigure}
\caption{Default probability in our model and in the \textit{reference model} under different SSP scenarios.}
\label{fig:comparison_default_probability}
\end{figure}

The behaviour of default probability mirrors the pattern of optimal emission, as reported in Figure~\ref{fig:comparison_default_probability}. Over the horizon on which it is computed, the model with lower emission has the higher default probability, except under SSP4-6.0. Under the stringent emission trajectories, this effect is reinforced by the absence of free allowance allocation in our model, as the firm pays for all its emission, whereas in the \textit{reference model} it pays only for emission above the benchmark. We evaluate the default probability up to 2065, not the full transition period $[2015, 2100]$ available in the IIASA SSP Scenario Database \citep{Riahi2017, Rogelj2018, gmd-12-1443-2019}. Under SSP2-4.5 and SSP4-6.0, the emission cap binds only after 2065, so that the optimal emission in our model coincides with that without climate policy over the horizon. However, since the firm value in \eqref{eq:optimal_firm_value} is forward-looking, the carbon cost after 2065 is already reflected in the default probability before 2065, which explains why our model gives a higher default probability than the \textit{reference model} towards 2065 under SSP4-6.0. Moreover, under SSP1-1.9 and SSP1-2.6, the default probability in our model is high at the shortest horizons and then decreases. Since $\Sigma^{i,i} (t)$ vanishes as $t \to 0$, the default threshold $L^i (t)$ tends to the initial firm value without carbon pricing, whereas the firm value in our model is reduced by the present value of the future carbon compliance costs, so that $\DP_t^i$ tends to one as $t \to 0$. Thus, we conclude that the \textit{reference model} penalizes the firm for deviating from the scenario, whereas our model makes it meet the scenario. Transition risk arises exactly where the emission cap binds, sharply in the most stringent scenarios and not at all in the scenarios without policy introduction.

\subsection{Comparison between the linearized case and the general case}
\label{sec:comparison_linearized_general}

In the linearized case, the approximation of the optimal emission and the equilibrium carbon price is based on the approximation of the marginal value of log-production $\partial_p \Vbar^i$ in \eqref{eq:marginal_value_approx}, where the marginal revenue is evaluated along the reference path $\pbar$. Thus, the choice of $\pbar$ matters. In the comparison with the \textit{reference model}, we choose a constant reference path $\pbar = 0$, which ignores the expected growth of the production. This choice tends to underestimate $\partial_p \Vbar^i$ and hence the optimal emission. In this section, we compare the linearized case with the general case of our model. Moreover, we introduce a consistent reference path such that $\pbar_t^i = \log (\Eb [\ee^{\hp_t^i}])$ in the linearized case, which bypasses the heavy PDE computation in the general case. However, $\hp_t^i$ depends on the optimal emission $\hbgami$, which in turn depends on $\partial_p \Vbar^i$. Thus, we still need to apply the safeguarded Newton and fixed point iteration to find the equilibrium carbon price.  

In the linearized case, the optimal normalized log-production $\ee^{\hp_t^i}$ is log-normal, so the consistent reference path is given explicitly as follows
\eqlnostar{eq:consistent_ref_path}{
\pbar_t^i = \log \Bigl( \Eb \bigl[\ee^{\hp_t^i} \bigr] \Bigr) 
= m^i (t,0) + \frac12 \Sigma^{i,i} (t),
\qquad t \ge 0.
}
The marginal value of log-production $\partial_p \Vbar^i$ in \eqref{eq:marginal_value_approx} is then given explicitly by
\eqlnostar{eq:marginal_value_consistent}{
\partial_p \Vbar^i \approx \bar{\Psi}_t^i := \int_t^\infty \ee^{-(r + b^i) (u-t)}\, S_u^i P_0^i \exp \biggl(m^i (u,0) + \frac12 \Sigma^{i,i} (u) \biggr) \dd u.
}
The approximation $\bar{\Psi}^i$ of the marginal value of log-production $\partial_p \Vbar^i$, as well as $\partial_p \Vbar^i$ itself in the general case, accounts for the expected growth of production and may be higher than the constant marginal value of log-production in \eqref{eq:marginal_value_constant} for most firms in cell $g$. To make a fair comparison, we specify the capacity bound $\eta_{e}^{i, \mathrm{max}}$ in \eqref{eq:domain_control} as follows
\eqlnostar{eq:capacity_bound}{
\eta_{e, t}^{i, \mathrm{max}} = v \frac{\bigl(\varphi_{e, \theta, t}^{i, \mathrm{ref}} \bigr)^+}{2 \beta_{e,\theta, t}^i},
\qquad
\varphi_{e, \theta, t}^{i, \mathrm{ref}} := \frac{\overline{\Pi}^i c_{e, \theta, t}^i}{r+b^i} - \alpha_{e, \theta, t}^i,
\qquad
v \ge 1.
}
The parameter $v$ can be interpreted as the maximal energy usage of firm $i$ from energy source $e$, relative to its optimal emission at a zero carbon price in the linearized case with the constant reference path.

In this numerical test, we assume that there are $10, 000$ firms in the global industrial sector, with each firm $i$ having the same initial production level $P_0^i = 1$. All firm-level parameters are assumed to be constant and specified as in Table~\ref{tab:parameter_settings}, and parameters related to the numerical implementation are given in Table~\ref{tab:parameter_numerical_procedure}. The transition horizon, from 2015 to 2100, is discretized using $200$ equally spaced time nodes, including both endpoints. In the linearized case with the constant reference path $\pbar^i$, the value of $\pbar^i$ is sampled as $\pbar^i \sim \Nc (0.5, 0.15^2)$ for each firm $i$. In the linearized case with the consistent reference path, the coefficient of the capacity bound $v$ in \eqref{eq:capacity_bound} is set to $v = 1.5$. As in Section~\ref{sec:comparison_ref_model}, each SSP emission pathway is scaled to start from the aggregate emission $\Gamma_0^g$ at a zero carbon price in the linearized case with the constant reference path.

\begin{table}[H]
\centering
\begin{threeparttable}
\caption{Parameter settings.}
\label{tab:parameter_settings}
\small
\begin{tabular}{llc}
\toprule
Parameter & Notation & Value \\
\midrule
Number of energy sources             & $|\Ec^i|$                & $\Uc\{1, \dots, 10\}$ \\
Initial production                   & $P_0^i$                  & $1$ \\
Interest rate                        & $r$                      & $0.02$ \\
Linear energy cost                   & $\alpha_e^i$             & $\Uc(0, 0.02)$ \\
Quadratic energy cost                & $\beta_e^i$              & $\Uc(0.5, 1.5)$ \\
Emission factor                      & $\theta_e^i$             & $1$ \\
Production drift                     & $a^i$                    & $\Uc(0.01, 0.1)$ \\
Mean-reversion speed                 & $b^i$                    & $\Uc(0.1, 0.5)$ \\
Energy efficiency                    & $c_e^i$                  & $\Uc(0.01, 0.15)$ \\
Volatility                           & $\sigma^i$               & $\Uc(0.15, 0.5)$ \\
Common factor                        & $\rho^i$                 & $0.3$ \\
Selling price                        & $S^i$                    & $\Uc(0.5, 2.0)$ \\
Free allowance allocation            & $A_t^i$                  & $0$ \\
Downside demand shock                & $q_t^g$                  & $1$ \\
Physical risk and adaptation         & --                       & excluded \\
\bottomrule
\end{tabular}
\begin{tablenotes}[flushleft]
\footnotesize
\item Firms whose long-term normalized log-production at a zero carbon price exceeds $\log(5)$ are redrawn.
\end{tablenotes}
\end{threeparttable}
\end{table}

\begin{table}[H]
\centering
\begin{threeparttable}
\caption{Parameters in the numerical procedure.}
\label{tab:parameter_numerical_procedure}
\small
\begin{tabular}{llc}
\toprule
& Parameter & Value \\
\midrule
PDE solver 
& Domain margin              & $8\sigma^i/\sqrt{2b^i}$ \\
& Grid points                & 801 \\
& Picard tolerance           & $10^{-8}$\\
& Maximal Picard iterations  & 50 \\
\midrule
Fixed point iteration
& $\vartheta$                & 0.5\\
& $\epsilon$                 & $10^{-3}$\\
& Maximal iterations         & 100\\
\midrule
Simulation 
& $N$                        & 50\\
& $N'$                      & 500\\
& $R$                        & $1,000$\\
\bottomrule
\end{tabular}
\end{threeparttable}
\end{table}

\begin{figure}[H]
\centering
\begin{subfigure}[t]{0.45\linewidth}
\centering
\includegraphics[width=\linewidth]
{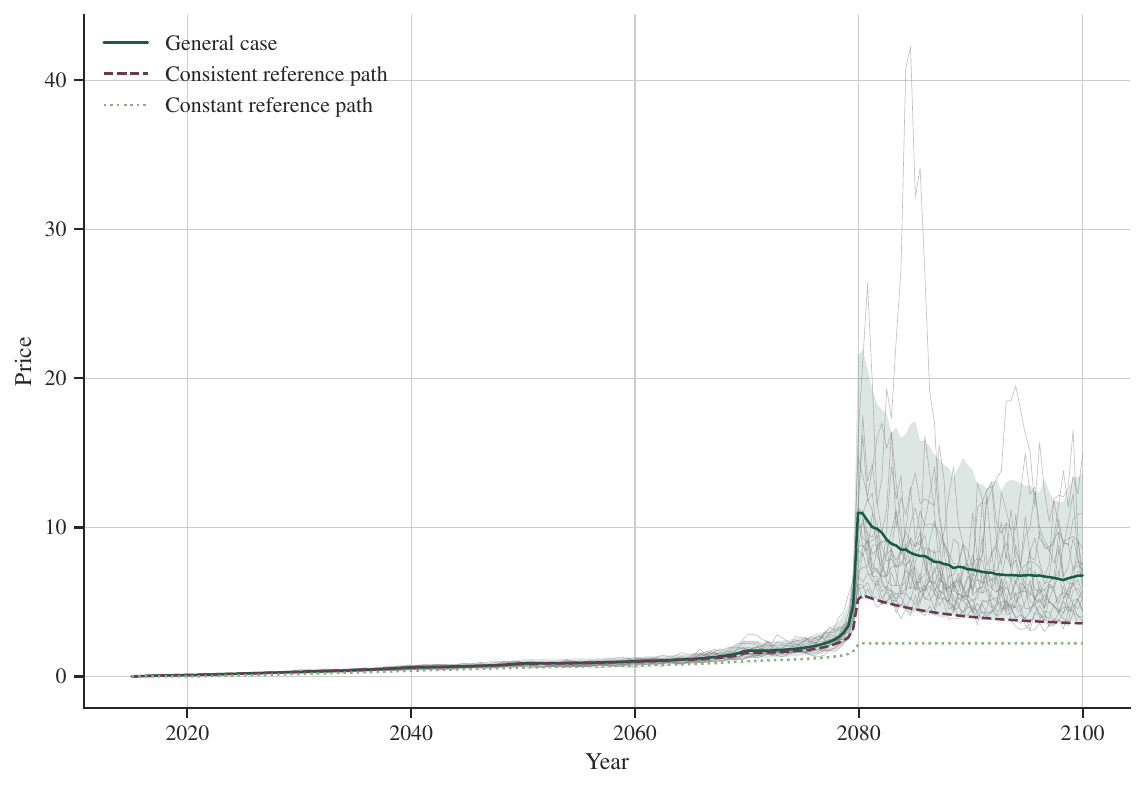}
\caption{SSP1-1.9.}
\label{fig:scaled_carbon_price_ssp119}
\end{subfigure}
\hfill
\begin{subfigure}[t]{0.45\linewidth}
\centering
\includegraphics[width=\linewidth]
{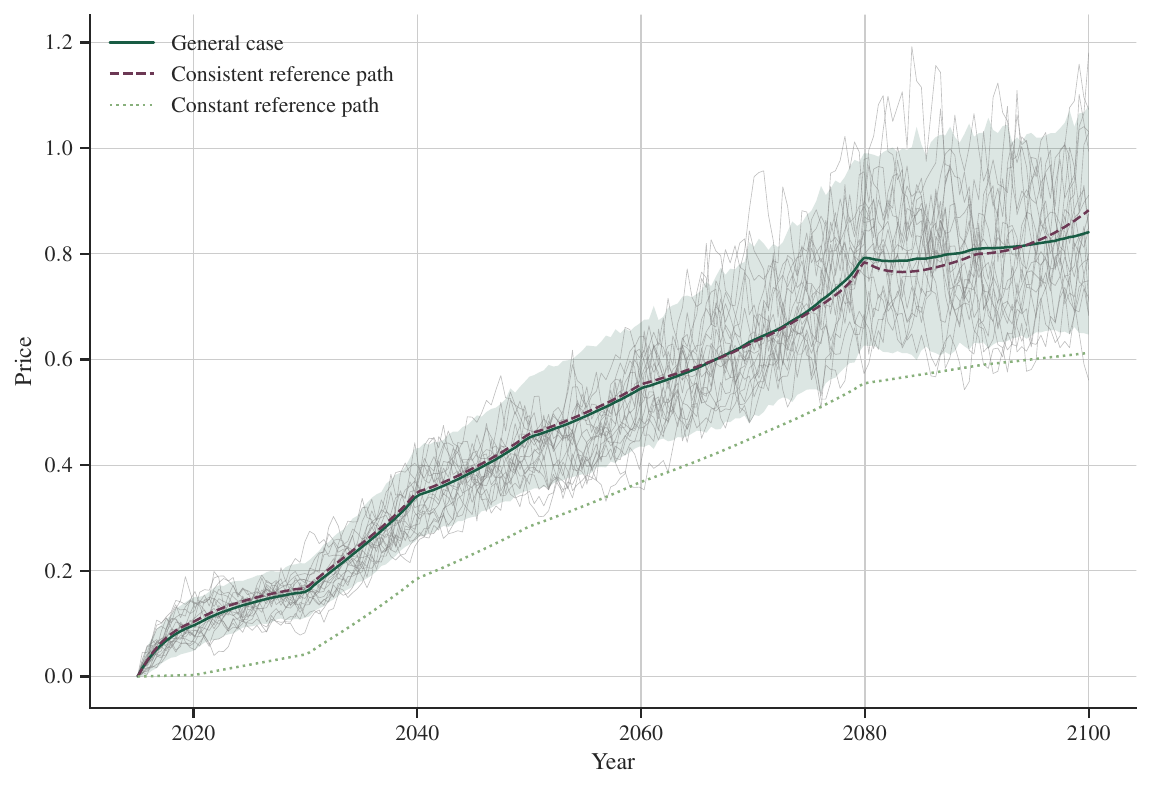}
\caption{SSP1-2.6.}
\label{fig:scaled_carbon_price_ssp126}
\end{subfigure}
\par\vspace{0.5cm}
\begin{subfigure}[t]{0.45\linewidth}
\centering
\includegraphics[width=\linewidth]
{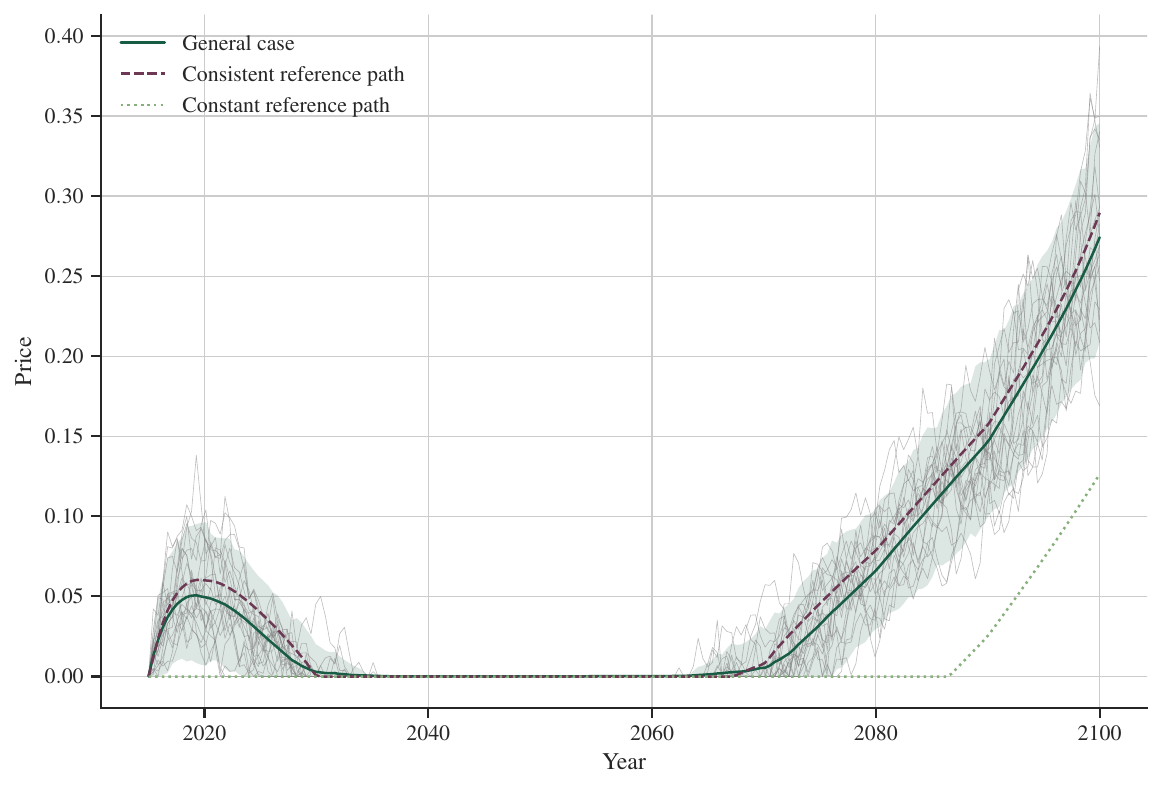}
\caption{SSP2-4.5.}
\label{fig:scaled_carbon_price_ssp245}
\end{subfigure}
\hfill
\begin{subfigure}[t]{0.45\linewidth}
\centering
\includegraphics[width=\linewidth]{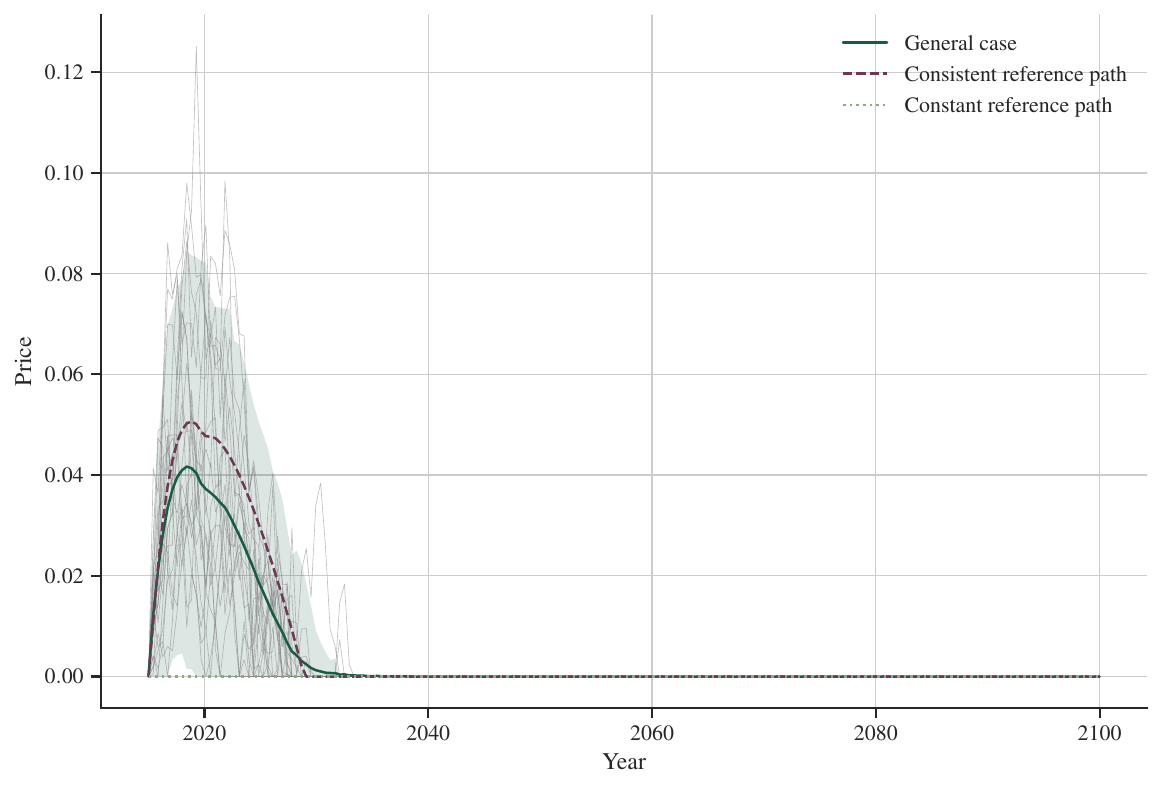}
\caption{SSP3-7.0 and SSP5-8.5.}
\label{fig:scaled_carbon_price_ssp370}
\end{subfigure}
\caption{Equilibrium carbon price in both the general and the linearized case under different SSP scenarios.}
\label{fig:scaled_carbon_price_ssp}
\end{figure}

The resulting equilibrium carbon price is reported in Figure~\ref{fig:scaled_carbon_price_ssp}, where the shaded area represents the 5\% and 95\% quantiles of the equilibrium carbon price over all simulated paths. The constant reference path ignores the expected production growth, so that, once the cap binds, its carbon price is much lower than in the other two cases, and it stays at zero under SSP3-7.0 and SSP5-8.5. In the other two cases, the carbon price is positive during the first fifteen to twenty years under SSP2-4.5, SSP3-7.0, and SSP5-8.5, as the cap is scaled to the emission at $t=0$, and the emission that the firms want grows faster than the cap while the production converges to it long-term level. Moreover, the linearized case with the consistent reference path captures most of the carbon price of the general case, and is slightly higher under these three scenarios. Indeed, in the general case, the capacity bound in \eqref{eq:capacity_bound} limits the increase of the emission when the production is high, which then does not fully offset its decrease of the emission when the production is low. Note that the carbon price under SSP1-1.9 jumps around 2080, when the cap reaches zero, and is then given by the choke price, the highest marginal value of emissions across all firm-energy pairs with positive capacity. In the general case, the choke price exceeds that of the linearized case with the consistent reference path, since the highest marginal value is taken over the realized production states.

\begin{figure}[H]
\centering
\begin{subfigure}[t]{0.45\linewidth}
\centering
\includegraphics[width=\linewidth]{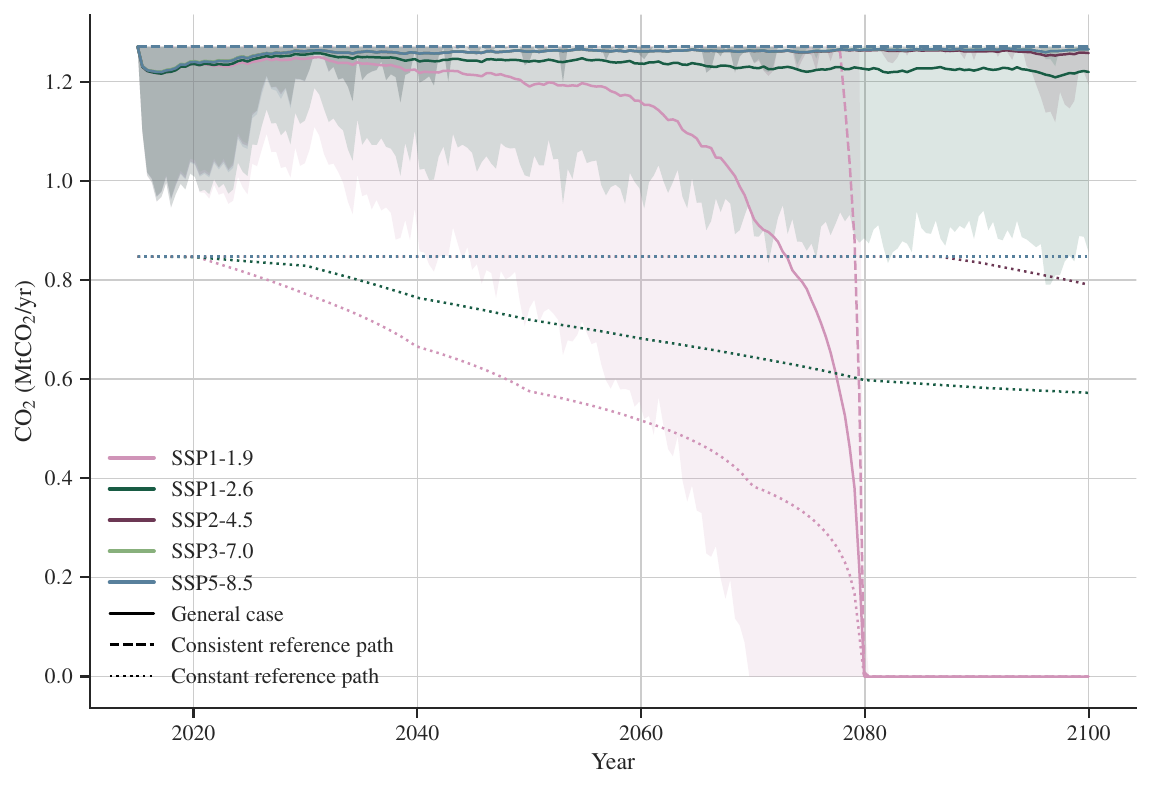}
\caption{Optimal emission of the firm with the highest production.}
\label{fig:scaled_firm_emissions_highest}
\end{subfigure}
\hfill
\begin{subfigure}[t]{0.45\linewidth}
\centering
\includegraphics[width=\linewidth]{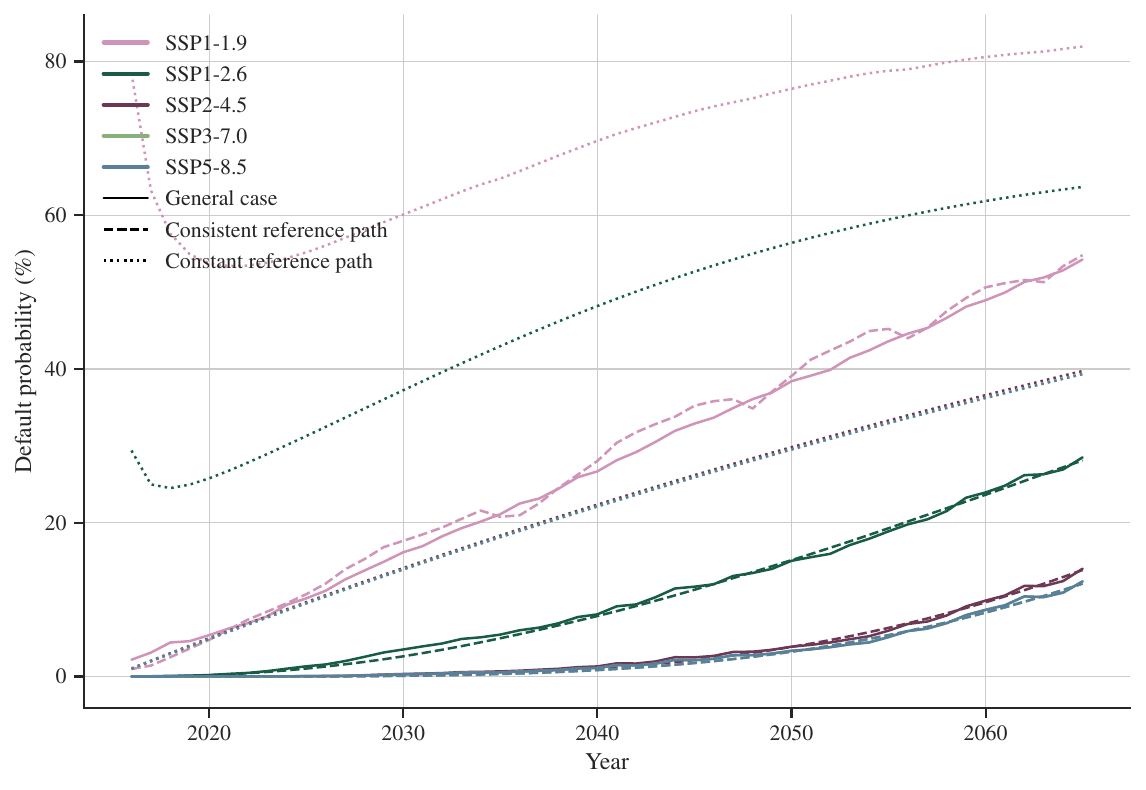}
\caption{Default probability of the firm with the highest production.}
\label{fig:default_probability_highest}
\end{subfigure}
\par\vspace{0.5cm}
\begin{subfigure}[t]{0.45\linewidth}
\centering
\includegraphics[width=\linewidth]{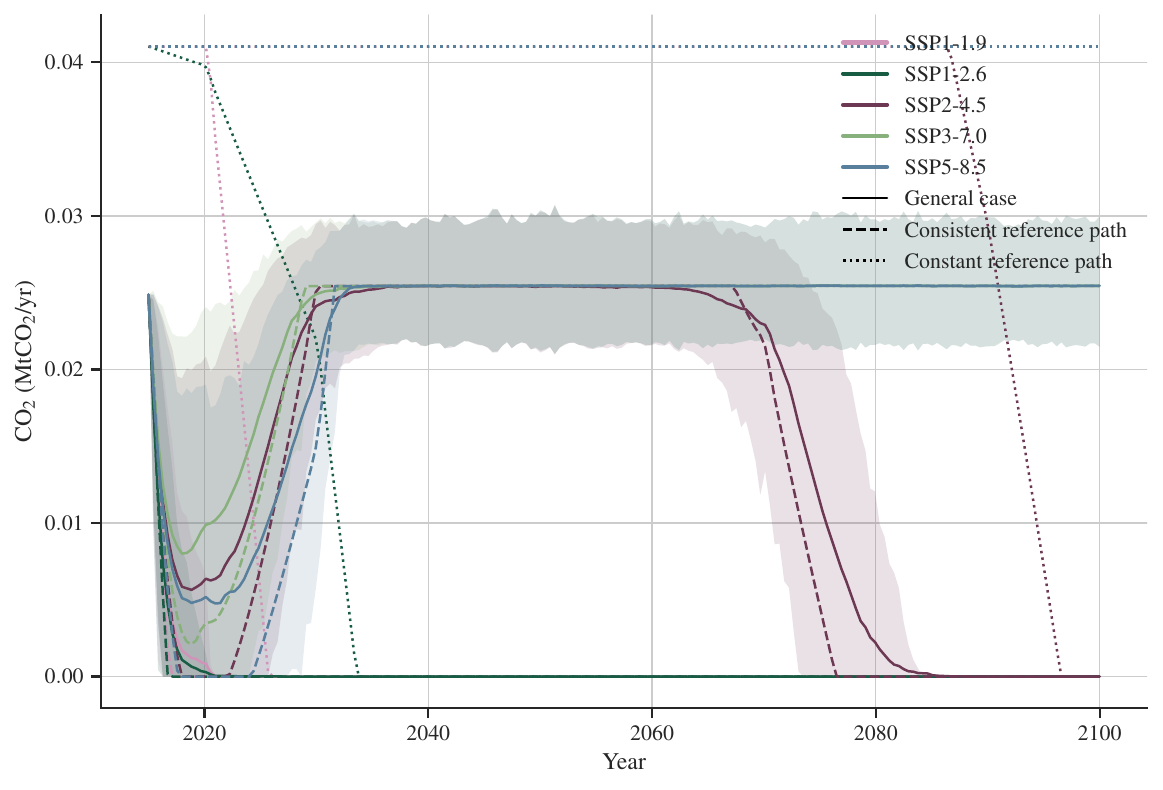}
\caption{Optimal emission of the firm with the lowest production.}
\label{fig:scaled_firm_emissions_lowest}
\end{subfigure}
\hfill
\begin{subfigure}[t]{0.45\linewidth}
\centering
\includegraphics[width=\linewidth]{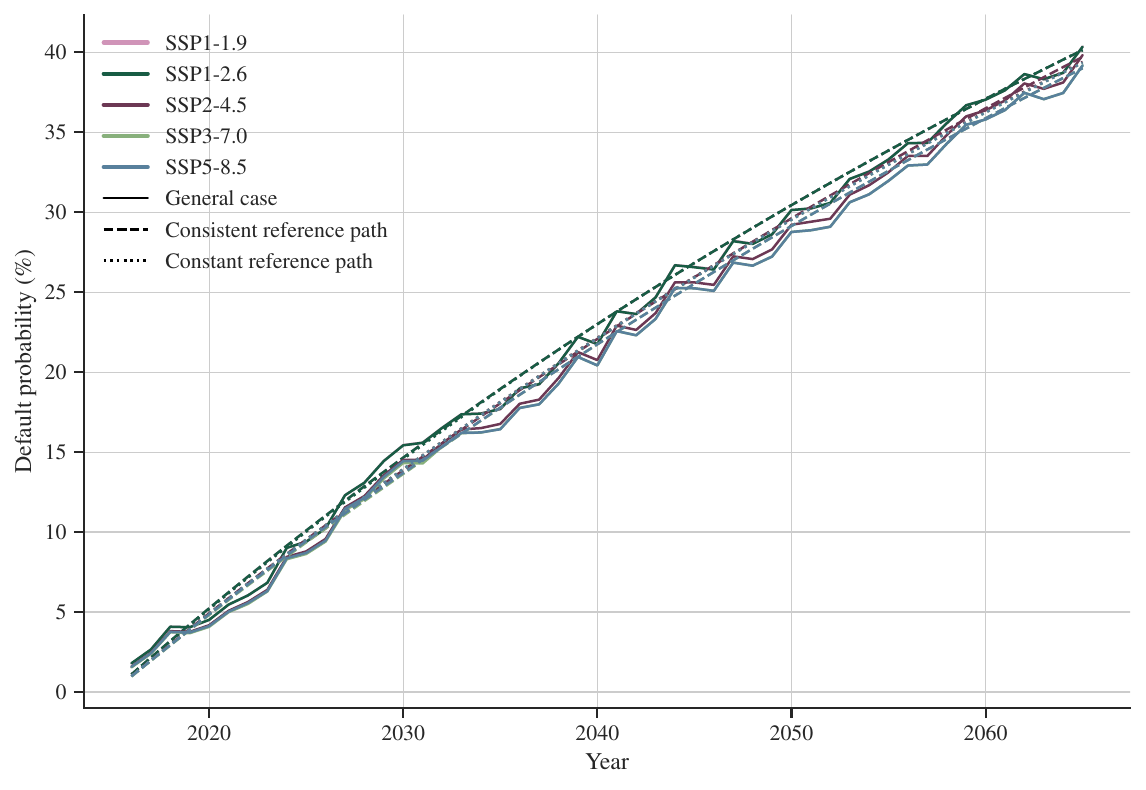}
\caption{Default probability of the firm with the lowest production.}
\label{fig:default_probability_lowest}
\end{subfigure}
\caption{Optimal emission and default probability of selected firms in both the general and linearized case under different SSP scenarios.}
\label{fig:emissions_default_probability}
\end{figure}

Since the aggregate optimal emission is capped by the SSP emission pathway, we instead report the optimal emission of the firms with the highest and lowest normalized log-production at the end of the transition horizon in the general case, as well as their default probabilities in Figure~\ref{fig:emissions_default_probability}. For the firm with the highest production, the constant reference path underestimates the benefit of energy use, which leads to a lower emission, a lower production, and a higher default probability than in the other two cases. However, the ranking of the optimal emission is reversed for the firm with the lowest production. Since energy use brings this firm little additional production, its optimal emission in the linearized case with the consistent reference path and in the general case falls during the first decade, when the carbon price is positive in all scenarios, and reaches zero early under SSP1-1.9 and SSP1-2.6. Under the other three scenarios, it returns to the level without climate policy once the carbon price falls to zero, before falling to zero again after 2070 under SSP2-4.5. Moreover, since the optimal emission is nearly zero or remains small, the difference in the carbon price has little effect on the firm's cash flows, which explains the almost identical default probability across the scenarios in Figure~\ref{fig:default_probability_lowest}.

\section{Calibration}
\label{sec:calibration_climate_risk}

In this section, we perform a calibration of our model, including both the linearized and general case, to the European power sector under the EU ETS, where the cap, emission, and generation by fuel are disclosed at the firm level. Since the data of physical losses and adaptation are only coarsely disclosed, we focus on the calibration of transition risk and leave the physical risk for future work. We first calibrate the model using the data ranging from 2002 to 2024, and then use the calibrated model to see whether the carbon price matches the European Union Allowance (EUA) carbon permit spot and futures prices, whether the default probability matches the term structure implied by the credit default swap (CDS) market, and how the default probability of the European power sector changes under different NGFS scenarios.

\subsection{Calibration procedure}
\label{sec:calibration_procedure_climate_risk}

The data of the European power sector are collected from various sources, including Trucost, Compustat Global, etc. Details of the data sources are reported in Table~\ref{tab:data_sources_calibration_climate_risk}. After data cleaning and preprocessing, we obtain a dataset of 126 firms in the European power sector, each of which reports its own emission data rather than only Trucost estimates. In this dataset, we keep the firms under the EU ETS, which removes the 21 firms under the UK and Swiss ETSs. We then remove 14 grid operators and 17 firms whose generation is not their main business, and keep the firms that report generation by fuel in at least four years between 2018 and 2024, as the EUA price became a material cost after the 2018 market stability reserve reform. We further drop the six subsidiaries whose parent company is itself in the sample. The resulting 35 firms account for two-thirds of the combustion emissions in the European Economic Area (EEA) Union Registry, and the emission cap is scaled by that share. 

\begin{table}[H]
\centering
\caption{Data sources for the calibration to the European power sector.}
\label{tab:data_sources_calibration_climate_risk}
\small
\setlength{\tabcolsep}{4pt}
\begin{tabular}{@{}p{3.4cm}p{6.2cm}p{4.6cm}@{}}
\toprule
Source & Data & Parameters \\
\midrule
Trucost & Scope 1 emissions, generation and emissions by fuel & $P_t^i$, $\gamma_e^i$, $\theta_e^i$ \\
Compustat Global & Sales, EBITDA, exchange rates & - \\
Markit & 5-year CDS spreads and 1--10-year curves & $\zeta_{\mathrm{ref}}^i$ \\
LSEG & WACC, bond spreads, EUA futures, GB power prices & - \\
EEA Union Registry & Verified emissions and free allocation, combustion installations & $\tGam_0$, $A_0$ \\
Directive (EU) 2023/959 & Linear reduction factor & Legislated cap path \\
NGFS Phase V & Electricity emissions, generation by fuel and demand by scenario & Scenario cap paths, $q_t^g$ \\
World Bank & Coal, gas and oil prices & $\alpha_{e,\theta}$ \\
ICAP & EUA prices & $\Pbar^g$ \\
Eurostat & Non-household electricity prices by country & $S_t^i$\\
IPCC 2006 & Default emission factors & Cross-check of $\theta_e^i$ \\
\bottomrule
\end{tabular}
\end{table}
 
The production $P_t^i$ of firm $i$ in the European power sector is its physical generation, normalized by its average over 2018 to 2024, and averaging over the whole period prevents the 2020 demand drop and the 2022 fuel-price spike from setting the reference state. Thus, $p=0$ is the calibration point. The production dynamics in \eqref{eq:log_production_sde} are then estimated by discretizing the SDE annually with firm fixed effects on the generation panel, and the fossil fuel input as the energy regressor. That is, we regress
\eqlnostar{eq:log_production_estimate}{
p_{t+1}^i - p_t^i = a^i - b^i p_t^i + c^i x_t^i + \epsilon_{t+1}^i,
}
where $x_t^i$ is the fossil fuel input of firm $i$ relative to its average over 2018 to 2024. The pooled slope on lagged output gives the value of $b^i$, the standard deviation of the residuals of each firm gives the value of $\sigma^i$, and, since the residuals of two firms have correlation $\rho^i \rho^j$ by \eqref{eq:combined_bm}, the square root of the mean pairwise residual correlation gives the common value of $\rho^i$. Since the panel is short, the fixed-effects estimate of $b^i$ is biased upward \citep{beggsBiasesDynamicModels1988}, and we use the split-panel jackknife \citep{dhaeneSplitpanelJackknifeEstimation2015} to correct it.
The energy-related term in \eqref{eq:log_production_sde} is estimated from the physical identity of a generator. According to the data, the total generation is fossil plus clean generation, so the fossil contribution to log output at the fixed point is minus the log of the clean share. Thus, we have
\eqstar{
\sum_{e \in \Ec^i} \frac{c_{e, \theta}^i \bar{\gamma}_e^i}{b} = - \log (1-s^i),
}
where $\bar{\gamma}_e^i$ is the average emission of energy source $e$ for firm $i$ between 2018 and 2024, and $s^i$ is the share of the fossil generation. The left-hand side is allocated to fuels by their generation shares, so that one MWh of fossil electricity carries the same weight in output whatever the fuel. The drift $a^i$ is then given by $- \sum_{e \in \Ec^i} c_{e, \theta}^i \bar{\gamma}_e^i$, which keeps the production at its mean over the calibrated time horizon under the observed emissions, and every effect of the transition on production is captured by $c$. 

Since the quadratic energy cost $\beta_{e, \theta}^i$ in \eqref{eq:energy_cost} is not directly observable, we estimate it by the FOC in \eqref{eq:foc}, which requires the marginal value of log-production $\partial_p \Vbar^i$. We calibrate $\beta_{e, \theta}^i$ in the linearized case with the constant reference path $\pbar^i = 0$, and denote by $\Pi_{\mathrm{eff}}^i$ the revenue scaled with the firm's output, treating it as the one free level of the firm. By \eqref{eq:constant_marginal_revenue} and \eqref{eq:marginal_value_constant}, we have
\eqstar{
\sum_{e \in \Ec^i} c_{e, \theta}^i \bar{\gamma}_e^i \frac{\Pi_{\mathrm{eff}^i}}{r+b} = \sum_{e \in \Ec^i} \bigl(\alpha_{e, \theta}^i + \Pbar \bigr) \Bigl( 1- \frac{1}{\varepsilon} \Bigr)\bar{\gamma}_e^i,
}
where $\varepsilon$ is the cost elasticity of emission, estimated by regressing the log emissions on the log of that cost with firm-by-fuel fixed effects. Only the elasticity of coal is significantly estimated, since gas and oil plants set the electricity price and their emissions follow the demand. Hence, we apply the coal elasticity to every fuel and keep the implied elasticities of the gas and oil pairs as a test. The resulting quadratic energy cost $\beta_{e, \theta}^i$ is then given by
\eqstar{
\beta_{e, \theta}^i = \frac{1}{2 \bar{\gamma}_e^i} \Bigl( \frac{c_{e, \theta}^i \Pi_{\mathrm{eff}}^i}{r+b} - \alpha_{e, \theta}^i - \Pbar \Bigr),
}
and the capacity bound in \eqref{eq:domain_control} is given by \eqref{eq:capacity_bound} with $v = 1.25$, instead of $v=1.5$ in Section~\ref{sec:comparison_linearized_general}, as the fixed point iteration in Algorithm~\ref{alg:equilibrium_carbon_price} does not converage for a larger $v$ in the calibration.

Two features of the power sector that the model does not price are introduced in the calibration. The first one is the replacement of the lost fossil output by clean generation. Let $\Upsilon_t^{i, X}$  denote the firm's optimal emission under scenario $X$. We replace the downside demand shock $q_t^g$ in \eqref{eq:demand_shock} by a firm-specific factor $q_t^i$, under which a share $\delta$ of the fossil output removed by the scenario is recovered through the demand adjustment, that is, 
\eqstar{
\log \bigl( q_t^i \bigr) = \delta \bar{c}^i \frac{\bigl(\Upsilon_0^i - \Upsilon_t^{i, X} \bigr)^+}{b}, \qquad \delta \in [0,1],
}
where $\bar{c}^i$ is the emission-weighted average of the energy efficiency $c_{e, \theta}^i$ over the fossil fuels. The share $\delta$ is estimated from the same panel by the regression
\eqstar{
p_t^i - p_{t-1}^i = a^i - b p_{t-1}^i + \delta z_{t-1}^i + \epsilon_t^i,
}
with firm fixed effects and standard errors clustered by firm, where $z_t^i := \sum_{e \in \Ec^i} c_{e, \theta}^i \gamma_{e, t}^i$.
The second feature is the pass-through of the carbon compliance cost of the marginal gas plant to the wholesale electricity price, which is earned on the clean output outside the controlled state. We thus multiply the revenue of firm $i$ by a factor $1 + \omega^i \psi \theta_m f_t \Pbar_t^g$, where $\omega^i$ is the clean generation per unit of $\Pi_{\mathrm{eff}}^i$, $\theta_m$ is the emission intensity of the marginal gas plant, and $f_t$ is the share of the fossil fleet still operating. The pass-through factor $\psi$ is the coefficient of the carbon price in a regression of the wholesale electricity price on the carbon price and the gas cost per MWh, divided by $\theta_m$, on the semi-annual Eurostat country panel of 2019 to 2021, with the monthlyGB day-ahead price of 2015 to 2020 as a check. The year from 2022 cannot be used, since the carbon and gas prices rose together, and the carbon coefficient then absorbs the gas cost and implies a pass-through above one. The literature value of 0.8, in line with the average pass-through about 80\% estimated by \citet{fabraPassThroughEmissionsCosts2014} for the Spanish market and the high pass-through reported by \citet{sijmCO2CostPassthrough2006}, is kept as the upper sensitivity. 

\subsection{Calibration results}
\label{sec:calibration_results_climate_risk}

\begin{figure}[H]
\centering
\begin{subfigure}[t]{0.45\linewidth}
\centering
\includegraphics[width=\linewidth]
{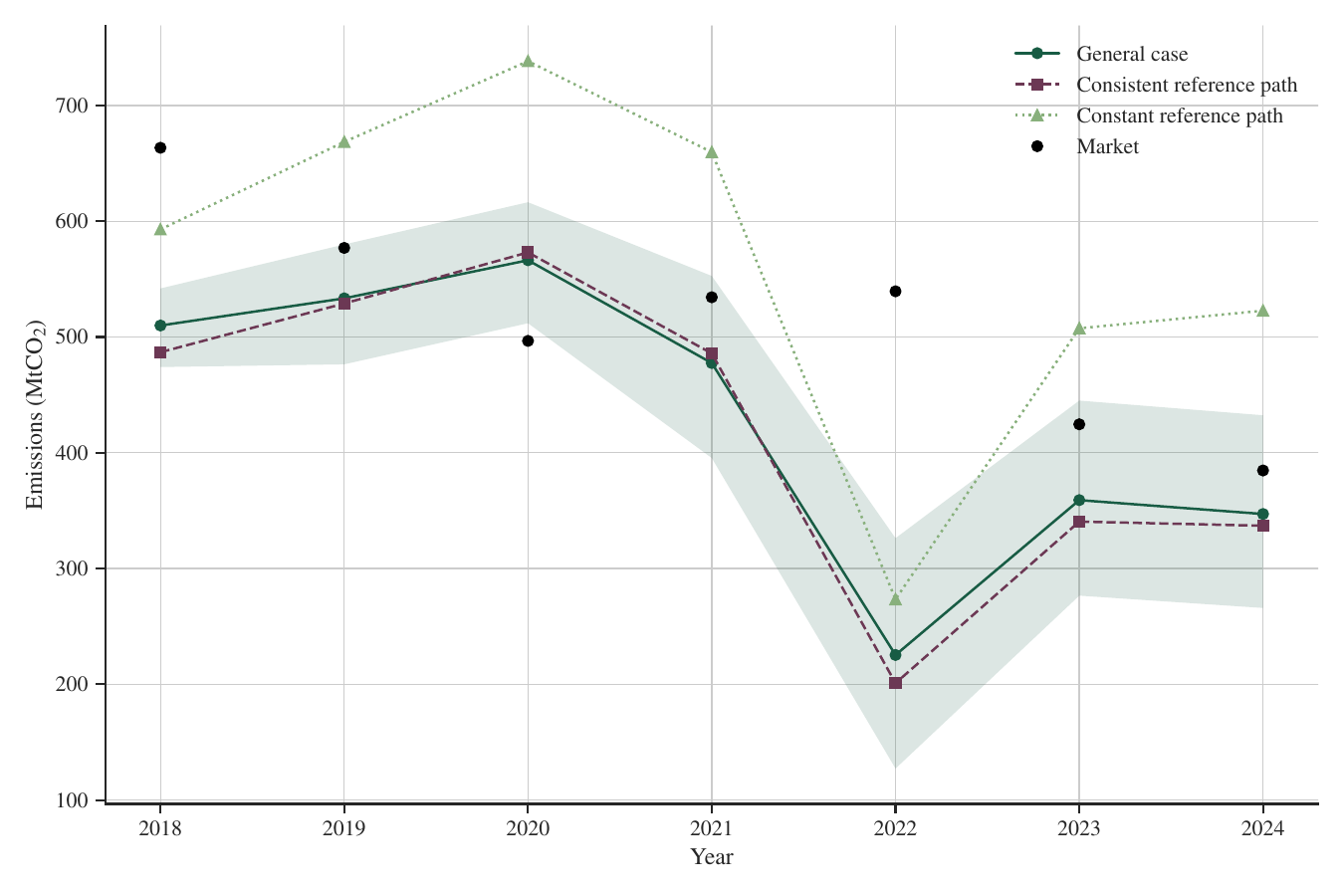}
\caption{Aggregated emission.}
\label{fig:emission_calibration}
\end{subfigure}
\hfill
\begin{subfigure}[t]{0.45\linewidth}
\centering
\includegraphics[width=\linewidth]
{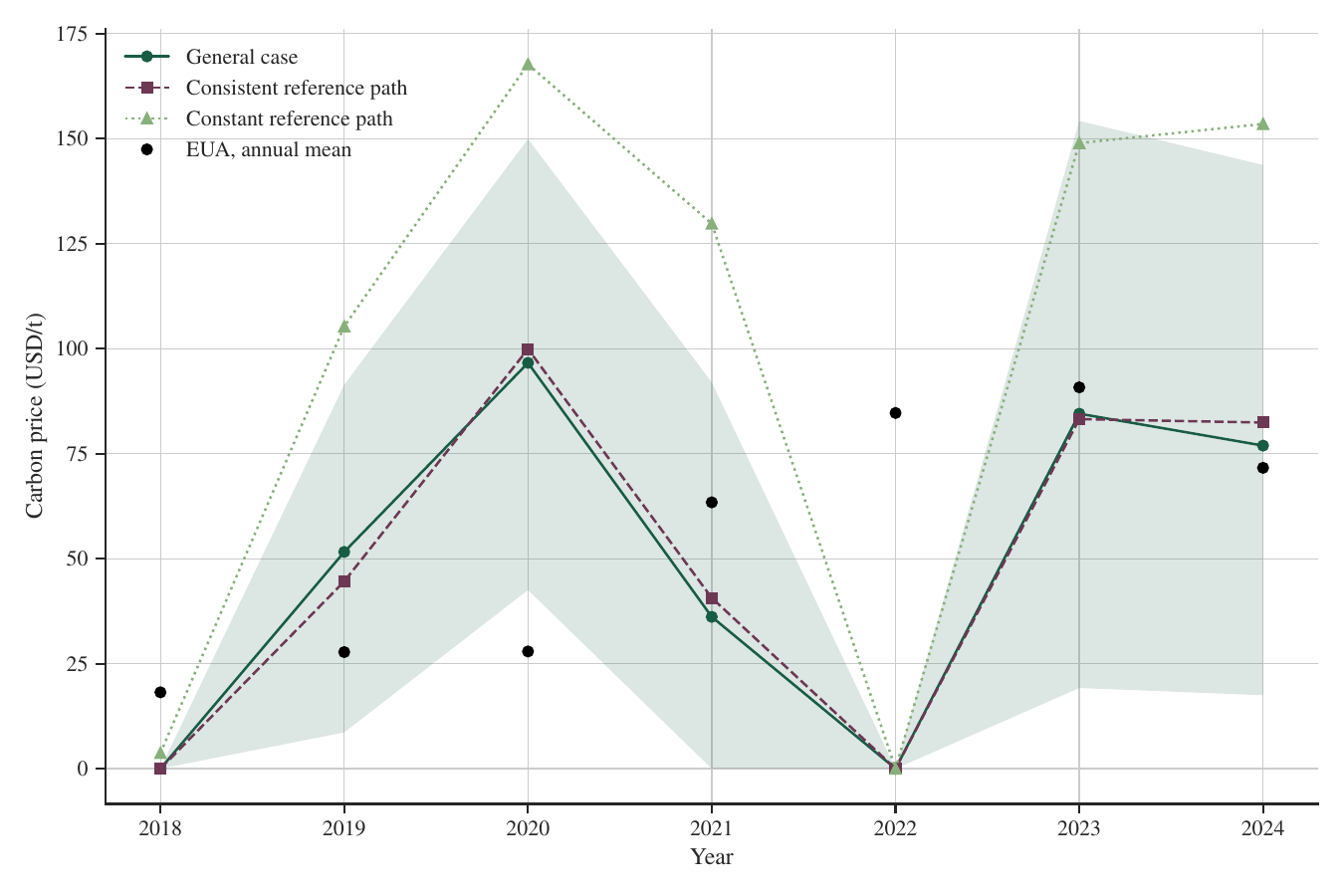}
\caption{Carbon price.}
\label{fig:carbon_price_calibration}
\end{subfigure}
\caption{Calibrated aggregated optimal emission and the equilibrium carbon price in both the general and the linearized case between 2018 and 2024.}
\label{fig:emission_carbon_price_calibration}
\end{figure}

The calibrated parameters are reported in Table~\ref{tab:calib_params}. Evaluated at the observed carbon and fuel prices, the calibrated supply curve reproduces the registry's emissions with a mean absolute error of 22\% in the linearized case with the consistent reference path and 20\% in the general case, against 29\% with the constant reference path, as shown in Figure~\ref{fig:emission_calibration}. Correspondingly, we present the carbon price that clears the registry's emissions under both the linearized and general case in Figure~\ref{fig:carbon_price_calibration}, with the band of 5\% and 95\% quantiles of the simulated carbon price in the general case. We can see that both the calibrated emission and the carbon price lose their accuracy in 2020 and 2022 due to the COVID-19 pandemic and the Russia-Ukraine war, respectively. Another cause lies in the FOC in \eqref{eq:foc}, where the fuel cost of the period is set against a forward-looking value of output, in which a one-year movement of the electricity price weighs about a fifth. To identify the cause of these failures, we perform three diagnostic tests, each of which changes one element of the calibration.
\begin{enumerate}
\item [(a)] Use the fuel cost smoothed with the forward-looking weight of the selling price.
\item [(b)] Let the electricity price enter the marginal value of log-production in the period itself.
\item [(c)] Use the GB day-ahead price as the electricity price.
\end{enumerate}
The three tests indicate that the failure is structural rather than due to the data. When the fuel cost and the revenue are treated symmetrically in test (a), the carbon price of the model follows the changes of the EUA price from year to year, although at a lower level. Letting the revenue of the period enter the marginal value in test (b) only moves the failure to the years of cheap fuel, and replacing the electricity price in test (c) does not improve the fit. Hence, the failure comes from setting the fuel cost of the period against a forward-looking value of output. In 2022, the gas price rose much faster than this value, so that no gas plant is profitable at a positive carbon price, and the opposite holds in 2019 and 2020. Removing this mismatch requires a production technology in which the current energy input enters the current output, which we leave for future research, and we keep the baseline calibration.
Although the calibration results are not satisfactory, we still use the calibrated parameters to see whether the forward-looking carbon price matches the EUA December futures prices from 2025 to 2030. As shown in Figure~\ref{fig:carbon_price_prediction}, the linearized case with the consistent reference path and the general case underestimate the EUA December futures prices, while the linearized case with the constant reference path overestimates them. However, all three cases can capture the trend of change.

\begin{table}[H]
\centering
\begin{threeparttable}
\caption{Calibrated parameters of the European power sector under the EU ETS.}
\label{tab:calib_params}
\footnotesize
\setlength{\tabcolsep}{1.6pt}
\begin{tabular}{llc@{\hspace{0.5cm}}llc}
\toprule
Mean reversion & $b$ & 0.17 & Cost elasticity, coal & $\varepsilon$ & $-0.51$ (0.17) \\
Volatility, generation / EBITDA & $\sigma^i$ & 0.17 / 0.31 & Implied elasticity, gas / oil & & $-0.53$ / $-0.97$ \\
Drift & $a^i$ & $-0.09$ & Effective revenue, $\times$ wholesale value & $\Pi_{\mathrm{eff}}^i$ & 3.4 \\
Common factor & $\rho$ & 0.28 & Marginal value of a tonne, USD/t & $c_{e,\theta}^i\partial_p\Vbar^i$ & 335 / 671 / 366 \\
Discount rate & $r$ & 0.028 & Replacement share & $\delta$ & 0.70 (0.09) \\
Default intensity & $\zeta_{\mathrm{ref}}^i$ & 0.0112 & Pass-through & $\psi$ & 0.43 (0.17) \\
\bottomrule
\end{tabular}
\begin{tablenotes}[flushleft]
\footnotesize
\item Values are firm medians.
\end{tablenotes}
\end{threeparttable}
\end{table}

\begin{figure}[H]
\centering
\begin{subfigure}[t]{0.3\linewidth}
\centering
\includegraphics[width=\linewidth]
{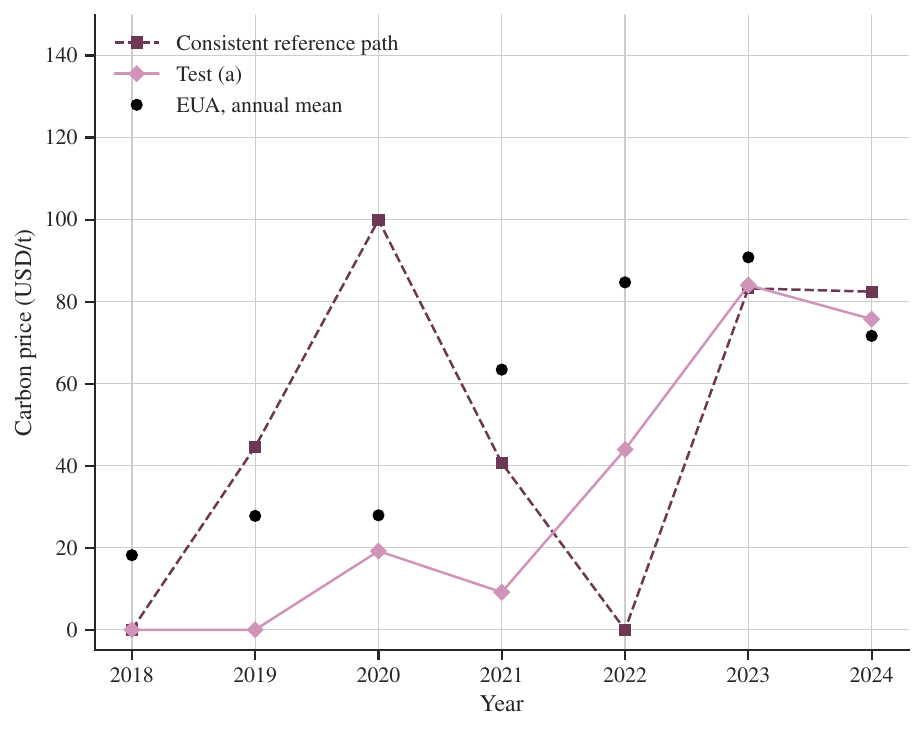}
\caption{Test (a).}
\label{fig:carbon_price_a_calibration}
\end{subfigure}
\hfill
\begin{subfigure}[t]{0.3\linewidth}
\centering
\includegraphics[width=\linewidth]
{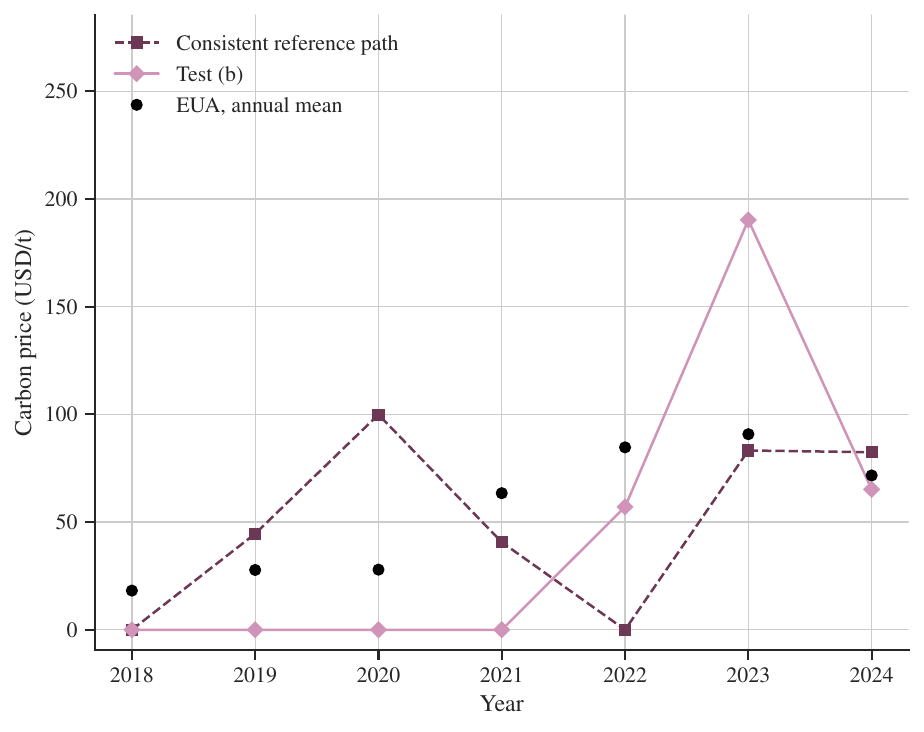}
\caption{Test (b).}
\label{fig:carbon_price_b_calibration}
\end{subfigure}
\hfill
\begin{subfigure}[t]{0.3\linewidth}
\centering
\includegraphics[width=\linewidth]
{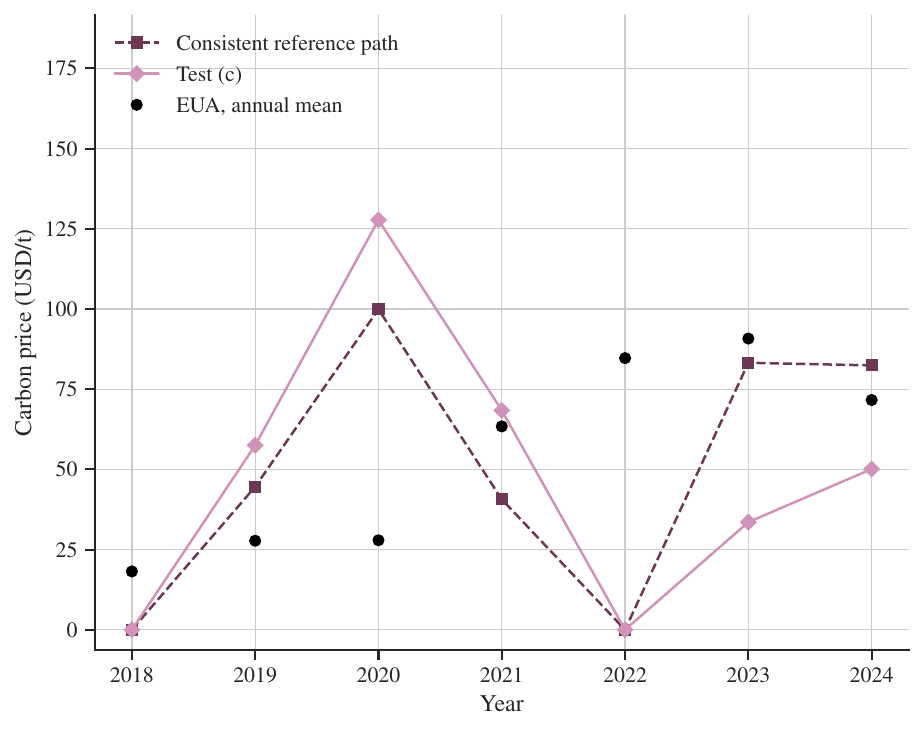}
\caption{Test (c).}
\label{fig:carbon_price_c_calibration}
\end{subfigure}
\caption{Three diagnostic tests of the carbon price calibration results against the linearized case with the consistent reference path.}
\label{fig:carbon_price_robust_calibration}
\end{figure}

\begin{figure}[H]
\centering
\includegraphics[width=0.7\linewidth]
{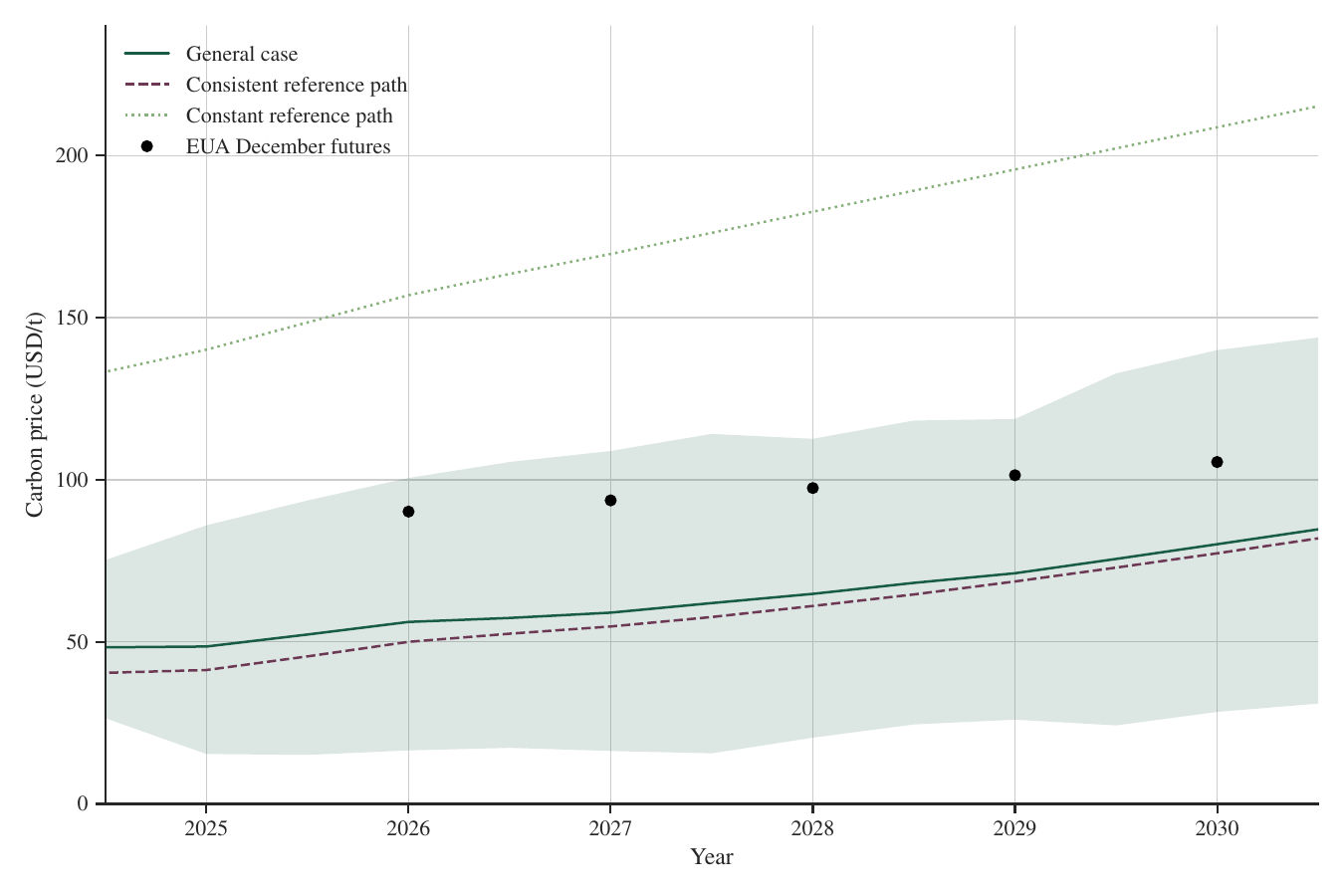}
\caption{Carbon price under the legislated cap path between 2025 and 2030.}
\label{fig:carbon_price_prediction}
\end{figure}

\begin{figure}[H]
\centering
\includegraphics[width=0.7\linewidth]
{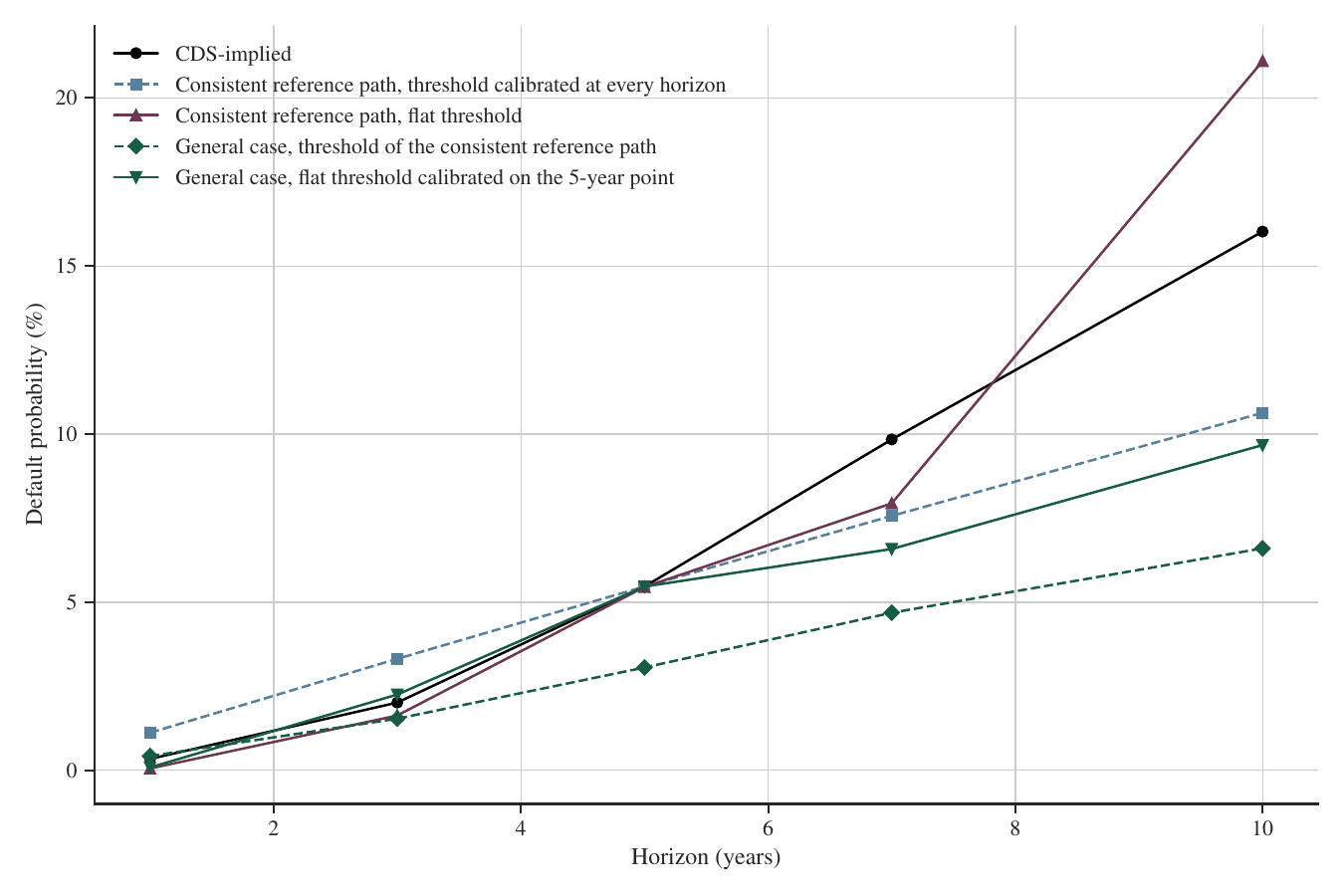}
\caption{Median default probability under the legislated cap path.}
\label{fig:default_probability_prediction}
\end{figure}

Since the calibration results show that the linearized case with the consistent reference path captures the average level of the EUA spot prices over 2018 to 2024 and the trend of the futures prices, we use the calibrated parameters in this case to check the accuracy of the default probability. The default probability $\DP_t^i$ remains the one defined in \eqref{eq:default_probability}. In the structural model of \citet{mertonPricingCorporateDebt1974}, it is the probability of default on a debt maturing at $t$, so that it can be compared with the cumulative default probability at tenor $t$ implied by the CDS spreads, in the same way as the term structure of expected default frequencies is compared with cumulative default rates. Since it ignores a default before $t$, it is a lower bound of the first-passage probability. Two thresholds are used in this section. The default threshold $L^i(t)$ in \eqref{eq:default_threshold_approx} is calibrated at every horizon $t$ under the legislated cap path, with the constant default intensity $\zeta_{\mathrm{ref}}$ implied by the 5-year CDS spread for the 12 firms with a CDS curve, and set to the cell median for the other 23 firms. Under this threshold, the default probability under the legislated cap path reproduces $1-\ee^{-\zeta_{\mathrm{ref}}t}$ by construction, so that it cannot serve as a validation of the model. For the validation, we therefore set $L^i(t)$ at its 5-year value and hold it flat over the horizon, and compare the resulting default probability at the other tenors with the CDS-implied one. Figure~\ref{fig:default_probability_prediction} shows that the default probability with the flat threshold is close to the CDS-implied one up to seven years, while the constant-intensity curve delivered by the threshold calibrated at every horizon is flatter than the CDS curve. However, at ten years, the default probability with the flat threshold is much higher than the CDS-implied one. This is because the reference output and emissions follow the cap towards its exit in 2040, so that the firm value declines deterministically through the 2030s against a debt that the flat threshold holds fixed, which forces the default probability to rise steeply between seven and ten years. 

\begin{figure}[H]
\centering
\begin{subfigure}[t]{0.45\linewidth}
\centering
\includegraphics[width=\linewidth]
{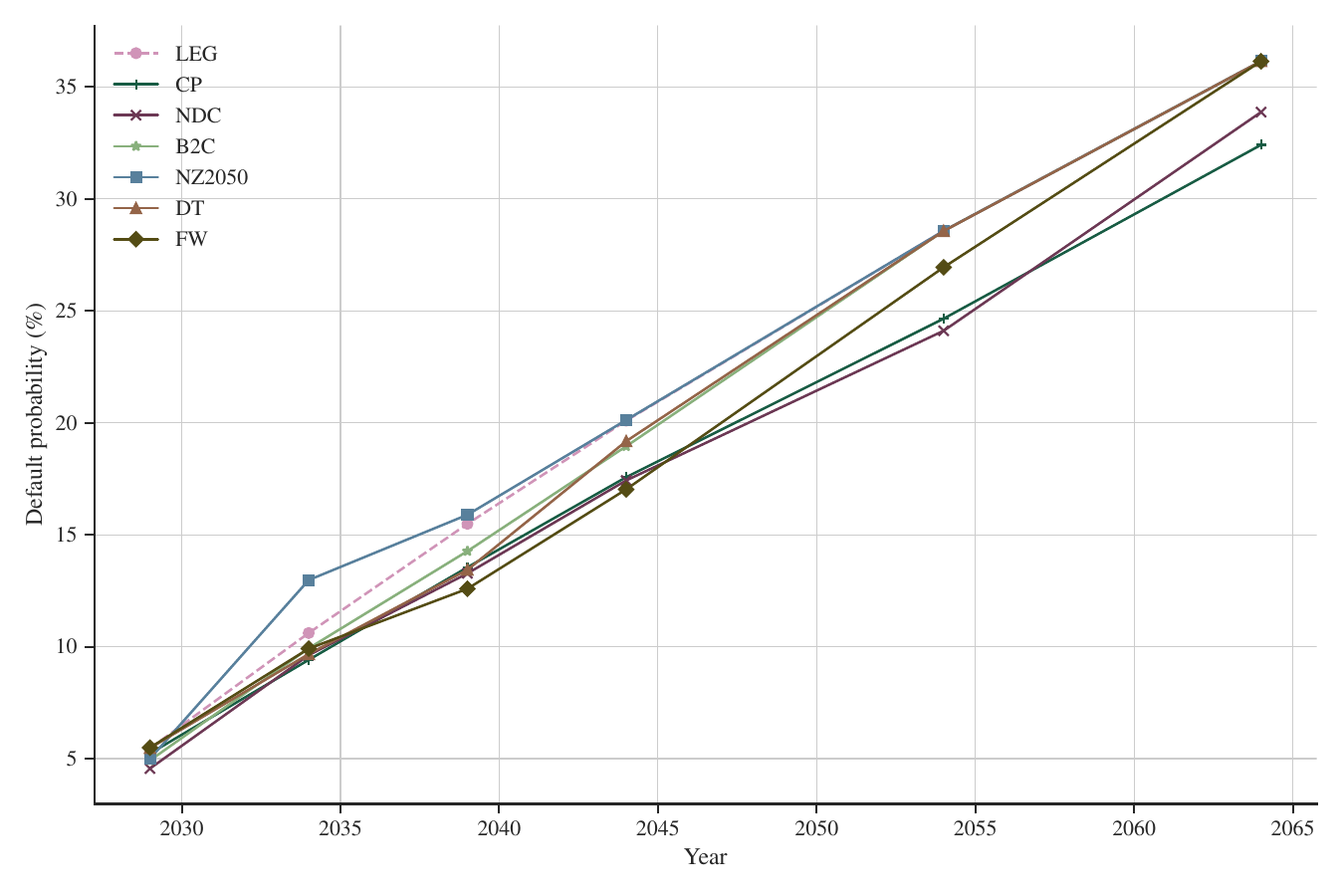}
\caption{The linearized case with the consistent reference path.}
\label{fig:default_probability_scenarios_linearized}
\end{subfigure}
\hfill
\begin{subfigure}[t]{0.45\linewidth}
\centering
\includegraphics[width=\linewidth]
{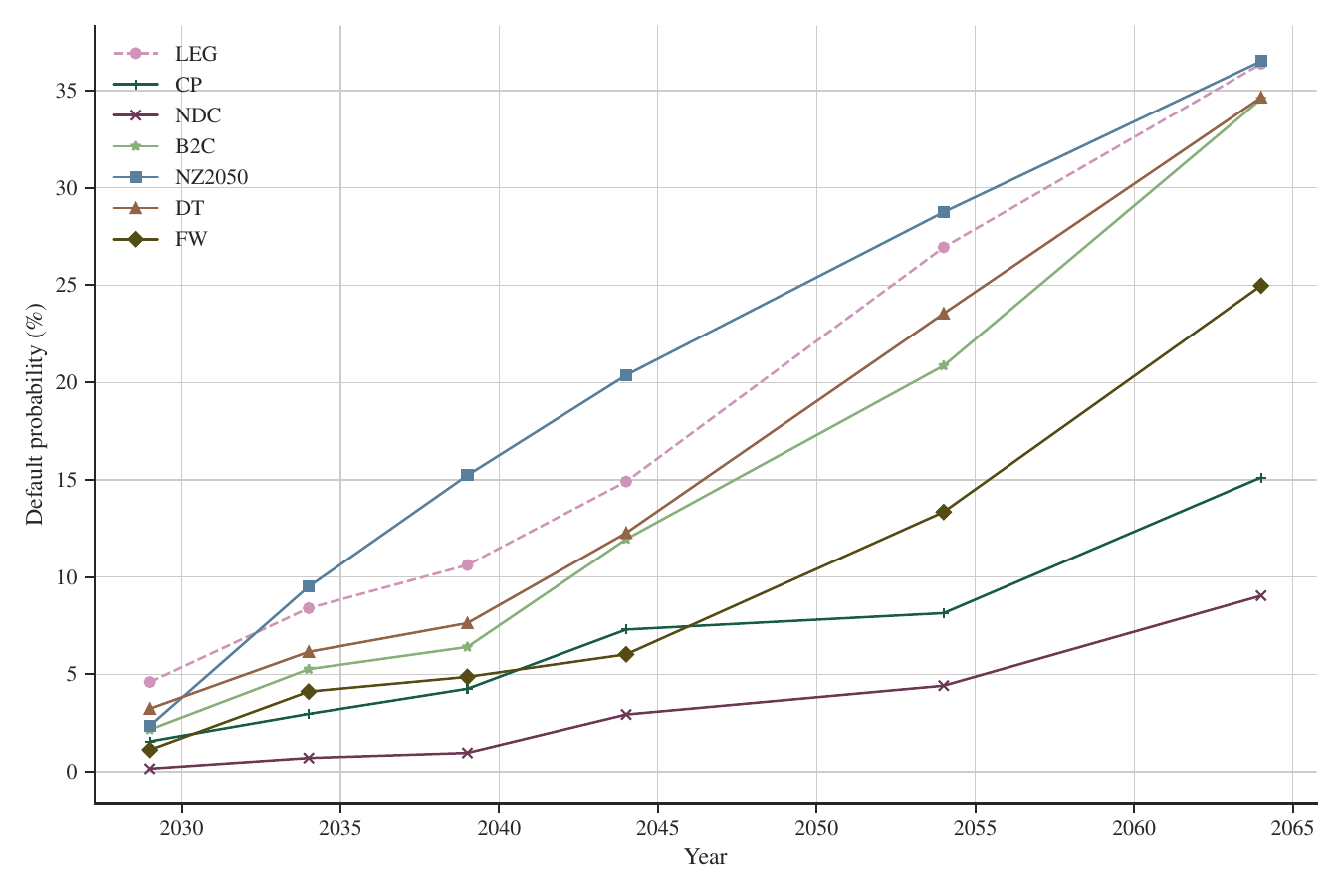}
\caption{The general case.}
\label{fig:default_probability_scenarios_general}
\end{subfigure}
\caption{Median default probability under different NGFS scenarios.}
\label{fig:default_probability_scenarios}
\end{figure}

\begin{figure}[H]
\centering
\begin{subfigure}[t]{0.45\linewidth}
\centering
\includegraphics[width=\linewidth]
{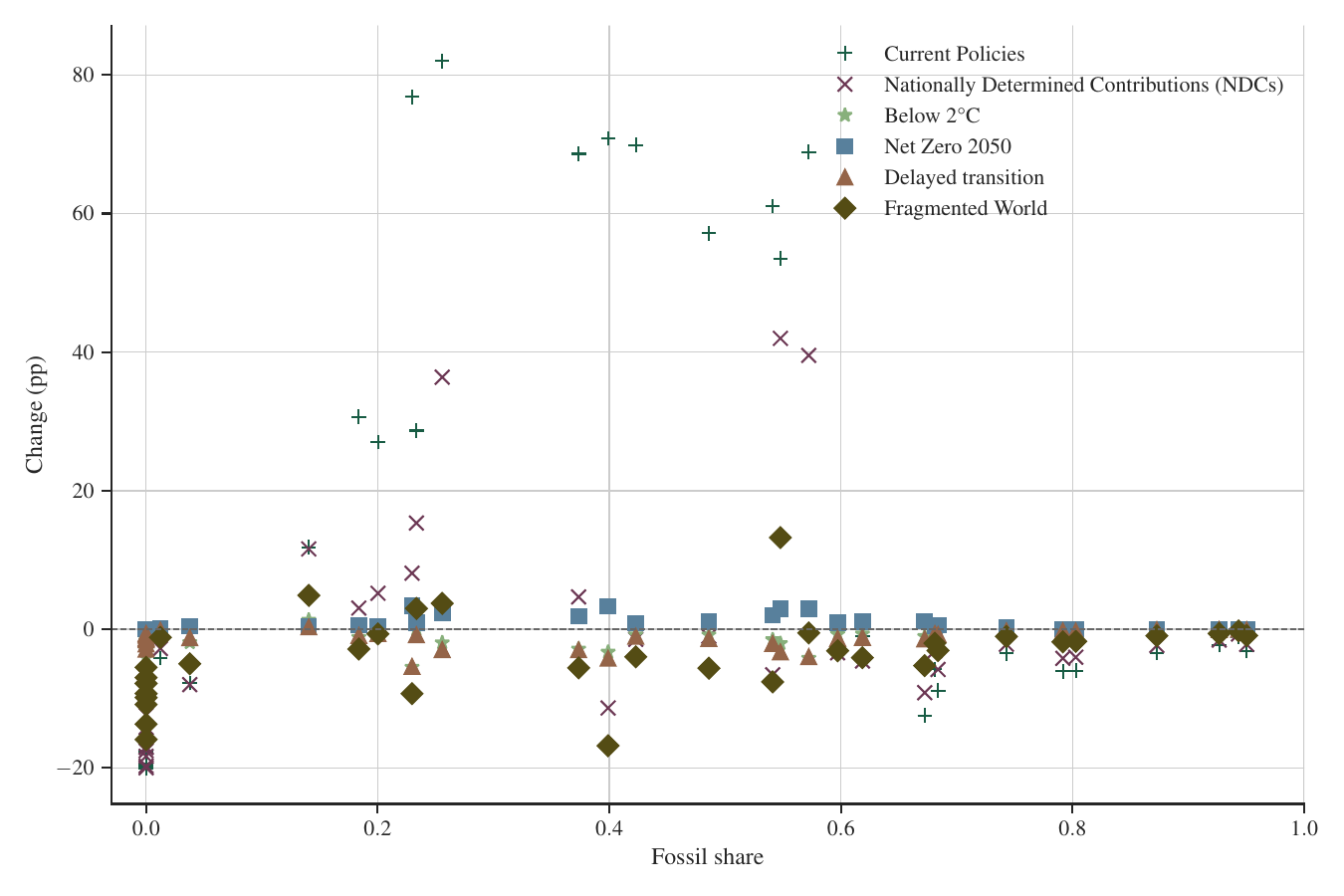}
\caption{The linearized case with the consistent reference path.}
\label{fig:change_default_probability_scenarios_linearized}
\end{subfigure}
\hfill
\begin{subfigure}[t]{0.45\linewidth}
\centering
\includegraphics[width=\linewidth]
{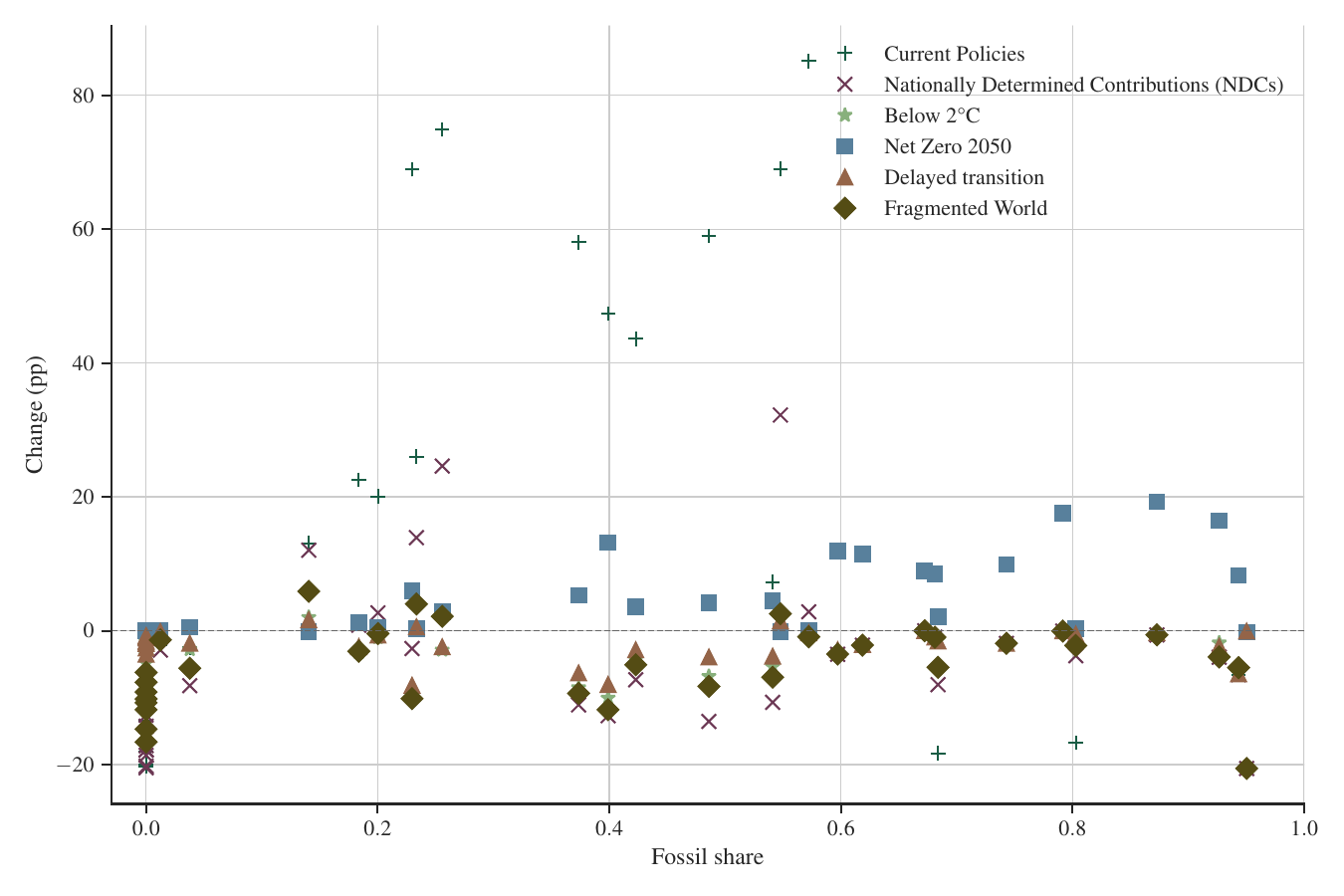}
\caption{The general case.}
\label{fig:change_default_probability_scenarios_general}
\end{subfigure}
\caption{Change of 20-year default probability against the legislated cap path under different NGFS scenarios.}
\label{fig:change_default_probability_scenarios}
\end{figure}

\begin{figure}[H]
\centering
\includegraphics[width=0.7\linewidth]
{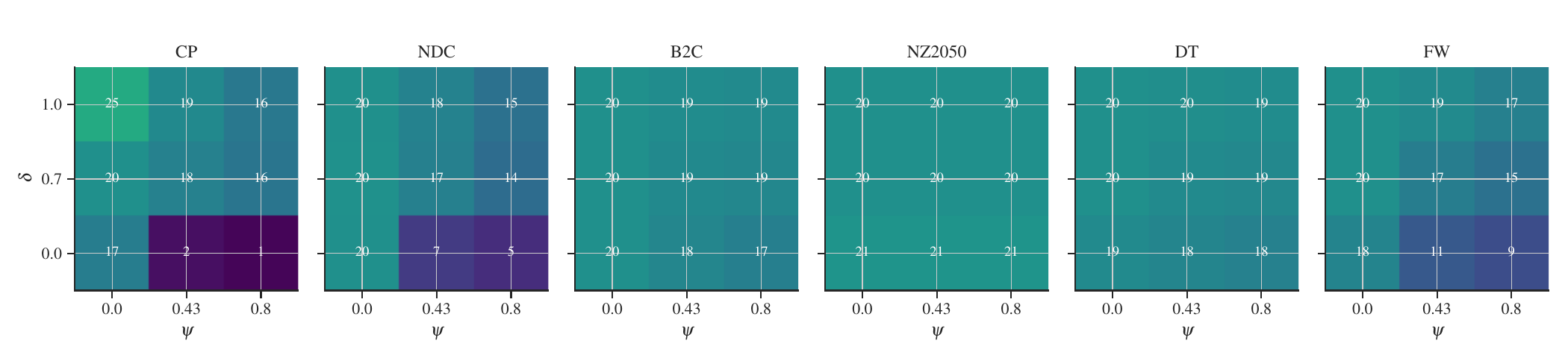}
\caption{Median default probability at twenty years in the linearized case with the consistent reference path on the grid of $(\delta, \psi)$.}
\label{fig:default_probability_parameter_grid}
\end{figure}

The calibration results have shown that the model reproduces the average level of the carbon price and its response to the cap, but not its variation from year to year. In the linearized case with the consistent reference path, the carbon price responds to the cap, whereas the median default probability responds much less (Table~\ref{tab:calib_scenarios}). The carbon price differs across the scenarios already in 2030 and diverges further by 2040, while the median default probability at twenty years stays within about three percentage points of the Legislated cap path in every scenario. Since $\DP_t^i$ in \eqref{eq:default_probability} depends only on the cash flows after time $t$ and on the distribution of the production state at time $t$, whose dependence on past emissions decays at the rate $b$, the scenarios that have retired the fossil fleet by 2044 approach the Legislated cap path at twenty years and coincide with it at thirty years, and the scenarios that keep part of the fleet look slightly safer, because a running plant keeps its margin while the exit replaces only seven tenths of it. However, at ten years, Net Zero 2050 lies above the Legislated cap path, since its cap falls to zero within a decade and the 2030s carry the cost of the exit. The difference shown in Figure~\ref{fig:default_probability_scenarios} is a matter of timing, not of level. Moreover, we find that a slow transition is worth to a firm what it keeps, the margin of its running plants and the carbon rent in the electricity price, less what it forgoes, the replacement that the exit would have brought. Thus, the sign of the effect is set by whether the firm's fossil margin exceeds the value of the output that would replace it, positive for the renewable producers and the coal fleets, negative for the gas-fired integrated utilities, as shown in Figure~\ref{fig:change_default_probability_scenarios}.
Figure~\ref{fig:default_probability_parameter_grid} varies $\delta$ and $\psi$ in the linearized case with the consistent reference path. Given a positive pass-through, what the model leaves open is therefore the size of the sector-level effect, not its sign, and that size turns on the two parameters that the model does not price.

\begin{table}[H]
\centering
\begin{threeparttable}
\caption{Equilibrium carbon price and default probability under different NGFS scenarios.}
\label{tab:calib_scenarios}
\small
\setlength{\tabcolsep}{5pt}
\begin{tabular}{lcccccc}
\toprule
& \multicolumn{2}{c}{Carbon price, USD/t} & \multicolumn{4}{c}{Median $\DP_t^i$, \%} \\
\cmidrule(lr){2-3} \cmidrule(lr){4-7}
Scenario & 2030 & 2040 & 5 years & 10 years & 20 years & 30 years \\
\midrule
\multicolumn{7}{l}{\textit{Linearized case with the consistent reference path}} \\
Legislated cap & 77 & 860 & 5.5 & 10.6 & 20.1 & 28.6 \\
Current Policies & 79 & 89 & 5.2 & 9.4 & 17.6 & 24.7 \\
NDCs & 134 & 123 & 4.6 & 9.6 & 17.4 & 24.1 \\
Below 2$^\circ$C & 91 & 223 & 4.9 & 9.9 & 19.0 & 28.6 \\
Net Zero 2050 & 199 & 805 & 5.0 & 13.0 & 20.1 & 28.6 \\
Delayed Transition & 63 & 239 & 5.5 & 9.7 & 19.2 & 28.6 \\
Fragmented World & 69 & 125 & 5.5 & 9.9 & 17.0 & 26.9 \\
\midrule
\multicolumn{7}{l}{\textit{General case}} \\
Legislated cap & 80 & 1092 & 4.6 & 8.4 & 14.9 & 26.9 \\
Current Policies\tnote{*} & 56 & 59 & 1.6 & 3.0 & 7.3 & 8.1 \\
NDCs & 148 & 125 & 0.2 & 0.7 & 2.9 & 4.4 \\
Below 2$^\circ$C & 100 & 192 & 2.2 & 5.3 & 11.9 & 20.9 \\
Net Zero 2050 & 199 & 946 & 2.4 & 9.5 & 20.4 & 28.8 \\
Delayed Transition & 71 & 227 & 3.2 & 6.2 & 12.3 & 23.5 \\
Fragmented World & 82 & 122 & 1.1 & 4.1 & 6.0 & 13.3 \\
\bottomrule
\end{tabular}
\begin{tablenotes}[flushleft]
\footnotesize
\item $\delta = 0.7$, and $\psi = 0.43$.
\item [*] Algorithm~\ref{alg:equilibrium_carbon_price} does not converge under Current Policies, and the values are those of the last iteration.
\end{tablenotes}
\end{threeparttable}
\end{table}

\section{Conclusion}
\label{sec:conclusion_climate_risk}

In this chapter, we extended the framework of \citet{bourgeyBridgingSocioeconomicPathways2024,bourgeyEfficientSharedSocioeconomic2026} to an ETS, in which the carbon price is determined by the emission cap and the emissions of all firms in a sector-region cell. We solved the stochastic control problem of the firm by a verification argument, proved the existence of the equilibrium carbon price, and characterized the cases where the cap is slack, binding, or zero. In the linearized case, the equilibrium carbon price coincides with the anticipated one, and the default probability is given in closed form. Numerically, we showed that the ETS makes the aggregate emission follow the cap once it binds, and that the linearized case with a consistent reference path reproduces most of the carbon price of the general case. We then calibrated the model to 35 European power producers under the EU ETS. The calibrated model reproduces the average level of the EUA price over 2018 to 2024, but not its variation from year to year, which comes from setting the energy cost of the period against a forward-looking value of output. Under the NGFS scenarios, a slow transition lowers the default probability of renewable producers and coal fleets and raises that of some gas-fired utilities, so that the sign of its effect is set by the energy mix of the firm. The extension to a production technology in which the current energy input enters the current output, and the calibration of the physical risk, are left for future research.

\section*{Acknowledgements}
This research was partially supported by the Natural Sciences and Engineering Research Council of Canada (NSERC) Discovery Grant RGPIN-2026-05644.

\printbibliography

\appendix
\section*{Appendix}
\label{sec:appendix_climate_risk}

\section{Proof of Theorem~\ref{thm:verification_of_optimal_emission}}
\label{sec:proof_of_thm_verification_of_optimal_emission}

To prove this theorem, it suffices to show that 
\eqstar{
\Jc^i \bigl(\hbgam^i \bigr) - \Jc^i \bigl(\bgam^i \bigr) 
= \Eb \biggl[\int_0^\infty \ee^{-r t}\, \Bigl(
\Hc_t^i \Bigl( \hp_t^i, \hbgam^i \bigl(t, \hp_t^i \bigr), \Pbar_t^g, 0 \Bigr)
- \Hc_t^i \Bigl( p_t^i, \bgam^i \bigl(t, p_t^i \bigr), \Pbar_t^g, 0 \Bigr)
\Bigr) \dd t \biggr]
\ge 0.
}
Since $\hbgam^i (t, p_t^i)$ maximizes the Hamiltonian $\Hc_t^i$ in HJB equation \eqref{eq:hjb_equation} at $(t, p_t^i)$ and $\bgam_t^i \in \Bc_t^i$, we have, for each $t \ge 0$, 
\eqstar{
\Delta \Hc_t^i := \Hc_t^i \bigl( p_t^i, \hbgam^i (t,p_t^i), \Pbar_t^g, \partial_p \Vbar (t, p_t^i) \bigr) 
- \Hc_t^i \bigl( p_t^i, \bgam^i (t, p_t^i), \Pbar_t^g, \partial_p \Vbar (t, p_t^i) \bigr) 
\ge 0.
}
Applying Itô's formula to the value function $\Vbar^i$ in \eqref{eq:value_func} with the value $\hp_t^i$, we obtain
\eqstar{
\dd \Vbar^i \bigl(t, \hp_t^i \bigr) &= 
\partial_t \Vbar^i \bigl(t, \hp_t^i \bigr) \dd t 
+ \partial_p \Vbar^i \bigl(t, \hp_t^i \bigr) \dd \hp_t^i 
+ \frac12 \partial_p^2 \Vbar^i \bigl(t, \hp_t^i \bigr) \bigl(\dd \hp_t^i \bigr)^2\\
&= \Bigl(\partial_t \Vbar^i \bigl(t, \hp_t^i \bigr) 
+ \frac{1}{2} (\sigma^i)^2 \partial_p^2 \Vbar^i \bigl(t, \hp_t^i \bigr) \Bigr) \dd t
+ \mu_t^i \Bigl(\hp_t^i, \hbgam^i \bigl(t, \hp_t^i \bigr) \Bigr) \partial_p \Vbar^i \bigl(t, \hp_t^i\bigr) \dd t 
+ \sigma^i \partial_p \Vbar^i \bigl(t, \hp_t^i\bigr) \dd W_t^i.
}
Applying the same procedure to $\Vbar^i$ with the value $p_t^i$, we obtain 
\eqstar{
\dd \Bigl(\ee^{-r t}\, \Bigl(\Vbar^i \bigl(t, \hp_t^i \bigr) - \Vbar^i \bigl(t, p_t^i \bigr) \Bigr) \Bigr) 
&= \ee^{-r t}\, \Bigl( - 
\Hc_t^i \Bigl( \hp_t^i, \hbgam^i \bigl(t, \hp_t^i \bigr), \Pbar_t^g, 0 \Bigr)
+ \Hc_t^i \Bigl( p_t^i, \bgam^i \bigl(t, p_t^i \bigr), \Pbar_t^g, 0 \Bigr)
+ \Delta \Hc_t^i \Bigr) \dd t \\
&\quad+ \ee^{-r t}\, \sigma^i  \Bigl(\partial_p \Vbar^i \bigl(t, \hp_t^i\bigr) - \partial_p \Vbar^i \bigl(t, p_t^i \bigr) \Bigr) \dd W_t^i.
}
Integrating over $[0, T]$ and taking the expectation, we have
\eqstar{
\Jc_T^i \bigl(\hbgam^i \bigr) - \Jc_T^i \bigl(\bgam^i \bigr) 
&= - \Eb \Bigl[\ee^{-r T}\, \Bigl(\Vbar^i \bigl(T, \hp_T^i \bigr) - \Vbar^i \bigl(T, p_T^i \bigr) \Bigr) \Bigr]
+ \Eb \Bigl[\int_0^T \ee^{-r t}\,\Delta \Hc_t^i \dd t \Bigr],
}
where $\Jc_T^i$ denotes the objective function in \eqref{eq:obj_func} with the time integral truncated at $T$, and stochastic integrals has zero expectation by Assumption~\ref{assump:log_production}.
By the transversality condition in \eqref{eq:transversality_condition}, the first term on the right-hand side converges to zero as $T \to \infty$, and $\Jc_T^i$ converges to $\Jc^i$ by Assumption~\ref{assump:integrability} and the dominated convergence theorem. Since $\Delta \Hc_t^i \ge 0$ for all $t \ge 0$, we conclude that
\eqstar{
\Jc^i \bigl(\hbgam^i \bigr) - \Jc^i \bigl(\bgam^i \bigr) = \Eb \Bigl[\int_0^\infty \ee^{-r t}\,\Delta \Hc_t^i \dd t \Bigr] \ge 0,
}
and $\hbgam^i$ is indeed the solution to the stochastic control problem in \eqref{eq:optimization_problem}.

\section{Proof of Theorem~\ref{thm:existence_uniqueness_carbon_price}}
\label{sec:proof_of_thm_existence_uniqueness_carbon_price}

Since $\beta_{e,\theta,t}^i > 0$, the unconstrained maximizer in \eqref{eq:unconstrained_max_emission_func} is affine and strictly decreasing in $\Pc^g$. Moreover, as a projection of an affine function onto $[0, \eta_{e,t}^{\mathrm{max}}]$, each $\hgamei$ is continuous in $\Pc^g$, and so is $\hGam^g$. The optimal emission strategy $\hgamei$ is then non-increasing in $\Pc^g$. For any $\Pc \ge 0$, we have
\eqstar{
0 \le \hGam^g (t, \bp,\Pc^g)
\le \sum_{i \in \Ic^g} \sum_{e \in \Ec^i} \eta_{e,t}^{\mathrm{max}}
< \infty.
}
As $\Pc^g \rightarrow \infty$, each unconstrained maximizer tends to negative infinity and is truncated to zero. Hence, $\lim_{\Pc \rightarrow \infty} \hGam^g(t, \bp, \Pc^g) = 0$.

Suppose that $\hGam^g(t, \bp, 0) < \tGam_t^g$. Then, $\Pc^{g, *}(t, \bp)=0$ satisfies the market-clearing conditions in \eqref{eq:clearing_condition}, and the cap is slack. Since $\hGam^g(t, \bp, \Pc^g)$ is non-increasing in $\Pc^g$, we have
\eqstar{
\hGam_t^g(\bp, \Pc^g)
\le \hGam_t^g(\bp, 0)
< \tGam_t^g,
}
for any $\Pc^g \ge 0$. The complementary slackness rules out any strictly positive equilibrium price. Consequently, the equilibrium carbon price is uniquely given by $\Pc^{g, *} (t, \bp) = 0$.
Suppose instead that $0 < \tGam_t^g < \hGam^g(t, \bp, 0)$. The aggregate optimal emission satisfies
\eqstar{
\hGam^g (t, \bp, 0) > \tGam_t^g,
\qquad
\lim_{\Pc^g \rightarrow \infty} \hGam_t^g (\bp, \Pc^g) = 0 
< \tGam_t^g.
}
By the intermediate value theorem, there exists at least one
$\Pc^{g, *} (t, \bp) > 0$, such that
\eqstar{
\hGam_t^g \bigl( \bp, \Pc_t^{g, *} (\bp) \bigr) =
\tGam_t^g.
}
This price satisfies \eqref{eq:clearing_condition}, and hence the cap binds. Moreover, the equilibrium carbon price cannot be zero in this case, since the aggregate optimal emission at $\Pc^{g, *} (t, \bp) = 0$ exceeds the cap. Suppose now that $\tGam_t^g = 0 < \hGam^g (t, \bp, 0)$. Since $\hGam^g \ge 0$, the market-clearing conditions in \eqref{eq:clearing_condition} reduce to $\hGam^g (t, \bp, \Pc^{g, *} (t, \bp)) = 0$, i.e., $\hgamei (t, p^i, \Pc^{g, *} (t, \bp)) = 0$, for every $i \in \Ic^g$ and $e \in \Ec^i$. By \eqref{eq:unconstrained_max_emission_func} and \eqref{eq:optimal_emission_func_per_source}, this holds if and only if $c_{e, \theta, t}^i \partial_p \Vbar^i (t, p^i) - \alpha_{e, \theta, t}^i \le \Pc^{g, *} (t, \bp)$ for all such pairs, that is, $\Pc^{g, *} (t, \bp) \ge \Pc^{g, c} (t, \bp)$. Moreover, $\hGam^g (t, \bp, 0) > 0$ implies that $c_{e, \theta, t}^i \partial_p \Vbar^i (t, p^i) - \alpha_{e, \theta, t}^i > 0$ for at least one such pair, so that $\Pc^{g, c} (t, \bp) > 0$.
Finally, suppose that there exists at least one pair $(i,e)$ at an equilibrium carbon price $\Pc^{g, *} (t, \bp)$, such that
\eqstar{
0 < \hgameii \bigl(t, p^i,\Pc^{g, *} (t, \bp) \bigr) < \eta_{e,t}^{\mathrm{max}}.
}
The optimal emission is then strictly decreasing in $\Pc^{g, *}(t, \bp)$ for this pair. Suppose another equilibrium price $\Pc' \neq \Pc^{g, *} (t, \bp)$ existed. By complementary slackness, any strictly positive equilibrium price yields aggregate optimal emission equal to $\hGam_t^g$. If one of the two prices is zero, $\Pc^{g, *} (t, \bp) = 0 < \Pc'$, then the market-clearing conditions and the monotonicity give $\tGam_t^g \ge \hGam^g(t, \bp, 0) \ge \hGam^g(t, \bp, \Pc') = \tGam_t^g$. Hence, in all cases, both prices yield aggregate optimal emission equal to $\tGam_t^g$. Since $\Pc^g \mapsto \hGam^g(t, \bp, \Pc)$ is non-increasing, this function would be constant on the interval between the two equilibrium carbon prices. This contradicts its strict decrease in $\Pc^{g, *}(t, \bp)$. Thus, the equilibrium carbon price is unique.

\end{document}